\documentclass[aps,prb,reprint,twocolumn,amsfonts,amssymb,amsmath,floatfix,tightenlines]{revtex4-1}

\usepackage{graphicx}
\usepackage{pdfpages}
\usepackage{relsize}
\usepackage{bm,times}
\usepackage[colorlinks=true,citecolor=blue,urlcolor=blue]{hyperref}
\numberwithin{equation}{section}

\makeatletter
\AtBeginDocument{\let\LS@rot\@undefined}
\makeatother

\def\A{\mathrm{A}}
\def\b{\mathrm{b}}
\def\B{\mathrm{B}}

\def\p{\partial}
\def\mb{\mathbf}

\def\dia{\mathrm{dia}}
\def\AL{\mathrm{AL}}
\def\Im{\mathrm{Im}}
\def\R{\mathrm{R}}
\def\Re{\mathrm{Re}}
\def\E{\mathrm{E}}
\def\H{\mathrm{H}}
\def\MT{\mathrm{MT}}
\def\VV{\mathrm{VV}}
\def\pair{\mathrm{pair}}
\def\Pauli{\mathrm{Pauli}}
\def\phikq2{\varphi_\mathbf{k-q/2}^2}

\def\phik{\varphi_\mathbf{k}}

\begin{document}
\title{Cuprate diamagnetism in the presence of a pseudogap: Beyond the standard fluctuation formalism}

\author{Rufus Boyack$^{1}$}
\author{Qijin Chen$^{1,2}$} 
\author{A. A. Varlamov$^{3}$}
\author{K. Levin$^{1}$}
\affiliation{$^1$James Franck Institute, University of Chicago, Chicago, Illinois 60637, USA}
\affiliation{$^2$Zhejiang Institute of Modern Physics and Department of Physics, Zhejiang University, Hangzhou, Zhejiang 310027, China}
\affiliation{$^3$ CNR-SPIN (Instituto Superconduttori, Materiali Innovativi e Dispositivi) Viale del Politecnico 1, I-00133, Rome, Italy}

\begin{abstract}
  It is often claimed that among the strongest evidence for preformed-pair physics in the cuprates 
  are the experimentally observed large values for the diamagnetic susceptibility and Nernst coefficient.
  These findings are most apparent in the underdoped regime, where a pseudogap is also evident. 
  While the conventional (Gaussian) fluctuation picture has been applied to address these results, 
  this preformed-pair approach omits the crucial effects of a pseudogap.  
  In this paper we remedy this omission by computing the 
  diamagnetic susceptibility and Nernst coefficient in the presence of a normal state gap. 
  We find a large diamagnetic response for a range of temperatures much higher than the transition temperature.
  In particular, we report semi-quantitative agreement with the measured
  diamagnetic susceptibility onset temperatures, over the entire range of hole dopings.
  Notable is the fact that at the lower critical doping of the superconducting dome, 
  where the transition temperature vanishes and the pseudogap onset temperature remains large, 
  the onset temperature for both diamagnetic and transverse thermoelectric transport coefficients tends to zero. 
  Due to the importance attributed to the cuprate diamagnetic susceptibility and Nernst coefficient, 
  this work helps to clarify the extent to which pairing fluctuations are a component of the cuprate pseudogap.
  \end{abstract}

\maketitle

\section{Introduction and overview of results}

Establishing the origin of the cuprate pseudogap is a longstanding
problem in the field of high-$T_c$ superconductivity~\cite{KeimerReview}. 
At its heart is the central issue of whether this pseudogap arises from precursor
superconductivity or from an alternative order parameter. 
In support of this latter viewpoint is an increasing number of experiments
showing evidence for (finite-range) charge-density-wave order~\cite{ghiringhelli,Comin}. 
With the application of a magnetic field this order appears to be stabilized~\cite{TaoWu}, 
although there is evidence the pseudogap itself remains intact.

On the other hand, there is also mounting support for the first viewpoint: 
the origin of the cuprate pseudogap is a precursor-pairing scenario. 
The conventional fluctuation formalism~\cite{VarlamovBook,GalitskiVarlamov}, 
used to support preformed-pair physics in the cuprates, provides a natural explanation
for the anomalously large diamagnetic susceptibility and large Nernst coefficient observed above $T_{c}$~\cite{Ong2000,Ong2010}.
However, this standard fluctuation theory is a weak-fluctuation approach that largely ignores the substantial
normal state gap, which is of fundamental interest here and observed in a variety of experiments.
As a result it is not expected to be valid in the doping regimes where such a gap is present.

This leads to the challenge addressed in this paper of going
beyond the weak-fluctuation formalism within a precursor-pairing approach. 
Here we compute the diamagnetic susceptibility and transverse thermoelectric coefficient
by applying a BCS--BEC crossover~\cite{Chen_99,Chen_2005,Chen_2014} scheme, above $T_c$.
This crossover scenario, built on a natural generalization~\cite{LeggettBook} of the BCS ground state, 
incorporates the variation from weak to strong attractive interactions between the underlying fermionic constituents. 
In this context, Leggett~\cite{LeggettNature} states in his summary article about the copper oxide superconductors: 
``The small size of the cuprate pairs puts us in the intermediate regime of the so-called BEC-BCS crossover". 
It is important to emphasize at the outset that the pseudogap phase for the $d$-wave cuprates 
is also well outside the BEC regime~\cite{Chen_99,Comment1}. 
Rather, the pseudogap phase represents an intermediate state between the BCS and BEC regimes.

There is a substantial body of literature on the diamagnetic susceptibility and Nernst coefficient in the cuprates. 
The early seminal experiments~\cite{Ong2000,Ong2006} first associated the Nernst response with vortex excitations. 
The diamagnetic susceptibility~\cite{Ong2010} was similarly
interpreted as reflecting some form of normal-state Cooper pairing.
More recent experimental emphasis has been on the inter-play of vortex
excitations with charge-density-wave order~\cite{OngPNAS,Taillefer}.
Although other alternatives have been contemplated~\cite{Nayak}, 
most of the theories addressing these experiments have been based on a preformed-pair formalism. 
This preformed-pair approach is associated with superconducting fluctuation contributions~\cite{VarlamovBook} 
to the diamagnetic~\cite{Patton} and Nernst~\cite{BehniaReview} responses.

However, in the context of transport the preformed-pair scenario has dealt
almost exclusively with a weak-fluctuation formalism~\cite{VarlamovBook}, 
considering only the lowest order fluctuation contributions to the electromagnetic (EM) response. 
In the absence of impurities, these consist of two density of states (DOS),
one Maki-Thompson (MT), and two identical Aslamazov-Larkin (AL) diagrams. 
For the diamagnetic susceptibility and the Nernst coefficient, 
it is found that the singular contribution arises from the
Aslamazov-Larkin diagrams~\cite{Aslamazov_1975,Ussishkin_2003,VarlamovBook}. 
These results can be equivalently derived from Gaussian-fluctuation theory,
which is associated with time dependent Ginzburg-Landau theory~\cite{Huse_2002}. 
There is also related work based on phase-only fluctuations~\cite{Ashvin_2007} within a two dimensional
BKT-like theory; phase fluctuations are thought to dominate their
amplitude counterparts in the vicinity of $T_{c}$~\cite{Sondhi}, 
and one presumes here that mobile vortices are the fundamental constituents.

\begin{figure*}
\includegraphics[width=4.5in,clip]{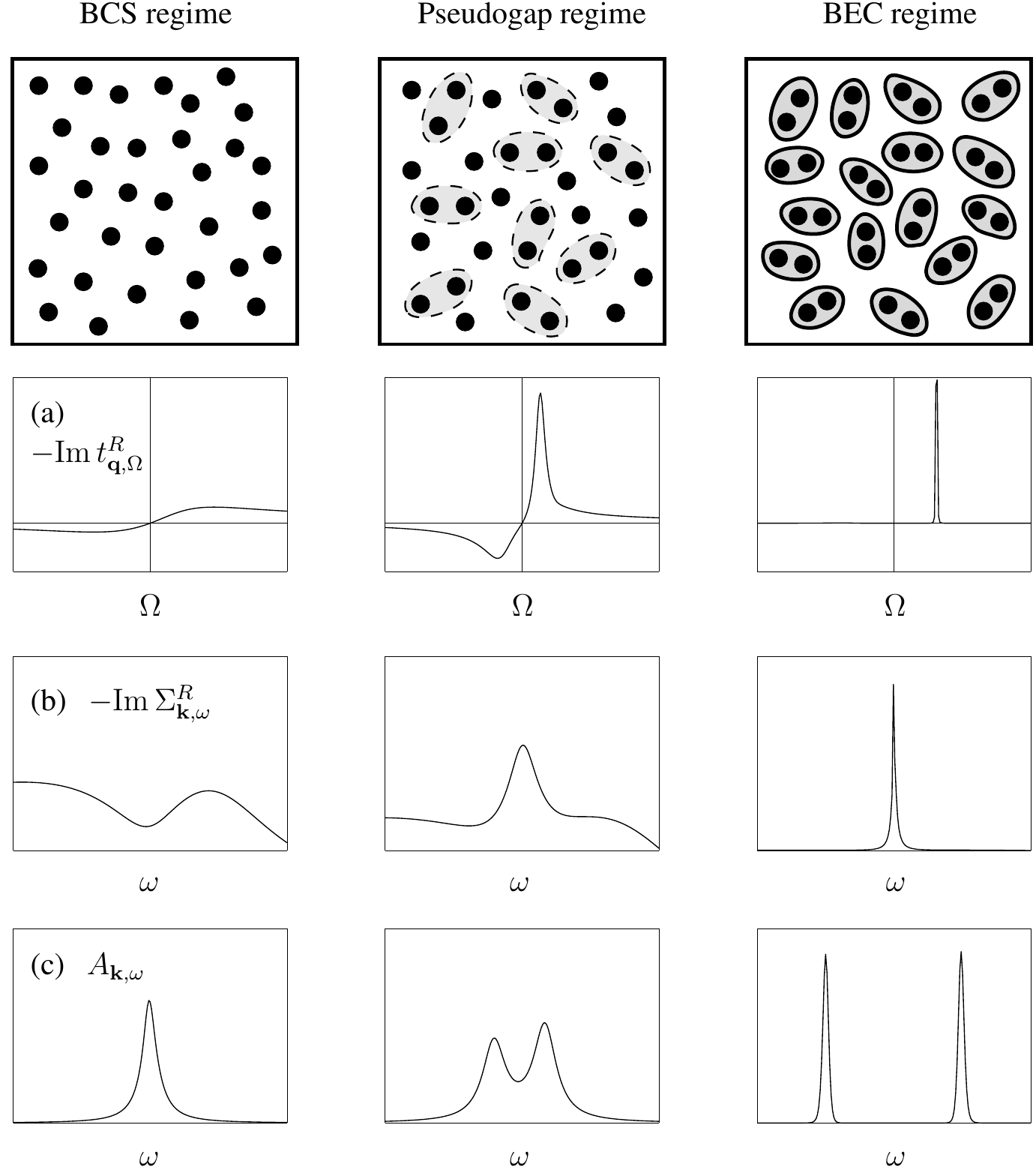}
\caption{Evolution from the weak-coupling BCS through the pseudogap to
  the strong-coupling BEC regimes for the excitations. The figure shows the corresponding 
  (row a) pair excitation spectrum $-\mathrm{Im}\ t(\Omega,\mb{q}=0)$, 
  (row b) imaginary part of the fermionic self energy $-\mathrm{Im}\ \Sigma(\omega,\mb{k})$, and
  (row c) fermionic spectral function $A(\omega,\mb{k})$ at the Fermi level for $T$ slightly above $T_{c}$. 
  This figure is taken from Ref.~\onlinecite{Maly1}.}
\label{fig:Maly}
\end{figure*}

Stronger pairing effects in transport have been included 
in differing contexts~\cite{Levchenko_2011,Wulin_2012,ShinaNernst}, 
all of which build on a fluctuation scenario. 
The authors of Ref.~\onlinecite{Levchenko_2011} introduced pseudogap self energy
effects in the standard Aslamazov-Larkin diagrams by correcting the so-called 
EM ``triangle" vertex, which represents an effective bosonic EM vertex.  
In Secs.~(\ref{sec:Diamagnetism_Approx}-\ref{sec:Heat_Diagrammatic})
of the paper this approach is discussed in more detail, where it is noted that 
correcting this vertex without simultaneously correcting the pair-propagator is inconsistent.
An alternative diagrammatic approach was studied in Ref.~\onlinecite{Wulin_2012}, 
which investigated the diamagnetic susceptibility associated with 
the fermionic quasi-particles in the presence of a pseudogap. 
This approach misses the essential physics of the bosonic fluctuations which, 
as Sec.~(\ref{sec:Diamagnetism_Approx}) shows,  are found to be the singular contribution.

Finally, the authors of Ref.~\onlinecite{ShinaNernst} introduced an extension of the Gaussian-fluctuation formalism~\cite{Huse_2002} by
computing the transport properties of independent, non-condensed bosons in contact with a Leggett-Caldeira particle bath. 
This bath leads to an inter-conversion with the bosons so that boson number is no longer fixed. 
The bath approach is a phenomenological treatment of transport in which the
reservoir yields finite lifetime effects, and simulates the role of paired fermions or composite bosons.

In contrast, in this paper we present a microscopic theory of electromagnetic and thermoelectric transport, 
based on a fluctuation formalism which more naturally includes the contribution of a pseudogap associated with fermion pairs. 
While the standard weak-fluctuation formalism relates in some ways to the physics of the present paper, 
we emphasize that widespread pseudogap effects are absent in the associated correlation functions; 
this is because they involve only non-interacting fermionic Green's functions.

These observations are illustrated in the top row of Fig.~(\ref{fig:Maly}),  which provides a more graphic physical
picture of the fluctuation-BCS, the pseudogap, and the BEC regimes. 
Below we refer to the fluctuation-BCS limit as the ``BCS limit". 
Strictly speaking, it goes beyond mean-field BCS theory and
serves as the basis for the conventional fluctuation picture. 
In the intermediate, or pseudogap regime, the system is fermionic with a
positive chemical potential $\mu \gg (\Delta_0, T_c)$, 
where $\Delta_0$ is the fermionic excitation gap at $T=0$. 
An important fact, however, is that at the onset of condensation there is a
non-zero gap (pseudogap) in the fermionic excitation spectrum.  
The distinction between BCS and BEC leads to different behavior of the
pair-propagator (or $t$-matrix)~\cite{Maly1}, $t(q)$, associated with composite bosons. 
Plotted in row (a) of Fig.~(\ref{fig:Maly}) is $\mathrm{Im}\ t(\Omega,\mathbf{q}=0)$ slightly above $T_c$,
illustrating the differences in the composite boson propagator in these regimes.

At small four-vector $q^{\mu}=(\Omega, \mb{q})$, the inverse (retarded) pair-propagator can be generically written as:
\begin{equation}
\label{eq:tmat_propagator}
t^{-1}(q)\approx Z[\kappa\Omega-\mb{q}^{2}/\left(2M_{\pair}\right)-\mu_{\pair}+i\Gamma\Omega].
\end{equation}
Here the coefficients $\kappa$ and $\Gamma$ are real and
dimensionless.  The real part defines an effective pair mass,
$M_{\pair}$, and a pair chemical potential,
$\mu_{\pair}\propto-t^{-1}(0)$, whereas the imaginary part,
$\propto\Gamma\Omega$, represents the diffusive contribution to the inverse
pair propagator~\cite{Comment0}. Our final results show that the
overall coefficient of proportionality, $Z$, is irrelevant; only the
ratio between $1/M_{\pair}$ and $\mu_{\pair}$ (as well as the ratio $\kappa/\Gamma$) appear.   
In the BCS (BEC) limit the parameter $\Gamma$, which reflects the damping of the pairs, 
is very large (small) compared to $\kappa$. 
Here we presume this damping derives from interactions with the fermions.
In the conventional fluctuation literature~\cite{VarlamovBook} the fluctuating Cooper pairs
are diffusive with a purely imaginary dispersion, so that $\kappa=0$ and $\Gamma\propto\pi/(8T_{c})$.
In general, $\kappa \neq 0 $ reflects particle-hole asymmetry.

From a microscopic point of view the pair propagator of the weak-fluctuation theory depends on
two bare Green's functions. However, in the presence of a pseudogap
one or more dressed Green's functions, which contain the pairing self
energy associated with the pseudogap, enters into the pair propagator. 
This leads to a different pair lifetime, mass, and chemical potential compared to the weak-fluctuation case. 
These distinctions then appear in response functions and in the associated transport coefficients.

It is useful in this overview section to present the central results of this paper for diamagnetic susceptibility, $\chi_{\dia}$:
\begin{equation}
\label{eq:chi_diamagnetism}
\chi_{\dia}=-\frac{T\left(2e\right)^2}{24\pi \hbar c^2}\sqrt{\frac{1/(2M_{\pair})}{\left|\mu_{\pair}\right|}},
\end{equation}
and similarly the transverse thermoelectric coefficient (related to the Nernst coefficient), $\widetilde{\alpha}_{xy}$:
\begin{equation}
\label{eq:alpha_thermoelectric}
\widetilde{\alpha}_{xy}=\frac{BTe^2}{12\pi\hbar^{2} c}\sqrt{\frac{1/(2M_{\pair})}{\left|\mu_{\pair}\right|}}
\,\left(\frac{3\kappa^2 + \Gamma^2}{\Gamma^2}\right).
\end{equation}
These expressions, obtained for three-dimensional (3D) systems,
are valid in the small $|\mu_{\pair}|$ limit: $(|\mu_{\pair}|\ll T_{c})$.  
The size of the diamagnetic susceptibility and transverse thermoelectric coefficient are determined by two key parameters: 
the pair mass $M_{\pair}$ (related to the inverse coherence length, often appearing as an inverse diffusion coefficient
in the weak-fluctuation literature) and the pair chemical potential $\mu_{\pair}$.

The rest of this paper is organized as follows. 
In Sec.~(\ref{sec:Pair_Propagator}) the pair propagator and associated properties 
of the normal state are characterized as the pairing varies from weak to strong attraction.
Sections~(\ref{sec:EM_Response}-\ref{sec:EM_Diagrammatic}) show how
our pseudogap formalism is implemented in the diamagnetic response,
while in Sec.~(\ref{sec:Diamagnetism_Approx}) the diamagnetic susceptibility is calculated in the small $|\mu_{\pair}|$ limit.  
The analogous calculations for the transverse thermoelectric coefficient are discussed in Sec.~(\ref{sec:Heat_Diagrammatic}). 
Our numerical results for the phase diagram and diamagnetic susceptibility onset temperature are then 
presented in Sec.~(\ref{sec:Numerics}) along with a comparison with experiment. 
Finally in Sec.~(\ref{sec:Conclusions}) our conclusions are outlined.

\section{Pair-propagator formalism}
\label{sec:Pair_Propagator}

In this section we give a brief overview of the pair-fluctuation
formalism underlying the work in this paper. 
For a more extensive and thorough review, see Refs.~\onlinecite{Chen_2005,Chen_2014}. 
At the heart of any calculation incorporating bosonic degrees of
freedom into diamagnetic susceptibility and general electromagnetic transport
is the explicit form of the pair propagator.
We emphasize that the BCS mean-field gap equation
provides important intuition about the form this fluctuation propagator should take. 
In the standard BCS mean-field theory the pairing gap parameter is exactly equal to the order parameter. 
More generally, a non-zero pairing gap will be present at the onset of condensation. 
Importantly, this pairing gap $\Delta$ must be continuous across $T_{c}$ in order to
properly describe a second order phase transition. 
This normal state, in which the pairing gap persists, represents the pseudogap phase.

To begin we first consider the BCS mean-field gap equation in the condensed phase~\cite{Chen_2005}:
\begin{equation}\label{eq:GE}
g^{-1}+\sum_{k}G(k)G_{0}(-k)\phik^2=0, \quad T\leq T_{c},
\end{equation}
where for the cuprates, the $d$-wave pairing interaction is given by
$V_\mathbf{k,k'} = g\phik\phik^\prime$ with $g<0$ and
$\phik = \cos k_x - \cos k_y$ with lattice constant $a=1$. 
The four-vector $k^{\mu}=(i\omega_{n},\mb{k})$, where $\omega_{n}$ is a fermionic Matsubara frequency, 
and the summation is defined by $\sum_{k}=T\sum_{i\omega_{n}}\sum_{\mb{k}}$. 
The bare single-particle Green's function, $G_{0}(k)$, is defined by $G_{0}^{-1}(k)=i\omega_{n}-\xi_{\mb{k}}$, 
where $\xi_{\mb{k}}=2t(2-\cos k_x - \cos k_y) + 2t_z(1-\cos k_z)-\mu$ is the
dispersion for a tight-binding model and $\mu$ is the fermion chemical potential. 
Here $t$ and $t_z$ are the in-plane and out-of-plane hopping matrix elements, respectively, with $t_z \ll t$ for the quasi-2D cuprates. 
In the continuum case, $\xi_{\mb{k}}=\mb{k}^2/(2m)-\mu$ with $m$ as the fermion mass. 
We set $\hbar=c=1$ and restore these units at the end of the calculation.

The full Green's function, $G(k)$, is determined from the bare Green's function and self energy, 
$\Sigma(k)$, through Dyson's equation: $G^{-1}(k)=G^{-1}_{0}(k)-\Sigma(k)$. 
In BCS mean-field theory, the self energy has the form
\begin{equation}
\label{eq:BCS_sigma}
\Sigma(k)=-\Delta^2G_{0}(-k)\phik^2=\Delta^2\phik^2/(i\omega_{n}+\xi_{\mb{k}}).
\end{equation}
The gap equation in Eq.~(\ref{eq:GE}) can be expressed as a generalized Thouless criterion~\cite{Mahan} 
for a pairing instability in the form $t^{-1}(q\rightarrow0)=0$.
This suggests that the (inverse) $t$-matrix appropriate to BCS theory is
\begin{equation}
\label{eq:Tmat}
t^{-1}(q)\equiv g^{-1}+\sum_{k}G(k)G_{0}(-k+q)\phikq2.
\end{equation} 
Here $q^{\mu}=(i\Omega_{m},\mb{q})$ (before analytic continuation),
where $\Omega_{m}$ is a bosonic Matsubara frequency.

It follows directly from the gap equation in Eq.~(\ref{eq:GE}) that
the associated $t$-matrix involves one bare and one dressed Green's function. 
This asymmetric form, while perhaps surprising, has been
derived in the literature~\cite{Kadanoff_Martin_1961} from a
microscopic approach by studying the equations of motion for the correlation functions. 
We emphasize that this $t$-matrix should be interpreted as the propagator 
for non-condensed fermion pairs associated with $q\neq0$.

At and below the condensation temperature the low momentum non-condensed pairs 
become gapless~\cite{Hugenholtz} and thus acquire zero chemical potential. 
Since $t^{-1}(q=0)\propto-\mu_{\pair}$, it follows that
\begin{equation}
\label{eq:Tmat_infty}
t(q=0)=\infty, \quad T \leq T_{c}.
\end{equation}
Thus the gap equation [Eq.~(\ref{eq:GE})] can be equivalently written
as a BEC condition:
\begin{equation}
\mu_{\pair}=0, \quad T \leq T_c,
\end{equation}
provided the self energy appearing in $G(k)$ is given by the usual BCS form [Eq.~(\ref{eq:BCS_sigma})].
All of this general formalism is consistent with the generic form for the
pair propagator in Eq.~(\ref{eq:tmat_propagator}).

Now we connect the physics below $T_{c}$ to that above $T_{c}$. 
In most $t$-matrix theories the associated fermionic self energy is
\begin{equation}
\Sigma(k)=\sum_{q}t(q)G_{0}(-k+q) \phikq2.
\label{eq:GG0_SelfEnergy}
\end{equation}
The quantity $t(q)$ is strongly peaked about $q=0$ as the transition
is approached from above because $|\mu_{\pair}|$ is small; 
this allows the normal state self energy to be written as
$\Sigma(k)\approx-\Delta^2G_0(-k)\phik^2$, with
\begin{equation}\label{eq:GE_aboveTc}
\Delta^2=-\sum_{q}t(q), \quad T \geq T_{c}. 
\end{equation}
With this result, the transition temperature $T_{c}$ can then be computed. 
This is determined as the temperature at which the normal
state value of $\Delta$, given in Eq.~(\ref{eq:GE_aboveTc}),
intersects with its value obtained at or below $T_{c}$, found from Eq.~(\ref{eq:GE}).

This physical picture is more complicated than in BCS mean-field theory 
because of the presence of a non-zero pseudogap at $T_c$,
which must be continuous at a second order phase transition. 
The parameters appearing in Eq.~(\ref{eq:tmat_propagator}), such as the
pair mass $M_{\pair}$, pair chemical potential $\mu_{\pair}$, and pair 
damping $\propto\Gamma$ can then be deduced from Eq.~(\ref{eq:Tmat}). 
It is crucial to include a self consistently determined fermionic chemical potential using the number equation $n=2\sum_{k}G(k)$. 
One can also define the pairing onset temperature $T^*$ most naturally 
as the temperature at which $\Delta$ vanishes, as determined, for example, from the mean-field gap equation. 
In this way a phase diagram for $T_{c}$ and $T^{*}$, as a function of band structure and interaction strength $g$, can be computed. 
This simultaneously yields the diamagnetic susceptibility and transverse thermoelectric coefficient 
via Eq.~(\ref{eq:chi_diamagnetism}) and Eq.~(\ref{eq:alpha_thermoelectric}).
These limiting forms are derived in Secs.~(\ref{sec:Diamagnetism_Approx}-\ref{sec:Heat_Diagrammatic}), 
while in Sec.~(\ref{sec:Numerics}) of the paper the complete diamagnetic susceptibility expression is numerically calculated.

Finally, it is useful to contrast these pseudogap effects with the
pair propagator for the more conventional weak-fluctuation theory.
Aslamazov and Larkin~\cite{Aslamazov_1975} have written down the
counterpart to Eq.~(\ref{eq:Tmat}) for the weak fluctuation case, which in the $d$-wave limit is given by
\begin{equation}
t^{-1}_{0}(q)\equiv g^{-1}+\sum_{k}G_{0}(k)G_{0}(-k+q)\phikq2.
\end{equation}
In the pair propagator all fermionic Green's functions are bare and no pseudogap is present. 
In contrast to the strong-pairing limit, the above $t$-matrix is associated with diffusive rather than propagating dynamics. 
Referring to Eq.~(\ref{eq:tmat_propagator}), the parameter $\kappa=0$, $|\mu_{\pair}|\propto(T-T_{c})$, 
$\Gamma\propto \pi/(8T_{c})$, and $1/(2M_{\pair})\propto D$ (the diffusion constant).

In the $\kappa\rightarrow0$ limit, instead of weakly-damped and
propagating non-condensed pairs, one has diffusive pair dynamics. 
This weak-attraction case, and its consequences for the fermionic
properties [via Eq.~(\ref{eq:GG0_SelfEnergy})], is presented in the first column in Fig.~(\ref{fig:Maly}).  
One can contrast the difference in behavior with that 
for the pseudogap case shown in the second column. 
Here the pairing strength has been increased relative
to the first column and the associated $t$-matrix acquires a
significant propagating term (second row) with broken particle-hole symmetry.

The third row of the second column shows that the fermionic self
energy, deduced from Eq.~(\ref{eq:GG0_SelfEnergy}), is reasonably well described by Eq.~(\ref{eq:BCS_sigma}). 
Furthermore, the fermionic spectral function in the last row now has a double-peaked form
associated with the presence of a normal state gap. 
The third column in Fig.~(\ref{fig:Maly}) is appropriate to the strong attraction case,
$\Gamma\ll\kappa$, where the system is in the BEC regime. 
We reiterate that this is well outside~\cite{Chen_99} the physical 
parameter range associated with the $d$-wave paired cuprates.

To maintain clarity in the equations, in the following sections we present our 
theoretical derivations for short range $s$-wave pairing in the 3D continuum with $\phik = 1$. 
However, our numerical results are for the quasi-2D $d$-wave case.

\section{Electromagnetic response}
\label{sec:EM_Response}

We begin with a discussion of diamagnetic susceptibility, which represents the orbital current response to an external magnetic field. 
Here we use linear response theory to derive the Kubo formula for diamagnetic susceptibility. 
In the presence of a weak and externally applied EM vector potential, $A^{\mu}(q)$, the EM current is $j^{\mu}(q)=K^{\mu\nu}(q)A_{\nu}(q)$. 
The response kernel is $K^{\mu\nu}(q)=P^{\mu\nu}(q)+(n/m)\delta^{\mu\nu}(1-\delta_{\mu,0})$, with $\mu$ and $\nu$ not summed over. 
Here $n$ is the particle number, determined from $n=2\sum_{k}G(k)$, and $P^{\mu\nu}(q)$ are the EM response functions given by~\cite{Schrieffer}:
\begin{equation}\label{eq:Response_Fermion}
P^{\mu\nu}(q)=2e^2\sum_{k}G(k_{+})\Gamma^{\mu}_{\E}(k_{+},k_{-})G(k_{-})\gamma^{\nu}_{\E}(k_{-},k_{+}).
\end{equation}
Here $e$ is the fermion charge. The bare EM vertex is $\gamma^{\mu}_{\E}(k_{+},k_{-})$ and the full EM vertex is
$\Gamma^{\mu}_{\E}(k_{+},k_{-})$~\cite{Comment3}, where $k_{\pm}=k\pm q/2$. 
The prefactor of 2 arises due to spin-degeneracy for a spin-$\tfrac{1}{2}$ system of fermions.

An important relation between the full Green's function and the full
EM vertex is the Ward-Takahashi identity (WTI)~\cite{Ryder}:
\begin{align}
q_{\mu}\Gamma^{\mu}_{\E}(k_{+},k_{-})&=G^{-1}(k_{+})-G^{-1}(k_{-}),\nonumber \\
&=q_{\mu}\gamma^{\mu}_{\E}(k_{+},k_{-})+\Sigma(k_{-})-\Sigma(k_{+}).
\end{align}
The bare WTI,
$q_{\mu}\gamma^{\mu}_{\E}(k_{+},k_{-})=G^{-1}_{0}(k_{+})-G^{-1}_{0}(k_{-})$,
is satisfied by the bare EM vertex
$\gamma^{\mu}_{\E}(k_{+},k_{-}) =(1,\textbf{k}/m)$. 
For a neutral (charged) system with a global $\mathrm{U}(1)$ symmetry, 
the corresponding conservation law is particle number (charge) conservation. 
The analysis here is for neutral superfluids. 
Satisfying the WTI is thus an important constraint which
enforces conservation of global particle number. 
Applying the WTI to the response kernel $K^{\mu\nu}(q)$ yields $q_{\mu}K^{\mu\nu}(q)=0$;
this is the statement of ``gauge invariance".

In the $q\rightarrow0$ limit, the WTI implies that
$\Gamma^{\mu}_{\E}(k,k)=\gamma^{\mu}_{\E}(k,k)-\partial\Sigma(k)/\partial k_{\mu}$.
Diagrammatically this relation asserts that the full EM vertex is
determined by performing all bare EM vertex insertions in the self energy diagram. 
In terms of components this expression becomes: $\Gamma^{0}_{\E}(k,k)=\partial G^{-1}(k)/\partial\omega$ 
and $\Gamma^{i}_{\E}(k,k)=-\partial G^{-1}(k)/\partial k^{i}$.

It is straightforward to derive diamagnetic susceptibility from these response functions. 
In the presence of a static external vector
potential the magnetic field is $\mb{B}=i\mb{q}\times\mb{A}$. 
Similarly the current can be written in terms of a divergence-free (orbital)
magnetization by $\mb{j}=i\mb{q}\times\mb{M}$. 
For convenience, $\mb{q}$ is directed along the $y$-axis: $\mb{q}=q^{y}\hat{\mb{y}}$. 
Using the definition of the EM current, and by taking the $q^{y}\rightarrow0$
limit in this expression, we then obtain
$\mb{M}\left(q^{y}\rightarrow0\right)=
-\left.\left[P^{xx}(q^{y})+n/m\right]/(q^{y})^2\right|_{q^{y}\rightarrow0}\mb{B}\left(q^{y}\rightarrow0\right)$.
From the definition of diamagnetic susceptibility,
$\chi_{\dia}=-\left.\left(\partial M^{z}/\partial B^{z}\right)\right|_{B^{z}\rightarrow0}$,
we then have the following Kubo formula for diamagnetic susceptibility~\cite{Vignale}:
\begin{equation}\label{eq:chi_Kubo_dia}
\chi_{\dia}=-\underset{\mb{q}\rightarrow0}{\lim}\left[\frac{P^{xx}(i\Omega_{m}=0,\mb{q})+n/m}{\mb{q}^{2}}\right]_{q^{x}=q^{z}=0}.
\end{equation}

Diamagnetic susceptibility is a transverse response to an applied vector potential;
that is, by taking the zero frequency limit first, and then the
momentum limits in the appropriate order, there is no longitudinal
contribution to the diamagnetic susceptibility of a uniform Fermi superfluid. 
Moreover, the Kubo formula in Eq.~(\ref{eq:chi_Kubo_dia})
also applies in the condensed phase of a uniform Fermi superfluid. 
This is because the collective mode contribution to
response in a uniform system is purely longitudinal in the zero
frequency, zero momentum limit, and therefore it gives no contribution to diamagnetic susceptibility. 
Above the superfluid phase transition temperature, $P^{xx}(0)=-n/m$; 
this identity enforces the physical constraint that there is no Meissner effect. 
As a consequence, the Kubo formula can then be written as
$\chi_{\dia}=-\underset{\mb{q}\rightarrow0}{\lim}\left.\left[P^{xx}(i\Omega_{m}=0,\mb{q})-P^{xx}(0)\right]/\mb{q}^{2}\right|_{q^{x}=q^{z}=0}.$

Another important contribution to magnetic susceptibility is paramagnetic susceptibility. 
Paramagnetism is the spin polarization response due to 
a spin imbalance caused by an external magnetic field. 
For a system of spin-$\tfrac{1}{2}$ fermions, the Kubo formula for paramagnetic susceptibility is~\cite{Mahan}
\begin{equation}\label{eq:chi_Kubo_Pauli}
\chi_{\Pauli}=-\underset{\mb{q}\rightarrow0}{\lim}\ \mu_{\B}^2P^{00}(i\Omega_{m}=0,\mb{q}),
\end{equation}
where $\mu_{\B}$ is the Bohr magneton. 
In a non-interacting fermionic system, the resulting (Pauli) paramagnetic susceptibility and (Landau) diamagnetic susceptibility
satisfy the well-known relation $\chi_{\dia}=-\tfrac{1}{3}\chi_{\Pauli}$.

\section{Diagrammatic analysis of response functions in pair-fluctuation theory}
\label{sec:EM_Diagrammatic}

We now build on our discussion in the introduction to incorporate strong-pairing fluctuations. 
There we motivated a specific choice for the composite boson propagator associated with non-condensed pairs. 
This is referred to below as the $GG_{0}$ pair-fluctuation theory. 
The self energy for this theory is
\begin{equation}
\label{eq:GG0_SelfEnergy1}
\Sigma(k)=\sum_{p}t(p)G_{0}(p-k)=\sum_{p}t(p+k)G_{0}(p).
\end{equation}
The inverse $t$-matrix is given by $t^{-1}(p)=g^{-1}+\Pi(p)$, with the pair susceptibility $\Pi(p)$ defined by
\begin{equation}
\Pi(p)=\sum_{l}G_{0}(p-l)G(l)=\sum_{l}G(p-l)G_{0}(l).
\end{equation}
Throughout this paper $k^{\mu}=(i\omega_{n},\mb{k})$, $l^{\mu}=(i\epsilon_{n},\mb{l})$ denote fermionic four-vectors, 
while $p^{\mu}=(i\varpi_{m},\mb{p})$ and $q^{\mu}=(i\Omega_{m},\mb{q})$ denote bosonic four-vectors.

In order to derive the full EM vertex, all bare EM vertex insertions
in the self-energy diagram must be performed~\cite{Ryder}. 
After summing all these bare EM vertex
insertions, there are in total three possible vertex insertions in the self-energy diagram: 
(1) a bare EM vertex can be inserted in the bare Green's function $G_{0}(p-l)$, 
(2) a full EM vertex can be inserted in the full Green's function in the pair susceptibility $\Pi(p+l)$, 
and (3) a bare EM vertex can be inserted in the bare Green's function in the pair susceptibility $\Pi(p+l)$. 
Thus the full EM vertex can be written schematically as
\begin{align}\label{eq:Full_vertex}
\Gamma^{\mu}_{\E}(k_{+},k_{-})&=\gamma^{\mu}_{\E}(k_{+},k_{-})+\MT^{\mu}_{\E}(k_{+},k_{-})\nonumber\\
&\quad+\AL_{\E,1}^{\mu}(k_{+},k_{-})+\AL_{\E,2}^{\mu}(k_{+},k_{-}).
\end{align}
The full EM vertex consists of the bare EM vertex, a Maki-Thompson-like 
vertex, and two distinct Aslamazov-Larkin-like vertices. 
These Feynman diagrams are analogous to those in the standard weak-fluctuation
theory~\cite{VarlamovBook} except that here, as appropriate, there are full rather than bare Green's functions.  
Note that these vertex corrections appear after making the above diagrammatic insertions: 
the MT diagram arises from a bare EM vertex insertion in the
bare Green's function appearing in the self energy, while the two AL diagrams
enter due to inserting bare or full EM vertices in the appropriate bare
or full Green function's in the pair susceptibility. 
In appendix~(\ref{sec:app_Full_EM_Vertex}) an explicit derivation of
these MT and AL diagrams is presented; their exact forms are given by
\begin{widetext}
\begin{align}
\label{eq:MT_vertex}\MT^{\mu}_{\E}(k_{+},k_{-})&=\sum_{p}t(p)G_{0}(p-k_{-})\gamma^{\mu}_{\E}(p-k_{-},p-k_{+})G_{0}(p-k_{+}),\\
\label{eq:AL1_vertex}\AL_{\E,1}^{\mu}(k_{+},k_{-})&=-\sum_{p}\sum_{l}t(p_{-})t(p_{+})G_{0}(p-k)G_{0}(p-l)G(l_{+})\Gamma^{\mu}_{\E}(l_{+},l_{-})G(l_{-}),\\
\label{eq:AL2_vertex}\AL_{\E,2}^{\mu}(k_{+},k_{-})&=-\sum_{p}\sum_{l}t(p_{-})t(p_{+})G_{0}(p-k)G(p-l)G_{0}(l_{+})\gamma^{\mu}_{\E}(l_{+},l_{-})G_{0}(l_{-}).
\end{align}
\end{widetext}

One can explicitly check that the full EM vertex satisfies the WTI. 
To do this, note that the MT and AL diagrams satisfy the following important identity:
$q_{\mu}\left[2\MT^{\mu}_{\E}(k_{+},k_{-})+\AL_{\E,1}^{\mu}(k_{+},k_{-})+\AL_{\E,2}^{\mu}(k_{+},k_{-})\right]=0$. 
This identity is proved in appendix~(\ref{sec:app_Full_EM_Vertex}), 
where it is derived from the definitions of the MT and AL diagrams in Eqs.~(\ref{eq:MT_vertex}-\ref{eq:AL2_vertex}). 
Using this identity, it follows that
$q_{\mu}\left[\Gamma^{\mu}_{\E}(k_{+},k_{-})-\gamma^{\mu}_{\E}(k_{+},k_{-})\right]=-q_{\mu}\MT^{\mu}_{\E}(k_{+},k_{-})$. 
From the MT vertex given in Eq.~(\ref{eq:MT_vertex}), along with the bare WTI, we then have
$q_{\mu}\left[\Gamma^{\mu}_{\E}(k_{+},k_{-})-\gamma^{\mu}_{\E}(k_{+},k_{-})\right]=\Sigma(k_{-})-\Sigma(k_{+})$,
so that
$q_{\mu}\Gamma^{\mu}_{\E}(k_{+},k_{-})=G^{-1}(k_{+})-G^{-1}(k_{-})$
and thus the full EM vertex satisfies the WTI.

While the formal expression for the full EM vertex can be written down, it is not in closed form due to the fact that this vertex
itself appears in the $\AL^{\mu}_{\E,1}$ diagram. 
We note that the lowest order $\MT$ and $\AL$ diagrams, which are obtained by
setting $\Gamma^{\mu}_{\E}\rightarrow\gamma^{\mu}_{\E}$, $G\rightarrow G_{0}$, and $t\rightarrow t_{0}$, 
in Eqs.~(\ref{eq:MT_vertex}-\ref{eq:AL2_vertex}), are consistent with
those diagrams which have appeared extensively in the weak-fluctuation literature~\cite{Maki_1968,Aslamazov_1968, Aslamazov_1975}.

The other important contribution to the lowest order EM response functions are the density of states (DOS) diagrams. 
These diagrams arise from substituting the bare EM vertex part of the full EM vertex
into the total response functions. Indeed, the bare EM vertex term in Eq.~(\ref{eq:Full_vertex}) 
gives a ``bubble" contribution to the total response functions in the form
$2e^2\sum_{k}G(k_{+})\gamma^{\mu}_{\E}(k_{+},k_{-})G(k_{-})\gamma^{\nu}_{\E}(k_{-},k_{+})$.

\sloppy By expanding the full Green's functions to second order in Dyson's equation: 
$G(k)\approx G_{0}(k)+G_{0}(k)\Sigma(k)G_{0}(k)$, the ``bubble" contribution becomes
$2e^2\sum_{k}[G_{0}(k_{+})\gamma^{\mu}(k_{+},k_{-})G_{0}(k_{-})$
$+G_{0}(k_{+})\gamma^{\mu}(k_{+},k_{-})G_{0}(k_{-})\Sigma(k_{-})G_{0}(k_{-})$
$+G_{0}(k_{+})\Sigma(k_{+})G_{0}(k_{+})\gamma^{\mu}(k_{+},k_{-})G_{0}(k_{-})]
\gamma^{\nu}(k_{-},k_{+})$,
which gives the lowest order diagram for non-interacting fermions, plus two additional DOS diagrams. 
Note that, this lowest order set of Feynman diagrams (the non-interacting response plus two DOS, one MT, and two AL diagrams) is not gauge-invariant. 
These diagrams satisfy the WTI to $\mathcal{O}\left(\Sigma\right)$, but violate it at $\mathcal{O}\left(\Sigma^2\right)$. 
The exact gauge-invariant full EM vertex, which satisfies the WTI, is given in Eqs.~(\ref{eq:Full_vertex}-\ref{eq:AL2_vertex}).

For an exact treatment of the EM response, at all temperatures, all diagrams must be considered. 
In order to make progress in computing the
diamagnetic susceptibility for the $GG_{0}$ pair-fluctuation theory, 
certain assumptions must be made and their validity correspondingly needs to be scrutinized. 
The following sections outline a set of approximations
enabling the diamagnetic susceptibility to be calculated analytically. 
The small parameter controlling these approximations will be discussed in further detail below.

\section{Approximate calculation of diamagnetic susceptibility in the small \texorpdfstring{$|\mu_{\pair}|$}{mu-pair} limit}
\label{sec:Diamagnetism_Approx}

This section derives the diamagnetic susceptibility for the $GG_{0}$
pair-fluctuation theory in the fairly extended regime above $T_{c}$,
where the bosonic chemical potential $\mu_{\pair}$ is small. 
The phase transition temperature, $T_{c}$, occurs when the pair chemical
potential vanishes: $\mu_{\pair}(T_{c})=0$. 
Thus the small parameter regime $|\mu_{\pair}|\ll T_{c}$ is what governs the various
approximations made within this calculation. 
In the typical weak-fluctuation physics~\cite{VarlamovBook}, this  parameter becomes
$\epsilon\equiv\mathrm{ln}\left(T/T_{c}\right)\approx\left(T-T_{c}\right)/T_{c}$.
This perturbative regime is necessarily limited to temperatures in
close proximity to $T_{c}$. By contrast, the constraint associated
with the pseudogap state ($|\mu_{\pair}|\ll T_{c}$) is less
restrictive; it is found to apply to considerably higher temperatures,
as is discussed in Sec.~(\ref{sec:Numerics}). 
As a consequence of this result, the temperature range where the diamagnetic susceptibility in the $GG_{0}$
pair-fluctuation theory is nearly singular is larger than the
corresponding range in the usual weak-fluctuation theory.

It should be noted that near condensation the pair propagator is not
so different from a modified free boson propagator, except that there
is no fixed number of (composite) boson particles. 
The propagator depends on the bosonic mass $m_{\b}=M_{\pair}$ and bosonic chemical
potential $\mu_{\b}=\mu_{\pair}$, which are determined
self-consistently from the underlying fermionic interactions. 
The pair chemical potential acts as an infra-red regulator and the
singular nature of the diamagnetic susceptibility is encapsulated by the limit $|\mu_{\pair}|\ll T_{c}$.

At $q=0$, the full response function satisfies $P^{xx}(0)=-n/m$. 
To compute the diamagnetic susceptibility from Eq.~(\ref{eq:chi_Kubo_dia}), 
the response function $P^{xx}(0,\mb{q})$ must then be expanded to $\mathcal{O}\left(\mb{q}^{2}\right)$.  
At all temperatures there will be contributions from the ``bubble", Maki-Thompson, and Aslamazov-Larkin diagrams. 
However, the AL diagrams have one more pair-propagator than the MT diagram 
(without expanding out the full Green's functions or full vertices that is). 
As discussed in the preceding paragraph, the near-singular nature of the diamagnetic susceptibility
arises due to the vanishing of the pair chemical potential. 
Since the AL diagrams contain one more pair propagator than the MT diagram, 
the degree of the singularity of the AL contribution to 
diamagnetic susceptibility (in 3D) is of a higher order than the MT contribution.  
Indeed, power counting arguments~\cite{Ussishkin_2003, VarlamovBook, Aslamazov_1975} 
indicate that near the condensation temperature the AL diagrams give singular contributions to diamagnetic susceptibility, 
whereas the MT diagram gives a non-singular diamagnetic response. 
For this reason, we omit calculating the ``bubble" and MT contributions to diamagnetic susceptibility~\cite{Comment2}. 
In the weak-fluctuation theory the contribution to diamagnetic susceptibility from the Aslamazov-Larkin diagrams is
also all that is considered~\cite{Aslamazov_1975} near the condensation temperature.

\begin{figure*}[t]
\centering\includegraphics[width=13cm,height=8cm,clip]{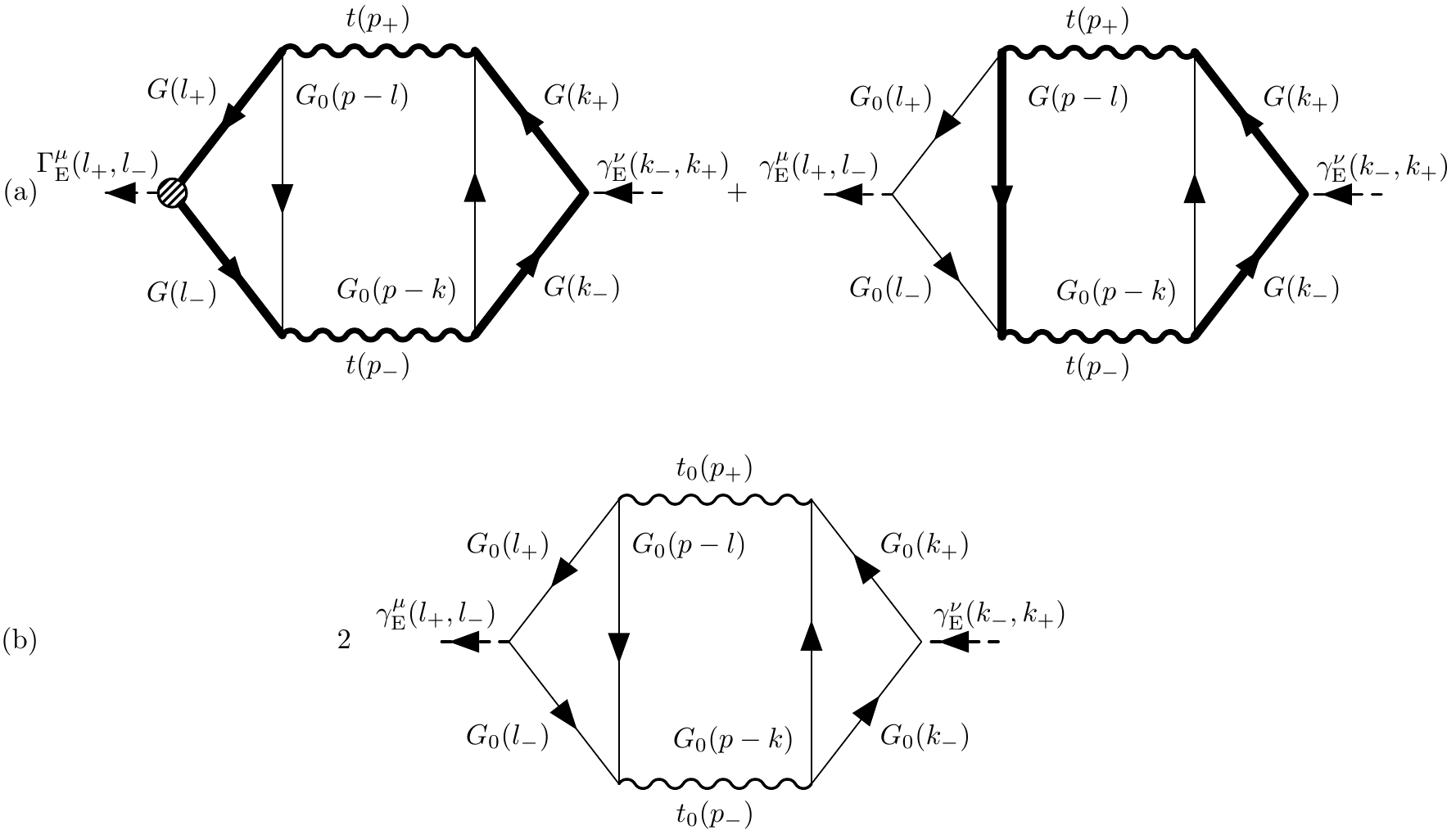}
\caption{Comparison of the Aslamazov-Larkin diagrams in 
  (a) $GG_{0}$ pair-fluctuation theory and (b) weak-fluctuation theory.
  These are the dominant diagrams that contribute to diamagnetic susceptibility, near the condensation regime. 
  Of importance is that it is the $GG_{0}$ pair-fluctuation theory whose Aslamazov-Larkin diagrams
  contain full Green's functions depending on the normal state pairing
  gap, whereas the weak-fluctuation theory contains only bare Green's functions.}
\label{fig:Aslamazov_Larkin_diagrams_comparison}
\end{figure*}

After ignoring the ``bubble" and MT contributions, the response function of interest now becomes
\begin{align}
P^{xx}(0,\mb{q})&\approx 2e^2\sum_{k}G(k_{+})[
\AL_{\E,1}^{x}(k_{+},k_{-})\nonumber\\
&\quad+\AL_{\E,2}^{x}(k_{+},k_{-})]G(k_{-})\gamma^{x}_{\E}(k_{-},k_{+}).
\end{align}
To contrast the $GG_{0}$ pair-fluctuation theory from the weak-fluctuation
theory~\cite{VarlamovBook,Aslamazov_1968,Aslamazov_1975}, 
in Fig.~(\ref{fig:Aslamazov_Larkin_diagrams_comparison}) the
Aslamazov-Larkin diagrams for both of these theories are shown. 
Of interest to note is that the weak-fluctuation theory has two
identical AL diagrams, containing only bare
Green's functions and thus no signature of a normal state pairing gap. 
In the $GG_{0}$ pair-fluctuation theory, however, for the
$\AL_{\E,1}$ diagram the left-most triangle vertex contains
two full Green's functions and one full EM vertex, while for the
$\AL_{\E,2}$ diagram this same triangle vertex contains one full Green's function. 
As a result, the $GG_{0}$ pair-fluctuation theory contains dressed 
Green's functions which depend on the normal state pairing gap.

The presence of these full Green's functions in the triangle vertex is
intimately connected to the form of the $t$-matrix. 
We emphasize (and discuss in more detail below) that choosing a particular form for the
pair-propagator constrains where bare and dressed Green's functions
can appear in the Aslamazov-Larkin diagram. 
All of this is fundamental to the goal of this paper, which is to include pseudogap
effects (as incorporated in dressed Green's functions) in a consistent
manner in the diamagnetic susceptibility.

Because calculating diamagnetic susceptibility requires expanding the response function to
$\mathcal{O}\left(\mb{q}^{2}\right)$, in the regime where $|\mu_{\pair}|$ is small
it is only the $\mb{q}$ dependence of the pair propagator which is significant. 
As a consequence, we ignore the $\mb{q}$ dependence of all the
Green's functions appearing in the approximate response function. 
The remaining $\mb{q}$ dependence occurring in $\AL_{\E,2}^{x}(k_{+},k_{-})$ is
then due to the two pair propagators in this vertex. 
The remaining $\mb{q}$ dependence occurring in
$\AL_{\E,1}^{x}(k_{+},k_{-})$ arises from two contributions: 
the two pair propagators that explicitly appear and the full EM
vertex occurring in the left-most triangle vertex.  
[See Fig.~(\ref{fig:Aslamazov_Larkin_diagrams_comparison}) for reference.]

In expanding the full EM vertex in $\AL_{\E,1}^{x}$ to $\mathcal{O}\left(\mb{q}^{2}\right)$, 
we ignore the MT contribution and only expand
to quadratic order the $\AL_{\E,1}^{x}$ and $\AL_{\E,2}^{x}$ terms. 
Thus, all pairs of pair propagators ($t$-matrices) are expanded to quadratic order. 
Finally, the structure of the $\AL_{\E,1}^{x}$ and $\AL_{\E,2}^{x}$ vertices at $q=0$ is needed. 
By differentiating the two equivalent expressions for the self energy appearing in 
Eq.~(\ref{eq:GG0_SelfEnergy1}) with respect to $k^{x}$, the following identities are obtained:
$\AL^{x}_{\E,1}(k,k)=\AL^{x}_{\E,2}(k,k)=-\MT^{x}_{\E}(k,k)$. 
Therefore, at $q=0$, the full EM vertex is
$\Gamma^{x}_{\E}(k,k)=\gamma^{x}_{\E}(k,k)+\AL_{\E,1}^{x}(k,k)$.

By iterating this relation in the full EM vertices appearing in
$\AL_{\E,1}$ and $\AL_{\E,2}$ diagrams, and expanding all pairs of
$t$-matrices to $\mathcal{O}\left(\mb{q}^{2}\right)$, the net result is a symmetric product 
of two bosonic vertices with two pair propagators expanded to quadratic order.  
The diamagnetic susceptibility thus becomes
\begin{equation}\label{eq:chi_approx}
\chi_{\dia}=e^2\sum_{p}\left[\Lambda^{x}_{\E,1}(p,p)\right]^2\left\{t(p)\frac{\partial^{2}t(p)}{\partial (p^{y})^2}
-\left[\frac{\partial t(p)}{\partial p^{y}}\right]^{2}\right\}.
\end{equation}
Here we have defined the bosonic EM vertex
$\Lambda^{x}_{\E,1}(p,p)=-\sum_{l}G_{0}(p-l)G^{2}(l)\Gamma^{x}_{\E}(l,l)$,
where the minus sign arises from the fermion loop in the Aslamazov-Larkin triangle vertex. 
In this form it is clear that the response function for the AL diagrams
reduces to a bosonic response function, with bosonic EM
vertices $\Lambda^{x}_{\E,1}(p,p)$ which are modified from a bosonic
bare EM vertex due to the underlying fermionic interactions.

Using the WTI the bosonic EM vertex $\Lambda^{x}_{\E,1}(p,p)$ can be
written in terms of derivatives of the pair susceptibility $\Pi(p)$ as
\begin{equation}\label{eq:boson_electric_vertex}
\frac{\partial\Pi(p)}{\partial p^{x}}=\sum_{l}G_{0}(p-l)G^{2}(l)\Gamma^{x}(l,l)=-\Lambda^{x}_{\E,1}(p,p).
\end{equation}
An equivalent expression is
$\Lambda^{x}_{\E,1}(p,p)=-\sum_{l}G(p-l)G_{0}^{2}(l)\gamma^{x}(l,l)=\Lambda^{x}_{\E,2}(p,p)$. 
Further details on this derivation are given in appendix~(\ref{sec:Boson_Electric_Vertex}). 
Inserting this result into Eq.~(\ref{eq:chi_approx}) then gives the diamagnetic susceptibility as
\begin{equation}\label{eq:chi_approx1}
\chi_{\dia}=e^2\sum_{p}\left[\frac{\partial\Pi(p)}{\partial p^{x}}\right]^2\left\{t(p)\frac{\partial^{2}t(p)}{\partial (p^{y})^2}
-\left[\frac{\partial t(p)}{\partial p^{y}}\right]^{2}\right\}.
\end{equation}

In the small $|\mu_{\pair}|$ limit, when performing the Matsubara frequency summation 
only the lowest order term in the frequency integral with 
bosonic frequency equal to zero needs to be retained~\citep{Aslamazov_1975}. 
In appendix~(\ref{sec:Diamagnetism_Approx_Matsubara}), the Matsubara frequency 
summation is carried out analytically and the preceding assumption is validated. 
Thus, we now have
\begin{equation}
\chi_{\dia}=Te^2\sum_{\mb{p}}\left[\frac{\partial\Pi(\mb{p})}{\partial p^{x}}\right]^2
\left\{t(\mb{p})\frac{\partial^{2}t(\mb{p})}{\partial (p^{y})^2}
-\left[\frac{\partial t(\mb{p})}{\partial p^{y}}\right]^{2}\right\}.
\end{equation}
Here $\Pi(\mb{p})\equiv\Pi(0,\mb{p})$ and $t(\mb{p})\equiv t(0,\mb{p})$.
To evaluate the form of the vertices involving the derivatives of the susceptibility, 
we use the definition $t^{-1}(p)=g^{-1}+\Pi(p)$, along with the approximate form of the pair propagator given in Eq.~(\ref{eq:tmat_propagator}), 
to obtain $\partial\Pi(\mb{p})/\partial p^{x}=\p t^{-1}(p)/\partial p^{x}=-Zp^{x}/M_{\pair}$.
The result for the bosonic EM vertex $\Lambda^{x}_{\E,1}$ is of the same form as one would expect for actual bosons, but with a modified mass.

We emphasize again that the composite boson EM vertex appearing in the
diamagnetic susceptibility is tightly constrained to the form of the $t$-matrix. 
Thus, one cannot assume a fixed form for the pair propagator, 
and then modify the Green's functions in the triangle
vertex appearing in the AL diagram~\cite{Levchenko_2011}, 
without also modifying the pair propagator.

Using the form of the bosonic EM vertex computed above, along with the
approximate form of the $t$-matrix in Eq.~(\ref{eq:tmat_propagator}),
and after performing integration by parts, the diamagnetic susceptibility reduces to
\begin{align}
\label{eq:chi_dia_AL}
\chi_{\dia}&=\frac{2Te^2}{3M_{\pair}}\sum_{\mathbf{p}}\left(\frac{p^{x}}{M_{\pair}}\right)^{2}\left[Zt(0,\mb{p})\right]^{3},\nonumber\\
&=-\frac{4Te^2}{9\pi^{2}}\int_{-\infty}^{\infty}dp\ \frac{p^{4}}{\left(p^{2}+2M_{\pair}\left|\mu_{\pair}\right|\right)^{3}}.
\end{align}
Note that $\mu_{\pair}=-|\mu_{\pair}|$ is negative, while $M_{\pair}$
is positive; this allows the spatial integral in the above expression to be computed. 
The $p$-integration is easily performed using a
closed contour integration in the upper half plane and evaluating the
residue at the pole $p=i\left(2M_{\pair}\left|\mu_{\pair}\right|\right)^{1/2}$.  
This gives the result presented in Eq.~(\ref{eq:chi_diamagnetism}) of the paper:
\begin{equation}\label{eq:chi_dia_AL2}
\chi_{\dia}=-\frac{T\left(2e\right)^2}{24\pi \hbar c^2}\sqrt{\frac{1/(2M_{\pair})}{\left|\mu_{\pair}\right|}}.
\end{equation}
Here the constants $\hbar$ and $c$ have been restored to ensure that $\chi_{\dia}$ is dimensionless. 
The diamagnetic susceptibility has been written in
this form to allow direct comparison with free bosonic transport. 
In appendix~(\ref{sec:app_Free_Bosons}) it is shown that, for free bosons, 
the diamagnetic susceptibility in the small chemical potential limit is
$\chi_{\b}=-\frac{T(e^{*})^2}{24\pi\hbar
  c^2}\sqrt{\frac{1/(2m_{\b})}{\left|\mu_{\b}\right|}}$.
Thus, in the small $|\mu_{\pair}|$ limit the diamagnetic susceptibility for the
$GG_{0}$ pair-fluctuation theory behaves like free bosons, but with
effective charge $e^{*}=2e$, mass $m_{\b}=M_{\pair}$, and chemical potential $\mu_{\b}=\mu_{\pair}$. 
The factor of 2 in the charge reflects the underlying internal fermionic constituents
of these composite bosons.

\section{Thermoelectric response}
\label{sec:Heat_Diagrammatic}

In this section we investigate thermoelectric response and the Nernst
coefficient in the presence of a normal state pseudogap.  
Here we follow the framework introduced in the previous sections in the analysis of the diamagnetic susceptibility. 
In contrast to the rather precise statements that were made about 
diamagnetic susceptibility in the pseudogap regime, for the Nernst response the situation is far more complex. 
Indeed, there is extensive controversy in the literature about this response function, 
even in the weak-fluctuation limit~\cite{Sergeev_1994,Sergeev_2008,Ussishkin_2003,Serbyn_2009}. 
There have also been attempts to study this quantity beyond the
weak-fluctuation limit~\cite{Ashvin_2007,ShinaNernst,Levchenko_2011}. 
To make progress, it will be useful to build on the more detailed and solid
understanding of diamagnetic susceptibility presented in the preceding sections.  
By again focusing on the central bosonic physics it is possible to express the Nernst coefficient 
in terms of bosonic response functions, with the parameters
$\mu_{\pair}$ and $M_{\pair}$ encapsulating pseudogap effects. 
The Nernst response in the weak-fluctuation limit will serve as a point of comparison.

The Nernst coefficient arises in transport by
applying a temperature gradient $(-\nabla T)_{x}$ in the presence of a
magnetic field $\mb{B}=B\hat{\mb{z}}$ and subsequently measuring the electric field
response $\mb{E}=E\hat{\mb{y}}$ (in the absence of a transport electric current).
This transport coefficient is defined by~\cite{Ussishkin_2003}
\begin{equation}
\nu_{N}=\frac{E}{\left(-\nabla T\right)_{x}B}=\frac{1}{B}\frac{\alpha_{xy}\sigma_{xx}-\alpha_{xx}\sigma_{xy}}{\sigma^{2}_{xx}+\sigma^{2}_{xy}}.
\end{equation}
For a particle-hole symmetric system (defined to mean a constant density of states near the Fermi
surface), $\sigma_{xy}=0$, so that $\nu_{N}=\alpha_{xy}/\left(B\sigma_{xx}\right).$ 
The Nernst coefficient is then reduced to calculating the transverse thermoelectric coefficient
$\alpha_{xy}$ and electrical conductivity $\sigma_{xx}$. 
For the $GG_{0}$ pair-fluctuation theory there is no particle-hole symmetry except in the BCS regime. 
Nevertheless, for the cuprates, which are of primary interest here, the transverse conductivity $\sigma_{xy}$
associated with the Hall effect is in general small: $\sigma_{xy} \ll \sigma_{xx}$~\cite{BehniaReview}.
Here we study only $\alpha_{xy}$ as an indication of the more complicated Nernst coefficient.

The Kubo formalism for thermal response is not as straightforward as it is for electric response.  
Indeed, the formulation of equilibrium linear response to a temperature change causes conceptual difficulties~\cite{Mahan}. 
One issue is that there is no unique definition of the heat-current vertex. 
Formally, the flow of heat corresponds to the flow of energy in the absence of the flow of matter~\cite{Kadanoff_Martin_1961, Mahan}. 
The heat current is thus equivalent to the energy current, and to derive the form
of the heat vertex we must investigate the consequences of energy conservation. 
As discussed in Sec.~(\ref{sec:EM_Response}),
the global $\mathrm{U}(1)$ particle number symmetry leads to a corresponding WTI. 
The same is true for energy conservation.
Indeed, this conservation law arises from the invariance of the Lagrangian of a theory under time translations. 
The corresponding WTI, which reflects the law of conservation of energy 
in terms of Green's functions, is~\citep{He_Levin_2014}
\begin{equation}\label{eq:energy_WTI}
q_{\mu}\Gamma^{\mu}_{\H}(k_{+},k_{-})=\omega_{-}G^{-1}(k_{+})-\omega_{+}G^{-1}(k_{-}),
\end{equation}
where $\omega_{\pm}\equiv\omega\pm\Omega/2$. Here $\Gamma^{\mu}_{\H}$ is the full heat  vertex. 
This form of the WTI  is not unique, and alternative forms can be derived by using
the equations of motion; for further details see Ref.~\onlinecite{He_Levin_2014}. 
The bare WTI for energy conservation,
$q_{\mu}\gamma^{\mu}_{\H}(k_{+},k_{-})=\omega_{-}G^{-1}_{0}(k_{+})-\omega_{+}G^{-1}_{0}(k_{-})$,
is satisfied by the bare heat vertex
$\gamma^{\mu}_{\H}(k_{+},k_{-})=\left(\gamma^{0}_{\H}(k_{+},k_{-}),\gamma^{i}_{\H}(k_{+},k_{-})\right)$,
where
$\gamma^{0}_{\H}(k_{+},k_{-})=\left(\mb{k}_{+}\cdot\mb{k}_{-}\right)/2m-\mu$
and
$\gamma^{i}(k_{+},k_{-})=\left[\omega_{+}k^{i}_{-}+\omega_{-}k^{i}_{+}\right]/(2m)$. 
The bare heat vertex
$\gamma^{i}_{\H}(k+q,k)=\left[i\omega_{n}(k^{i}+q^{i})+(i\omega_{n}+i\Omega_{m})k^{i}\right]/(2m)$
agrees with Ref.~\onlinecite{Kadanoff_Martin_1961}.

As in Sec.~(\ref{sec:EM_Diagrammatic}), the full heat vertex is found
by performing all possible bare heat vertex insertions in the self energy diagram. 
However, an additional vertex insertion arises from inserting 
the energy-momentum tensor interaction directly into the $t$-matrix. 
The final result is that the full heat vertex is 
\begin{align}\label{eq:Full_heat_vertex}
\Gamma^{\mu}_{\H}(k_{+},k_{-})&=\gamma^{\mu}_{\H}(k_{+},k_{-})+\MT^{\mu}_{\H}(k_{+},k_{-})+\lambda^{\mu}_{\H}(k_{+},k_{-})\nonumber\\
&\quad+\AL_{\H,1}^{\mu}(k_{+},k_{-})+\AL_{\H,2}^{\mu}(k_{+},k_{-}).
\end{align}
The Maki-Thompson, Aslamazov-Larkin, and $\lambda^{\mu}_{\H}$ heat-current vertices are 
\begin{widetext}
\begin{align}
\MT^{\mu}_{\H}(k_{+},k_{-})&=\sum_{p}t(p)G_{0}(p-k_{-})\gamma^{\mu}_{\H}(p-k_{-},p-k_{+})G_{0}(p-k_{+}),\label{eq:MT_heat_vertex}\\
\AL_{\H,1}^{\mu}(k_{+},k_{-})&=-\sum_{p}\sum_{l}t(p^{-})t(p^{+})G_{0}(p-k)G_{0}(p-l)G(l^{+})\Gamma^{\mu}_{\H}(l^{+},l^{-})G(l^{-}),\label{eq:AL1_heat_vertex}\\
\AL_{\H,2}^{\mu}(k_{+},k_{-})&=-\sum_{p}\sum_{l}t(p^{-})t(p^{+})G_{0}(p-k)G(p-l)G_{0}(l^{+})\gamma^{\mu}_{\H}(l^{+},l^{-})G_{0}(l^{-}),\label{eq:AL2_heat_vertex}\\
\lambda^{\mu}_{\H}(k_{+},k_{-})&=\sum_{p}g^{-1}\delta^{\mu 0}t(p^{+})t(p^{-})G_{0}(p-k_{+}).\label{eq:L_heat_vertex}
\end{align}
\end{widetext}
Here $\delta^{\mu0}$ is the Kronecker delta function, equal to unity
only for the time component ($\mu=0$) and zero otherwise. 
In appendix~(\ref{sec:WTI_Heat_Vertex}) an explicit calculation is
presented which shows that the full heat vertex, as determined by
Eqs.~(\ref{eq:Full_heat_vertex}-\ref{eq:L_heat_vertex}), satisfies the
WTI in Eq.~(\ref{eq:energy_WTI}).

Following Ref.~\onlinecite{Ussishkin_2003}, we consider the heat-current response to an applied electric field. 
The applied electric and magnetic fields are in the $\hat{x}$- and $\hat{z}$-directions, respectively, 
and the heat-current response is considered in the $\hat{y}$-direction. 
The correlation function of interest is then a heat-current-electric-current correlation function:
$P^{yx}_{\H\E}(i\Omega_{m},\mb{0})=2e\sum_{k}G(k_{+})\Gamma^{y}_{\H}(k_{+},k_{-})G(k_{-})\gamma^{x}_{\E}(k_{-},k_{+})$, where $\mb{q}=0$. 
The full heat vertex is determined in Eqs.~(\ref{eq:Full_heat_vertex}-\ref{eq:L_heat_vertex}), and since
only the $\hat{y}$-component is of interest, the vertex in Eq.~(\ref{eq:L_heat_vertex}) gives zero contribution.

Here we calculate the transverse thermoelectric coefficient only to linear order in magnetic field. 
This linearization results in performing all possible
($\hat{y}$-component) electromagnetic vertex insertions in the
heat-current-electric-current correlation function~\cite{Lee_1980}. 
The resulting correlation function is a three point correlation function: 
$\Lambda^{yyx}(i\Omega_{m},Q)$, where $\mb{Q}=Q\mb{\hat{x}}$ represents the momentum inserted into the
heat-current-electric-current correlation function. 
The transverse thermoelectric response, $\widetilde{\alpha}_{xy}$, can then be computed to linear
order in magnetic field $B$ using the definition
$\widetilde{\alpha}_{xy}=B\left[c\chi_{\dia}/\hbar-j_{y}/(EB)\right]$, where
the second term is determined from the Kubo formula~\cite{Ussishkin_2003}:
\begin{equation}\label{eq:thermoelectric_coeff}
\frac{j_{y}}{EB}=-\underset{\Omega,Q\rightarrow0}{\mathrm{lim}}\frac{1}{\Omega Qc}\mathrm{Re}
\left[\left.\Lambda^{yyx}(\Omega,Q)\right|_{i\Omega_{m}\rightarrow\Omega+i0^{+}}\right].
\end{equation}
The order of limits is crucial: first $Q\rightarrow0$, and then $\Omega\rightarrow0$.  
The need for including the magnetization
current in the above definition~\cite{Halperin_1997} is because they
contribute to the total microscopic current, and therefore must be
subtracted to obtain the transport current~\cite{Ussishkin_2003}.  
The parameter $\alpha_{xy}$ appearing in the Nernst coefficient is then
determined by $\alpha_{xy}=\widetilde{\alpha}_{xy}/T$.

The total number of EM vertex insertions is quite formidable, and an
exact theoretical treatment is challenging.  
In principle, if one inserts the full EM vertex into all the full Green's functions, the
bare EM vertex into all the bare Green's functions, and the
appropriate triangle vertices into all the $t$-matrices appearing in
$P^{yx}_{\H\E}(i\Omega_{m},\mb{0})$, then the full set of Feynman diagrams for the
heat-current response to an applied electric and magnetic field will be obtained. 
For the $GG_{0}$ pair-correlation theory in particular,
the various full Green's functions and full vertices present in the
response function means there will be a large number of diagrams to
consider, more so than in the weak-fluctuation case.

However, on the basis of the analysis performed in the previous section, 
and also from the near condensation calculations for the
weak-fluctuation theory~\citep{Ussishkin_2003}, it is expected that
only the AL diagrams with EM vertices inserted into the $t$-matrix give singular contributions. 
This is because such diagrams contain three $t$-matrix propagators, and thus in the small $|\mu_{\pair}|$ limit
they have a higher order in their degree of singularity than any other diagrams. 
See Ref.~\onlinecite{Ussishkin_2003} for the subtleties
involved in the power counting arguments related to the Nernst response.

Therefore, as an approximate calculation, we consider only the EM
vertex insertions in the $t$-matrices appearing in the two
AL diagrams that contribute to the heat-current-electric-current correlation function. 
There are two EM triangle vertices that can be inserted into each of the $t$-matrices
appearing in both $\AL_{\H,1}$ and $\AL_{\H,2}$. 
These arise from the $GG_{0}$ Green's functions appearing in the pair susceptibility, 
and thus either the full or bare EM vertex can be inserted into the
corresponding Green's function, which results in the two different types of EM triangle vertices.

Since the bosonic EM vertex $\Lambda_{\E,1}=\Lambda_{\E,2}$ is the same for both EM
triangle vertices appearing in $\AL_{\E,1}$ and $\AL_{\E,2}$
diagrams, this results in a symmetry factor of two. 
[For further details see appendix~(\ref{sec:Boson_Electric_Vertex})]. 
In addition there is another factor of two due to spin degeneracy for a system of spin-$\tfrac{1}{2}$fermions. 
Thus, the Nernst calculation is effectively reduced to calculating two AL
diagrams, plus their mirror images, with one corresponding bosonic
heat vertex ($\Lambda^{y}_{\H,1}$ or $\Lambda^{y}_{\H,2}$ depending on the diagram) 
and two bosonic EM vertices ($\Lambda^{y}_{\E,1}$, $\Lambda^{x}_{\E,1}$), 
multiplied by a symmetry factor of four.

There is extensive debate in the literature about the correct gauge-invariant approach to heat response~\cite{Sergeev_2008,Sergeev_2011}. 
Part of the issue concerns the appropriate diagrams to include, and how to ensure that gauge invariance is satisfied. 
For further discussion see also Refs.~\onlinecite{Varlamov_arxiv,Sergeev_arxiv}. 
Here we note that the full heat vertex presented in Eqs.~(\ref{eq:Full_heat_vertex}-\ref{eq:L_heat_vertex}) is consistent
with the WTI for energy conservation in Eq.~(\ref{eq:energy_WTI}).

Another issue under debate is the role of particle-hole asymmetry. 
In Ref.~\onlinecite{Sergeev_2008} it is claimed that the Nernst response
vanishes without particle-hole asymmetry. 
However, in Ref.~\onlinecite{Michaeli_2009} this claim is refuted. 
Indeed, for a normal Fermi metal that possesses particle-hole symmetry the Nernst coefficient is
(approximately) zero~\cite{Serbyn_2009,Michaeli_2009,Varlamov_arxiv}. 
In the weak-fluctuation case, however, the bosonic contribution to Nernst
response from the AL diagrams is found to be
significant~\cite{Serbyn_2009,Ussishkin_2003,Varlamov_arxiv}, even in
the absence of particle-hole asymmetry.

There is also contention in the Nernst literature~\cite{Sergeev_1994,Sergeev_2008,Ussishkin_2003,Serbyn_2009,Levchenko_2011}
concerning the specific form of the heat vertex appearing in the AL diagrams. 
This uncertainty is in contrast to the bosonic EM vertex, given in Eq.~(\ref{eq:boson_electric_vertex}). 
Following the EM vertex calculation, a similar analysis can be performed for the heat vertex. 
Since it is more involved, the derivation is presented in appendix~(\ref{sec:Boson_Heat_Vertex}). 
The result is that the sum of the heat triangle vertices for $\AL^{y}_{\H,1}$ and $\AL^{y}_{\H,2}$
reduces to a bosonic heat vertex, defined by:
$\Lambda^{y}_{\H,1}(p,p)+\Lambda^{y}_{\H,2}(p,p)\equiv\Lambda^{y}_{\H}(p,p)=-\varpi\left[\p t^{-1}(p)/\p p^{x}\right]$.

For comparison, the fermionic heat vertex obeys
$\Gamma^{y}_{\H}(k,k)=-\omega\left[\p G^{-1}(k)/\p k^{x}\right]$. 
Note, there is an additional factor of two compared to the EM case, which obeys:
$\Lambda^{y}_{\E,1}(p,p)+\Lambda^{y}_{\E,2}(p,p)\equiv\Lambda^{y}_{\E}(p,p)=2\left[\p
  t^{-1}(p)/\p p^{y}\right]$.
The Nernst literature~\cite{Sergeev_1994,Sergeev_2008,Ussishkin_2003,Serbyn_2009,Levchenko_2011} debates this factor of two;
in appendix~(\ref{sec:Boson_Heat_Vertex}) we provide our own
interpretation which makes the result less ambiguous. 
The point is that the heat and EM vertices, for fermions and bosons, are related by
$\Gamma^{y}_{\H}(k,k)=(\omega/e)\Gamma^{y}_{\E}(k,k)$, and
$\Lambda^{y}_{\H}(p,p)=(\varpi/e^{*})\Lambda^{y}_{\E}(p,p)$, 
where $e^{*}=2e$~\cite{Comment3}. 
Independent work~\cite{Narikiyo1_2011,Narikiyo2_2011} has also arrived at the same
conclusion, based on a similar derivation using the Ward-Takahashi identity.

Now we return to the calculation of the transverse thermoelectric coefficient. 
The previous analysis of the heat vertex means that the Nernst response is reduced to calculating one Aslamazov-Larkin diagram, 
plus its mirror image, with one bosonic heat vertex $(\Lambda^{y}_{\H})$ 
and two bosonic EM vertices $(\Lambda^{y}_{\E,1}, \Lambda^{x}_{\E,1})$,
multiplied by a symmetry factor of four. 
It is important to note that in combining the two heat vertices $\Lambda^{y}_{\H,1}$ and
$\Lambda^{y}_{\H,2}$ into one bosonic heat vertex $\Lambda^{y}_{\H}$
the number of diagrams that need to be computed has effectively been reduced by a factor of two. 
Thus, the three point correlation function that needs to be computed is
\begin{widetext}
\begin{align}
\Lambda^{yyx}(i\Omega_{m},Q)&=-4e^2\sum_{p}\biggl[\frac{Zp^{x}_{+}}{M_{\pair}}\left(\frac{Zp^{y}}{M_{\pair}}\right)^{2}\left(i\varpi_{m}+i\Omega_{m}/2\right)
t(i\varpi_{m}+i\Omega_{m}/2,\mb{p}_{+})t(i\varpi_{m},\mb{p}_{-})t(i\varpi_{m},\mb{p}_{+})\nonumber\\
&\hspace{1.55cm}+\frac{Zp^{x}_{-}}{M_{\pair}}\left(\frac{Zp^{y}}{M_{\pair}}\right)^{2}\left(i\varpi_{m}-i\Omega_{m}/2\right)
t(i\varpi_{m}-i\Omega_{m}/2,\mb{p}_{-})t(i\varpi_{m},\mb{p}_{+})t(i\varpi_{m},\mb{p}_{-})\biggr],
\end{align}
\end{widetext}
where $\mb{p}_{\pm}\equiv\mb{p}\pm\mb{Q}/2$.

Performing the Matsubara frequency summation, and then taking the limits
$Q\rightarrow0$, followed by $\Omega\rightarrow0$ in the Kubo formula given in Eq.~(\ref{eq:thermoelectric_coeff}), we obtain 
\begin{equation}\label{eq:thermoelectric_coeff2}
\frac{j_{y}}{EB}=-\frac{Te^2}{4\pi\hbar^2 c}\sqrt{\frac{1/(2M_{\pair})}{\left|\mu_{\pair}\right|}}
\,\left(\frac{\kappa^2+\Gamma^2}{\Gamma^2}\right).
\end{equation}
Here the constants $\hbar$ and $c$ have been restored. 
For further details of the calculation see appendix~(\ref{sec:Thermoelectric_Coefficient}). 
The transverse thermoelectric coefficient, $\widetilde{\alpha}_{xy}$, is then found by
combining Eq.~(\ref{eq:chi_dia_AL2}) and Eq.~(\ref{eq:thermoelectric_coeff2}) and using the definition
$\widetilde{\alpha}_{xy}=B\left[c\chi_{\dia}/\hbar-j^{y}/(EB)\right]$; 
this gives the result stated in Eq.~(\ref{eq:alpha_thermoelectric}) at the beginning of the paper:
\begin{equation}\label{eq:thermoelectric_coeff3}
\widetilde{\alpha}_{xy}=\frac{BTe^2}{12\pi\hbar^2 c}\sqrt{\frac{1/(2M_{\pair})}{\left|\mu_{\pair}\right|}}
\,\left(\frac{3\kappa^2+\Gamma^2}{\Gamma^2}\right).
\end{equation}
Similar results can be obtained from Ref.~\onlinecite{ShinaNernst}.
Just as for the diamagnetic susceptibility, the transverse thermoelectric coefficient 
in Eq.~(\ref{eq:thermoelectric_coeff3}) is large when $|\mu_{\pair}|\ll T_{c}$. 
In the $\kappa\rightarrow0$ limit,  Eq.~(\ref{eq:thermoelectric_coeff3}) reproduces the result in
the weak-fluctuation literature~\cite{Ussishkin_2003}. 
It is of interest to note that whereas diamagnetic susceptibility is insensitive to the parameter $\Gamma$, 
the transverse thermoelectric coefficient depends crucially on this parameter. 
The parameter $\Gamma$ serves as a regularization
for the transverse thermoelectric coefficient in the BEC limit, whereas for
diamagnetic susceptibility such a regularization is not required.

The ratio of the absolute magnetization to the transverse thermoelectric
coefficient has received a lot of interest~\cite{Ashvin_2007}; 
in the weak-fluctuation limit this ratio is exactly $2(\hbar/c)$:
$|Bc\chi_{\dia}|/\left(\hbar\widetilde{\alpha}_{xy}\right)=2$, and in
the phase-only fluctuation picture this ratio is obtained in the large temperature limit~\cite{Ashvin_2007}. 
From the results in Eq.~(\ref{eq:chi_dia_AL2}) and Eq.~(\ref{eq:thermoelectric_coeff3}),
we find this ratio to be
$|Bc\chi_{\dia}|/\left(\hbar\widetilde{\alpha}_{xy}\right)=2\left[1-3\kappa^{2}/\left(3\kappa^2+\Gamma^2\right)\right]$. 
In the weak-fluctuation limit $\kappa=0$, and we recover the standard result. 
More generally, the BCS limit is $\Gamma\gg\kappa$, so that
$|Bc\chi_{\dia}|/(\hbar\widetilde{\alpha}_{xy})\rightarrow2$, however,
the BEC limit is $\Gamma\ll\kappa$, so that
$|Bc\chi_{\dia}|/(\hbar\widetilde{\alpha}_{xy})\rightarrow(2/3)\left(\Gamma/\kappa\right)^2$. 
In the intermediate pseudogap regime, where both $\kappa,\Gamma\neq0$,
the ratio is in between these two limits; it decreases as the pairing strength increases.

In summary, the singular nature of the diamagnetic susceptibility 
and transverse thermoelectric coefficient shows the importance
of including fluctuating bosonic degrees of freedom.  
The next section presents numerical results for the diamagnetic susceptibility, 
which depends on the effects of the normal state gap through the parameters $M_{\pair}$ and $\mu_{\pair}$.

\section{Numerical Results}
\label{sec:Numerics}

\begin{figure}
\includegraphics[width=3in,clip]{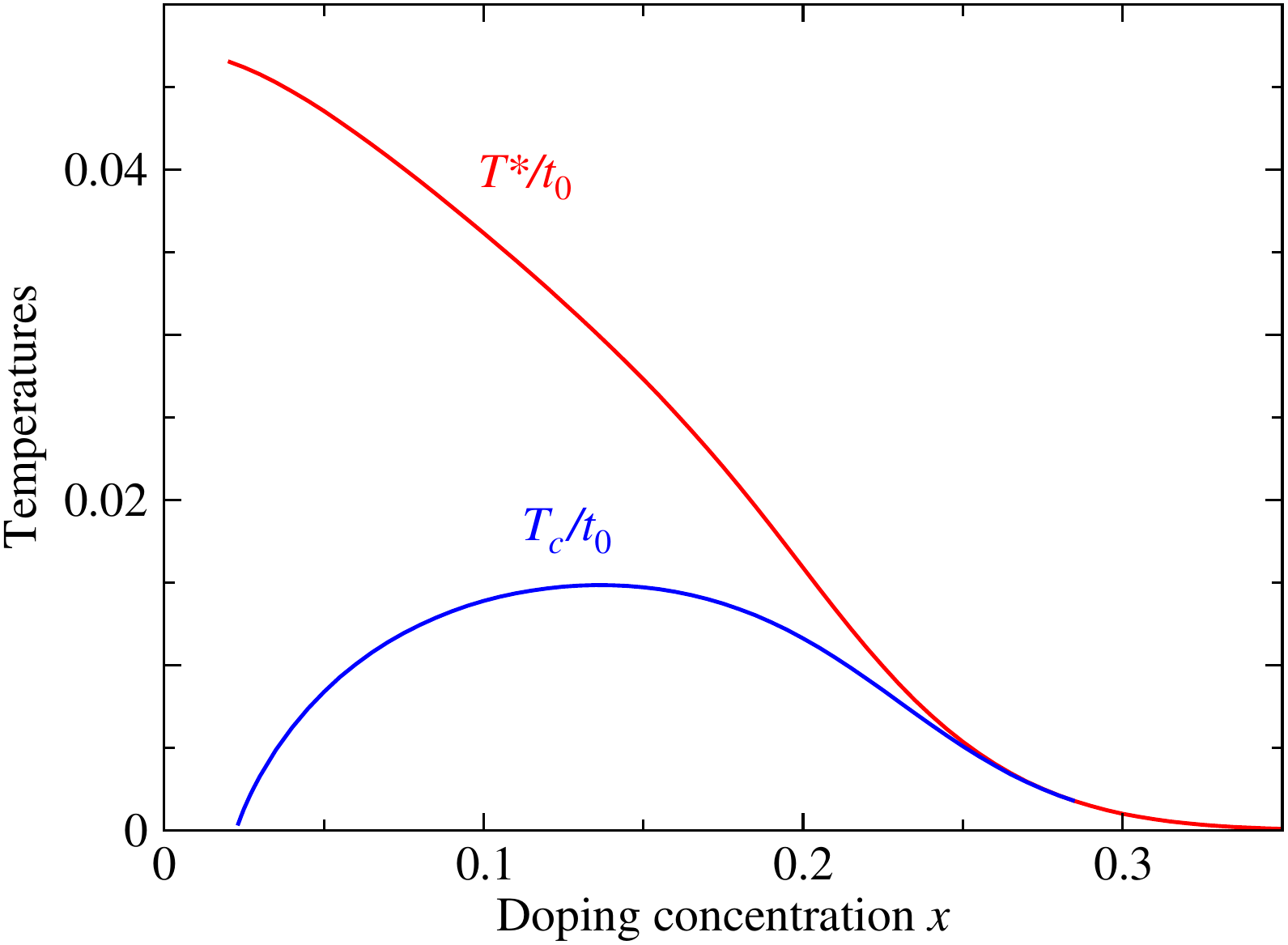}
\caption{Theoretically calculated cuprate phase diagram. Plotted are
  $T_c$ and $T^*$, in units of $t_0$, as functions of doping
  concentration $x$.}
\label{fig:Phase_diagram}
\end{figure}

We now present the results of our numerical calculations of cuprate
diamagnetic susceptibility, along with a comparison to experimental data. 
To compare between theory and experiment, it is first necessary to start with a
semi-quantitative understanding of the phase diagram. 
Section~(\ref{sec:Pair_Propagator}) outlined a procedure to
compute both $T_{c}$ and $T^{*}$ using the $t$-matrix of Eq.~(\ref{eq:Tmat}). 
The resulting phase diagram is shown in Fig.~(\ref{fig:Phase_diagram}), 
which plots both $T_{c}$ (blue) and $T^{*}$ (red) curves as functions of doping concentration, $x$. 
Qualitatively the horizontal axis is a measure of the
dimensionless interaction strength with stronger (weaker) interaction
effects on the left (right) side, reflecting underdoped (overdoped) cuprates. 
This figure generically indicates what occurs in the
weak-interaction regime (on the right), where $T_{c}\approx T^{*}$,
and the strong interaction regime (on the left), where $T_{c}$ and $T^{*}$ are anti-correlated.

This anti-correlation can be understood from the early work of
Nozi{\`e}res and Schmitt-Rink~\cite{NSR_1985} who showed that, on a lattice, 
as the attraction becomes stronger it becomes increasingly
difficult for pairs to hop, since they first need to unbind. 
This is responsible for the large pair mass. In the $d$-wave case the effects
are more extreme~\cite{Chen_99} as the pairs are more extended in size. 
At sufficiently strong attraction a superconductor-insulator transition is observed. 
This explains the behavior at the lower critical doping of the $T_{c}$ dome.

The phase diagram in Fig.~(\ref{fig:Phase_diagram}) is based on a
nearest-neighbor quasi-two-dimensional tight-binding band structure. 
The cuprate half bandwidth $4t$ for the in-plane dispersion
sets the scale for the units of energy.  
The anisotropy parameter is taken as $t_{z}/t=0.003$, 
in agreement with estimates for BSCCO and LSCO superconductors. 
Note that the $T_{c}$ curve depends on $t_{z}/t$ only logarithmically. 
Details of the band parameters are not
particularly important, provided they are chosen to capture the
generic effect that $T^{*}$ increases with under doping. 
Our calculations have included the doping concentration $x$ and
\textit{doping independent} interaction strength $g$. 
The $x$ dependence is included in the hopping integral $t$ in the form
$t\approx t_{0}x$, where $t_{0}$ is an energy scale characteristic of the parent compound. 
We choose the dimensionless ratio $-g/4t_0= 0.04725$ to optimize the fit to $T^*$.  
For the moment only the single parameter $t_0$ is left unspecified.

\begin{figure}
\centerline{\includegraphics[width=3in,clip]{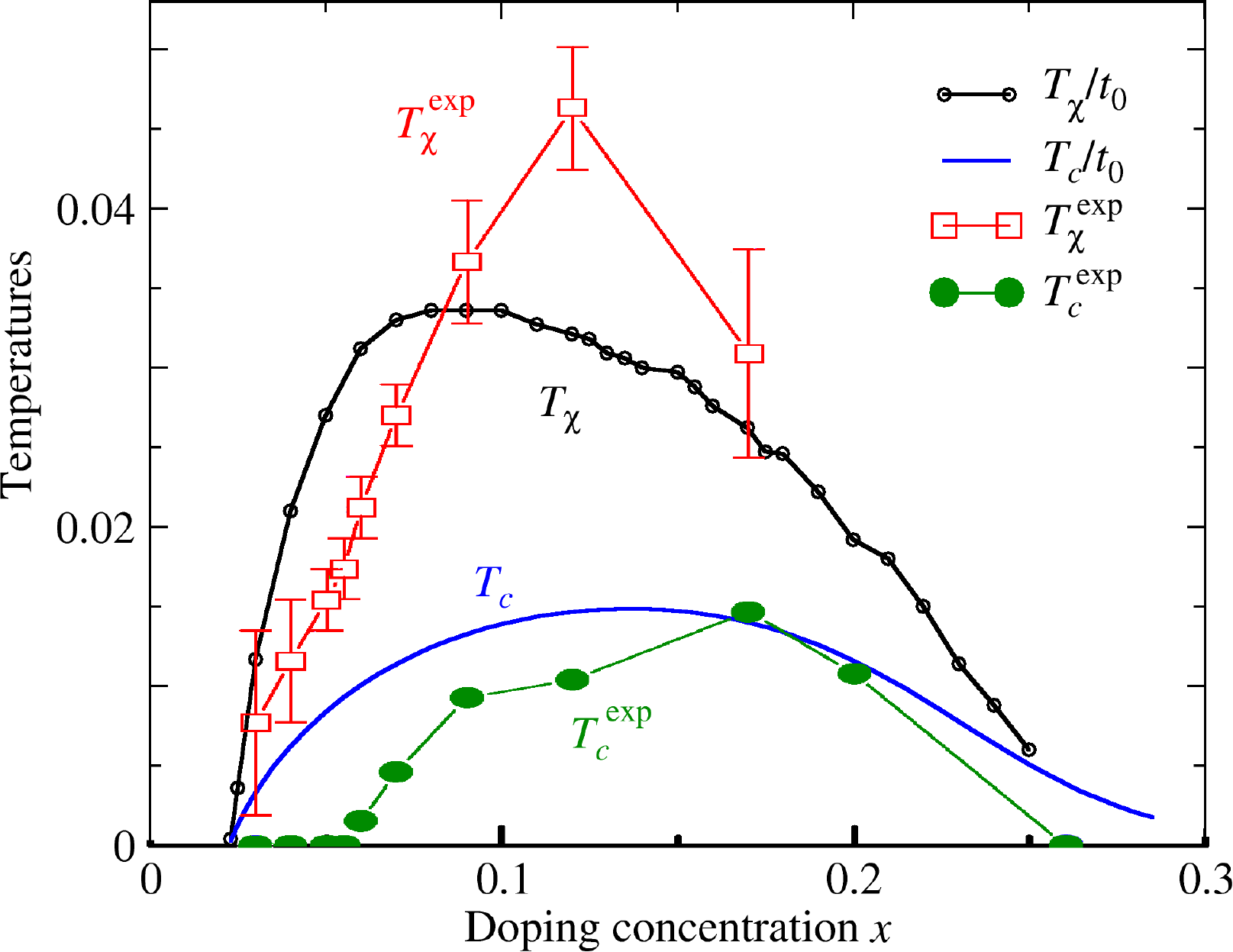}}
\caption{Doping dependence of the calculated diamagnetic
  susceptibility onset temperature $T_{\chi}$ (black), and $T_{c}$
  (blue), along with corresponding experimental data from
  Ref.~\onlinecite{Ong2010} for $T_{\chi}$ (red squares) and $T_c$
  (green discs). For both cases, the maximum of $T_{\chi}$ is skew
  towards the underdoped regime.}
\label{fig:Tchi}
\end{figure}

The plot in Fig.~(\ref{fig:Phase_diagram}) shows that $T_{c}$ vanishes
at a lower critical doping of $x=0.025$, which is slightly less than
the experimental value of $x=0.05$~\cite{Ong2010}.  
Nevertheless the overall shape as compared with experiment, shown later in the paper in
Fig.~(\ref{fig:Tchi}), for $T_{c}$ (and $T^{*}$), is reasonable.
While not shown in Fig.~(\ref{fig:Phase_diagram}), at each value of $x$ the magnitude 
of the pairing gap $\Delta$ (or pseudogap), at $T_{c}$ for example, 
shows an approximate proportionality to $T^{*}$.

The diamagnetic susceptibility has a singular inverse square root
dependence on the bosonic chemical potential, $\mu_{\pair}$, as
derived in Eq.~(\ref{eq:chi_dia_AL2}).  This pair chemical potential
itself varies with temperature and doping concentration.  It is
useful, then, to first study $\mu_{\pair}$ and compare with the
weak-fluctuation limit.  The combination $-Z\mu_{\pair}$ in the
strong-pairing theory can be viewed as equivalent to $N_{0}\epsilon$
in the weak-fluctuation theory, where $N_{0}$ is the fermionic density
of states at the Fermi surface and $\epsilon=\ln(T/T_{c})$.  The
parameter $Z$, which is the prefactor of proportionality in the
$t$-matrix defined in Eq.~(\ref{eq:tmat_propagator}), is associated
with the linear frequency contribution to the inverse $t$-matrix.

\begin{figure}
\includegraphics[width=3in,clip]{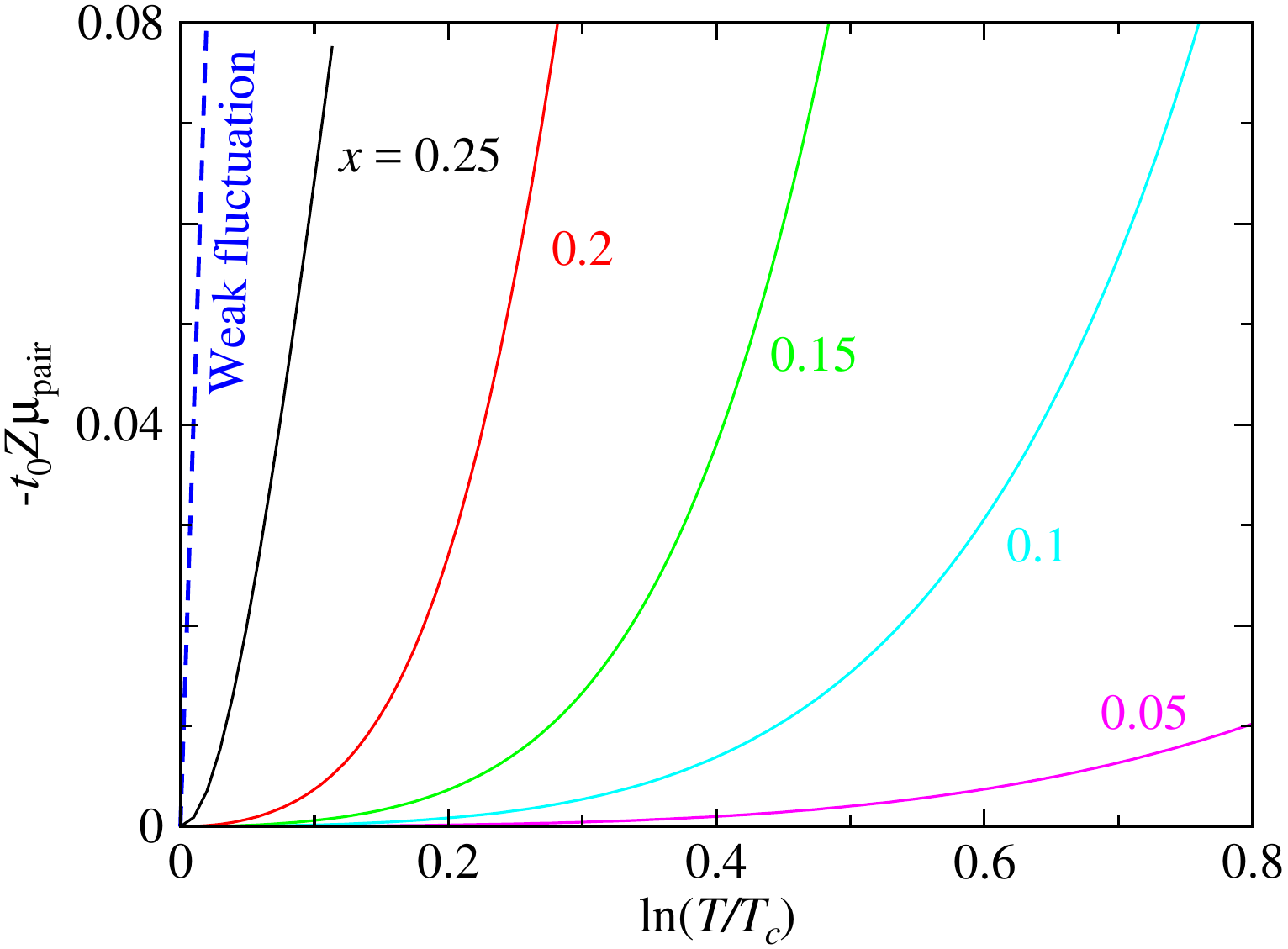}
\caption{The product $-t_{0}Z\mu_{\pair}$ as a function of
  $\epsilon=\ln (T/T_c)$ for various doping concentrations $x$ from
  overdoped $(x=0.25)$ to underdoped $(x=0.05)$. For comparison, the
  result from the weak-fluctuation formalism (blue dashed straight line) is also plotted.}
\label{fig:mupair}
\end{figure}

In Fig.~(\ref{fig:mupair}) we plot the product $-t_{0}Z\mu_{\pair}$,
as a function of $\ln (T/T_c)$, for different doping concentrations as
labeled, from the overdoped ($x=0.25$) to the strongly underdoped ($x=0.05$) limit.  
The blue dashed straight line is $N_{0}\epsilon$ in the weak-fluctuation theory.  
To make this comparison we have estimated the fermion mass on a quasi-2D lattice 
at the Fermi level using an in-plane Fermi wavevector $\mb{k} = 0.9(\pi/2, \pi/2)$ along
the nodal direction, which yields an effective fermion mass $m = 1/(0.31t)$.  
This leads to the association $t_0N_0 = t_0(m/\pi) = 4.1$ (for $x=0.25$), 
which sets the slope of the blue dashed line.

It is evident from Fig.~(\ref{fig:mupair}) that, as the magnitude of
the pseudogap increases from the overdoped to underdoped regime,
$-t_{0}Z\mu_{\pair}$ decreases rapidly for a given $\epsilon$. 
This means that in the underdoped regime there is a larger range of
temperatures where $|\mu_{\pair}|$ is effectively ``small". 
As a consequence, the strong-pairing fluctuation theory has a large 
diamagnetic susceptibility at temperatures higher than is the case for the diffusive, weak-fluctuation theory.

Now we are in a position to study the behavior of the normal state
diamagnetic susceptibility. In order to calculate $\chi_{\dia}$ for
the quasi-2D cuprates, in an extended range of temperatures above
$T_c$, we use Eq.~(\ref{eq:chib}) in
appendix~(\ref{sec:app_Free_Bosons}), with $e^*=2e$, $m_{\b}$ replaced
by the in-plane pair mass $M_\parallel$, and $\xi_{\mb{p}}$ replaced
by the appropriate anisotropic pair dispersion $\Omega_\mathbf{p}$~\cite{Comment0}. 
Of particular importance is the
onset temperature~\cite{Ong2010}, $T_{\chi}$. This is the temperature
at which the total magnetic susceptibility departs from the background contribution. 
We consider the background contribution to the total
magnetic susceptibility as arising from the (Pauli)
paramagnetic~\cite{noteonPauli} contribution, $\chi_{\Pauli}$, associated with the fermionic quasi-particles. 
The crucial contribution in this analysis is the diamagnetic susceptibility,
$\chi_{\dia}$, which is dominated by the bosonic pairing fluctuations.
The total magnetic susceptibility is then $\chi=\chi_{\Pauli}+\chi_{\dia}$.

From a theoretical point of view it is reasonable to view the dominant
background contribution to be based on $\chi_{\Pauli}$.  
The experimental background~\cite{Ong2010}, however, indicates that the
Pauli contribution is relatively insignificant compared to a much
larger Van Vleck paramagnetic term, $\chi_{\VV}$.  
This Van Vleck contribution is difficult to theoretically calculate from first principles. 
The experimental data suggests that $\chi_{\VV}$ is
approximately $1-2$ orders of magnitude larger than $\chi_{\Pauli}$.
Accordingly we adjust the vertical scale of our total magnetic
susceptibility to give an analogous effect to the experimentally
measured background term.

\begin{figure}\centering
\includegraphics[width=3in,clip]{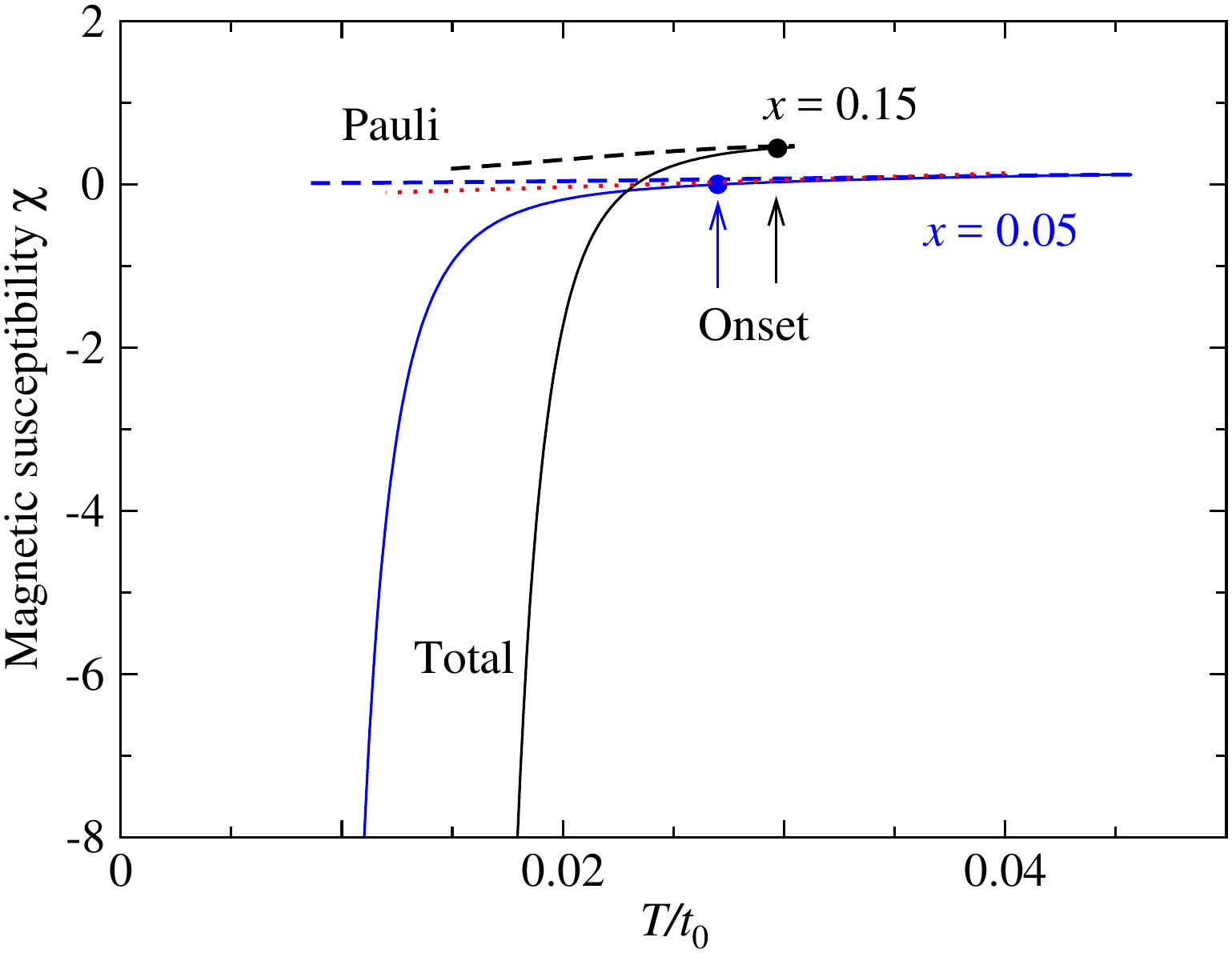}
\caption{Magnetic susceptibility above $T_c$ at optimal doping
  $x=0.15$ (black) and underdoping $x=0.05$ (blue).  The dashed lines
  are the Pauli paramagnetic susceptibility and the solid lines are
  the sum of the paramagnetic and diamagnetic contributions to
  magnetic susceptibility.  The solid dots indicate the temperature,
  $T_{\chi}$, at which the onset of diamagnetic susceptibility occurs.
  For the underdoped case, the red dotted line is a linear fit to the
  high temperature data.}
\label{fig:chi_dia}
\end{figure}

In Fig.~(\ref{fig:chi_dia}) we indicate this procedure. 
We focus on two representative examples for the optimal doping case $x = 0.15$
(black curves) and the underdoped case $x=0.05$ (blue curves) to
illustrate how the diamagnetic onset temperatures are determined. 
This onset is indicated in the figure by the colored dot. For the former
case, the onset is simply given by the departure temperature of the
total susceptibility (solid curve) from the Pauli background (dashed curve). 
For the strongly underdoped (blue curve) case, there is a
large temperature regime above $T_{\chi}$ where $\chi_{\dia}$ is small but nonzero. 
In such cases, for example $x=0.05$, we closely follow
the experimental procedure by fitting $\chi$ in this regime with a
(red dotted) straight line and then determine $T_{\chi}$ by where
$\chi$ departs from this line.

Summarizing our results, the extracted diamagnetic susceptibility
onset temperature, $T_{\chi}$, is plotted in Fig.~(\ref{fig:Tchi}) as
the open black circles, while the experimental data from
Ref.~\onlinecite{Ong2010} is shown in the open red squares. 
The theoretical and experimental transition temperatures are also plotted.
We determine the previously unspecified energy scale $t_{0}$ by
fitting the theoretical $T_{c}$ curve to the experimentally measured
$T_{c}$ near optimal doping $(x=0.15)$.

Our theoretically calculated diamagnetic susceptibility onset
temperatures are found to be in reasonable agreement with the
experimental data in the underdoped and overdoped cases. 
The theoretical plot has a peak in $T_{\chi}$ which is skew-symmetric
towards the underdoped regime; this is a feature also exhibited in the experimental data. 
The experimentally observed decrease in $T_{\chi}$
in the underdoped regime is a feature which is captured in the theoretical plot. 
This is an important theoretical finding because
this regime is outside the applicability of the weak-fluctuation theory.  
The theoretical $T_{\chi}$ vanishes simultaneously with
$T_{c}$ as the doping concentration approaches its lower critical value~\cite{Comment4}.

The peak in $T_{\chi}$ in the experimental data is, however, slightly higher than the peak in the theoretically predicted values.  
More experimental data and a better theoretical treatment of the background contribution would aid in this regard.  
Nonetheless, the prediction that there is a significant high temperature contribution to
diamagnetic susceptibility due to strong pair fluctuations, 
as expressed in Eq.~(\ref{eq:chi_dia_AL2}), is captured in the theoretical figures.

While a comparison between the theoretical and experimental Nernst coefficient has not been presented, 
it should be pointed out that the onset temperature of the 
transverse thermoelectric response is expected to be roughly the same as $T_{\chi}$.  
This follows from the simple proportionality between absolute magnetization and transverse thermoelectric coefficient:
$\propto 2\Gamma^{2}/(3\kappa^{2}+\Gamma^{2})$.  
We note that the parameters $\kappa$ and $\Gamma$ are weakly temperature dependent, 
so that $T_{\chi}$ is a reasonable estimate for this thermoelectric onset temperature.  
This observation appears consistent with experimental claims in Ref.~\onlinecite{Ong2010}.

\section{Conclusions}
\label{sec:Conclusions}

In this paper we have studied the effects of a normal state pseudogap on the  diamagnetic susceptibility 
and transverse thermoelectric coefficient, as applied to the hole doped cuprates. 
Strong support for a cuprate pseudogap deriving from a ``preformed-pair'' scenario 
comes from the anomalous enhancement in both these quantities.  
An essential addition to the literature then is a calculation of these transport coefficients,
which incorporates into the underlying
response theory the presence of a pseudogap itself. 
This paper achieves this goal, by using a strong-pairing fluctuation theory in
which the dominant contributions to the diamagnetic susceptibility and 
transverse thermoelectric coefficient come from modified Aslamazov-Larkin diagrams. 
This differs from the conventional weak-pairing fluctuation theory in which
these two transport coefficients were derived in the absence of a normal state gap. 
By incorporating longer lived and more stable fermion pairs, 
we find our calculations compare favorably with their
experimental counterparts over the broad range of hole doping concentrations.

These results are obtained through detailed diagrammatic calculations
which are tightly constrained by the Ward-Takahashi identity. 
They depend importantly on the associated form of the pair propagator,
which differs from its weak fluctuation analogue in large part because
the pairs have propagating rather than diffusive dynamics. 
We have emphasized in this paper that the calculation related to the
thermoelectric coefficient is not at the same level of rigor as
that for diamagnetic susceptibility, which from our perspective is quite
precise. Nevertheless there is a fair degree of confidence that, just
as for the diamagnetic susceptibility, the important parameter controlling the
singular behavior in the transverse thermoelectric coefficient  is of the form
$\sqrt{1/(2M_{\pair}\left|\mu_{\pair}\right|)}$.

More generally we note the similarity between the transverse thermoelectric
coefficient in Eq.~(\ref{eq:thermoelectric_coeff3}) and the
diamagnetic susceptibility in Eq.~(\ref{eq:chi_dia_AL2}). 
The first of these also depends on additional parameters $\Gamma$ and $\kappa$; 
while the former reflects the pair damping, 
the latter reflects the particle-hole asymmetry which accompanies long lived pairs. 
It is clear from the expressions in
Eq.~(\ref{eq:chi_dia_AL2}) and Eq.~(\ref{eq:thermoelectric_coeff3})
that the simple ratio of $2\hbar/c$ between the absolute magnetization and transverse thermoelectric coefficient, 
in the linear magnetic field regime, is only expected to be correct in the weak-fluctuation limit. 
As pairing becomes stronger, the transverse thermoelectric response 
becomes progressively larger than its diamagnetic counterpart. 
This is because the pairs become longer lived
so that $\Gamma$ becomes much smaller than $\kappa$.

The diamagnetic susceptibility and transverse thermoelectric coefficient are dependent on two key
parameters: the pair mass $M_{\pair}$ and the pair chemical potential $\mu_{\pair}$. 
In the cuprates we find both parameters vary with hole doping concentration, $x$.
They also both reflect, in slightly different ways, 
the two important temperatures $T^{*}$ (pairing onset
temperature) and $T_{c}$ (phase transition temperature). 
In the simplest terms, $M_{\pair}(T,x)$ is more directly reflective of $T_c(x)$
since we find the phase transition temperature vanishes when $M_{\pair}$ diverges. 
By contrast $\mu_{\pair}(T,x)$ is more directly reflective of
$T^*(x)$ since (as we have shown) a higher pairing onset temperature leads to a
stabilization of the pairs and to a reduction in their chemical potential.
In this way, both temperature scales play an important role in
establishing the behavior of the diamagnetic susceptibility and
Nernst coefficient in the high temperature superconductors.

\begin{acknowledgments}
This work was supported by NSF-DMR-MRSEC 1420709, NSF of China (Grant No.~11274267), and NSF of Zhejiang Province of China (Grant No.~LZ13A040001). 
It is our pleasure to thank Alexey Galda for many fruitful discussions and correspondence on this topic.
\end{acknowledgments}

\appendix
\numberwithin{equation}{section}
\numberwithin{figure}{section}
\begin{widetext}
\section{Obtaining the full EM vertex using the Ward-Takahashi identity}
\label{sec:app_Full_EM_Vertex}

\subsection{\texorpdfstring{$GG_{0}$}{GG0} pair-fluctuation theory}
In this section the Ward-Takahashi identity (WTI) is used to derive the full electromagnetic (EM) vertex for the $GG_{0}$ pair-fluctuation theory, which appears in Eqs.~(\ref{eq:Full_vertex}-\ref{eq:AL2_vertex}) of the main text. 
The WTI for the full EM vertex is~\cite{Ryder}
\begin{align}\label{eq:App_WTI}
q_{\mu}\Gamma_{\E}^{\mu}(k_{+},k_{-})&=G^{-1}(k_{+})-G^{-1}(k_{-}),\nonumber\\
&=q_{\mu}\gamma_{\E}^{\mu}(k_{+},k_{-})+\Sigma(k_{-})-\Sigma(k_{+}).
\end{align}
Here $k_{\pm}\equiv k\pm q/2$. The self energy for the $GG_{0}$ pair-fluctuation theory is $\Sigma(k)=\sum_{p}t(p)G_{0}(p-k)=\sum_{p}t(p+k)G_{0}(p)$,
where the $t$-matrix is defined through the pair susceptibility by $t^{-1}(p)=g^{-1}+\Pi(p)$, with $\Pi(p)=\sum_{l}G_{0}(p-l)G(l)=\sum_{l}G(p-l)G_{0}(l)$ the definition of the pair susceptibility. 
Throughout this paper $k, l$ denote fermionic four-momenta: $k^{\mu}=(i\omega_{n},\mb{k})$, 
$l^{\mu}=(i\epsilon_{n},\mb{l})$, where $\omega_{n}$ and $\epsilon_{n}$ are fermionic Matsubara frequencies, whereas $p, q$ denote bosonic four-momenta: 
$p^{\mu}=(i\varpi_{m},\mb{p})$, $q^{\mu}=(i\Omega_{m},\mb{q})$, where $\varpi_{m}$ and $\Omega_{m}$ are bosonic Matsubara frequencies. 
Using the two equivalent forms of the self energy given above, the self energy difference appearing in Eq.~(\ref{eq:App_WTI}) becomes
\begin{align}\label{eq:App_selfenergy_diff}
\Sigma(k_{-})-\Sigma(k_{+})&=\sum_{p}t(p)G_{0}(p-k_{-})\left[G^{-1}_{0}(p-k_{-})-G^{-1}_{0}(p-k_{+})\right]G_{0}(p-k_{+})\nonumber\\
&\quad+2\sum_{p}G_{0}(p)t(p+k_{+})\left[t^{-1}(p+k_{+})-t^{-1}(p+k_{-})\right]t(p+k_{-}).
\end{align}

By using the bare WTI, the term in square brackets on the first line is given by the contraction $q_{\mu}\gamma_{\E}^{\mu}(p-k_{-},p-k_{+})$. 
From the definition of the $t$-matrix, the difference of the two inverse $t$-matrices is
\begin{equation}\label{eq:APP_tmat_diff}
t^{-1}(p+k_{+})-t^{-1}(p+k_{-})=\Pi(p+k_{+})-\Pi(p+k_{-}).
\end{equation}
Using the two equivalent forms of the pair susceptibility, the pair susceptibility difference becomes
\begin{align}
2\left[\Pi(p+k_{+})-\Pi(p+k_{-})\right]&=-\sum_{l}G_{0}(l)G(p+k_{+}-l)\left[G^{-1}(p+k_{+}-l)-G^{-1}(p+k_{-}-l)\right]G(p+k_{-}-l)\nonumber\\
&\quad-\sum_{l}G(l)G_{0}(p+k_{+}-l)\left[G^{-1}_{0}(p+k_{+}-l)-G^{-1}_{0}(p+k_{-}-l)\right]G_{0}(p+k_{-}-l).
\end{align}
From the WTI, the first term in square brackets is the contraction $q_{\mu}\Gamma_{\E}^{\mu}(p+k_{+}-l,p+k_{-}-l)$, similarly 
the second term in square brackets is the contraction $q_{\mu}\gamma_{\E}^{\mu}(p+k_{+}-l,p+k_{-}-l)$. 
Inserting these results into Eq.~(\ref{eq:App_selfenergy_diff}) and Eq.~(\ref{eq:APP_tmat_diff}) then gives
\begin{align}
\Sigma(k_{-})-\Sigma(k_{+})&=\sum_{p}t(p)G_{0}(p-k_{-})q_{\mu}\gamma_{\E}^{\mu}(p-k_{-},p-k_{+})G_{0}(p-k_{+})\nonumber\\
&\quad-\sum_{p}G_{0}(p)t(p+k_{+})\sum_{l}G_{0}(l)G(p+k_{+}-l)q_{\mu}\Gamma_{\E}^{\mu}(p+k_{+}-l,p+k_{-}-l)G(p+k_{-}-l)t(p+k_{-})\nonumber\\
&\quad-\sum_{p}G(p)t(p+k_{+})\sum_{l}G(l)G_{0}(p+k_{+}-l)q_{\mu}\gamma_{\E}^{\mu}(p+k_{+}-l,p+k_{-}-l)G_{0}(p+k_{-}-l)t(p+k_{-}).
\end{align}
In the second and third lines, first let $p\rightarrow p-k$, and then after that let $l\rightarrow p-l$. 
Inserting the resulting expression into Eq.~(\ref{eq:App_WTI}), and solving for the full EM vertex, then gives the following result:
\begin{align}\label{eq:App_full_vertex}
\Gamma_{\E}^{\mu}(k_{+},k_{-})&=\gamma_{\E}^{\mu}(k_{+},k_{-})\nonumber\\
&\quad+\sum_{p}t(p)G_{0}(p-k_{-})\gamma_{\E}^{\mu}(p-k_{-},p-k_{+})G_{0}(p-k_{+})\nonumber\\
&\quad-\sum_{p}\sum_{l}t(p_{-})t(p_{+})G_{0}(p-k)G_{0}(p-l)G(l_{+})\Gamma_{\E}^{\mu}(l_{+},l_{-})G(l_{-})\nonumber\\
&\quad-\sum_{p}\sum_{l}t(p_{-})t(p_{+})G_{0}(p-k)G(p-l)G_{0}(l_{+})\gamma_{\E}^{\mu}(l_{+},l_{-})G_{0}(l_{-}).
\end{align}
This reproduces the full EM vertex given in Eqs.~(\ref{eq:Full_vertex}-\ref{eq:AL2_vertex}) of the main text. 
The first line is the bare EM vertex $\gamma_{\E}^{\mu}$, the second line is the Maki-Thompson vertex $\MT^{\mu}_{\E}$, the third line is the Aslamazov-Larkin vertex $\AL^{\mu}_{\E,1}$, 
and the fourth line is the other Aslamazov-Larkin vertex $\AL^{\mu}_{\E,2}$. The Feynman diagrams for the full EM vertex are given in Fig.~(\ref{fig:Full_vertex_diagrams_GG0}).

\begin{figure*}[t]
\centering\includegraphics[width=12cm,height=10cm,clip]{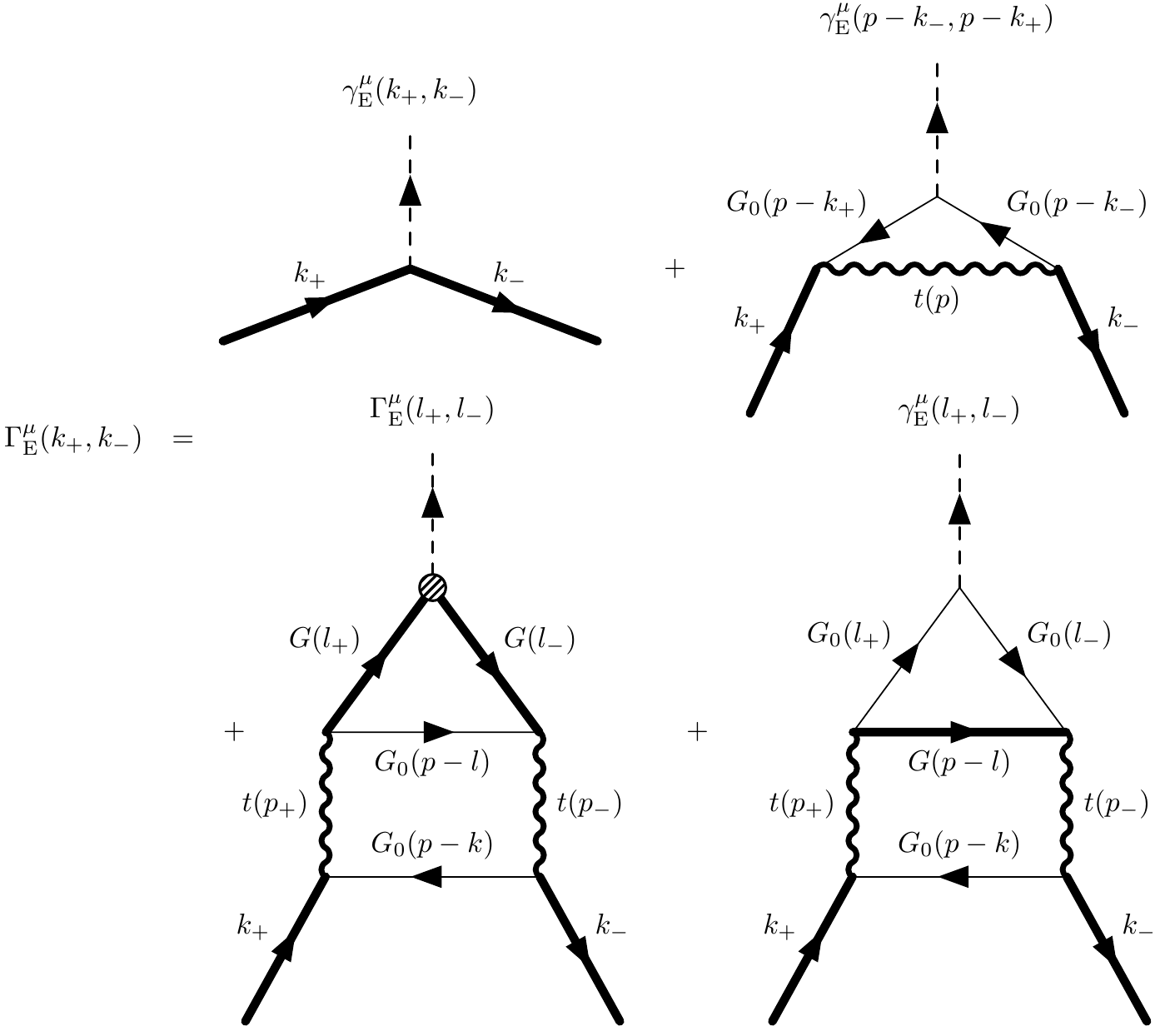}
\caption{Feynman diagrams for the full EM vertex in the $GG_{0}$ pair-fluctuation theory.}
\label{fig:Full_vertex_diagrams_GG0}
\end{figure*}

An important identity mentioned in Sec.~(\ref{sec:EM_Diagrammatic}) of the main text, which relates the Maki-Thompson vertex to the Aslamazov-Larkin vertices, 
is $q_{\mu}\left[2\MT^{\mu}_{\E}(k^{+},k^{-})+\AL_{\E,1}^{\mu}(k^{+},k^{-})+\AL_{\E,2}^{\mu}(k^{+},k^{-})\right]=0$.
This is proved directly as follows. By applying the bare WTI to the MT vertex in Eq.~(\ref{eq:App_full_vertex}), it follows that
\begin{align}\label{eq:app_MT_contraction}
q_{\mu}\MT^{\mu}_{\E}(k^{+},k^{-})&=\sum_{p}t(p)\left[G_{0}(p-k_{+})-G_{0}(p-k_{-})\right],\nonumber\\
&=\Sigma(k_{+})-\Sigma(k_{-}).
\end{align}
Similarly, by applying the bare and full WTIs to the AL vertices in Eq.~(\ref{eq:App_full_vertex}), it follows that
\begin{align}\label{eq:app_AL_contraction}
q_{\mu}\left[\AL^{\mu}_{\E,1}(k_{+},k_{-})+\AL^{\mu}_{\E,2}(k_{+},k_{-})\right]&=-\sum_{p}t(p_{-})t(p_{+})G_{0}(p-k)\sum_{l}G_{0}(p-l)
\left[G(l_{-})-G(l_{+})\right]\nonumber\\
&\quad-\sum_{p}t(p_{-})t(p_{+})G_{0}(p-k)\sum_{l}G(p-l)\left[G_{0}(l_{-})-G_{0}(l_{+})\right],\nonumber\\
&=-2\sum_{p}t(p_{-})t(p_{+})G_{0}(p-k)\left[\Pi(p_{-})-\Pi(p_{+})\right],\nonumber\\
&=-2\sum_{p}\left[t(p_{+})-t(p_{-})\right]G_{0}(p-k),\nonumber\\
&=-2\left[\Sigma(k_{+})-\Sigma(k_{-})\right].
\end{align}
From Eq.~(\ref{eq:app_MT_contraction}) and Eq.~(\ref{eq:app_AL_contraction}) the desired result follows:
$q_{\mu}\left[2\MT^{\mu}_{\E}(k^{+},k^{-})+\AL_{\E,1}^{\mu}(k^{+},k^{-})+\AL_{\E,2}^{\mu}(k^{+},k^{-})\right]=0$.

\begin{figure*}[t]
\centering\includegraphics[width=15cm,height=7cm,clip]{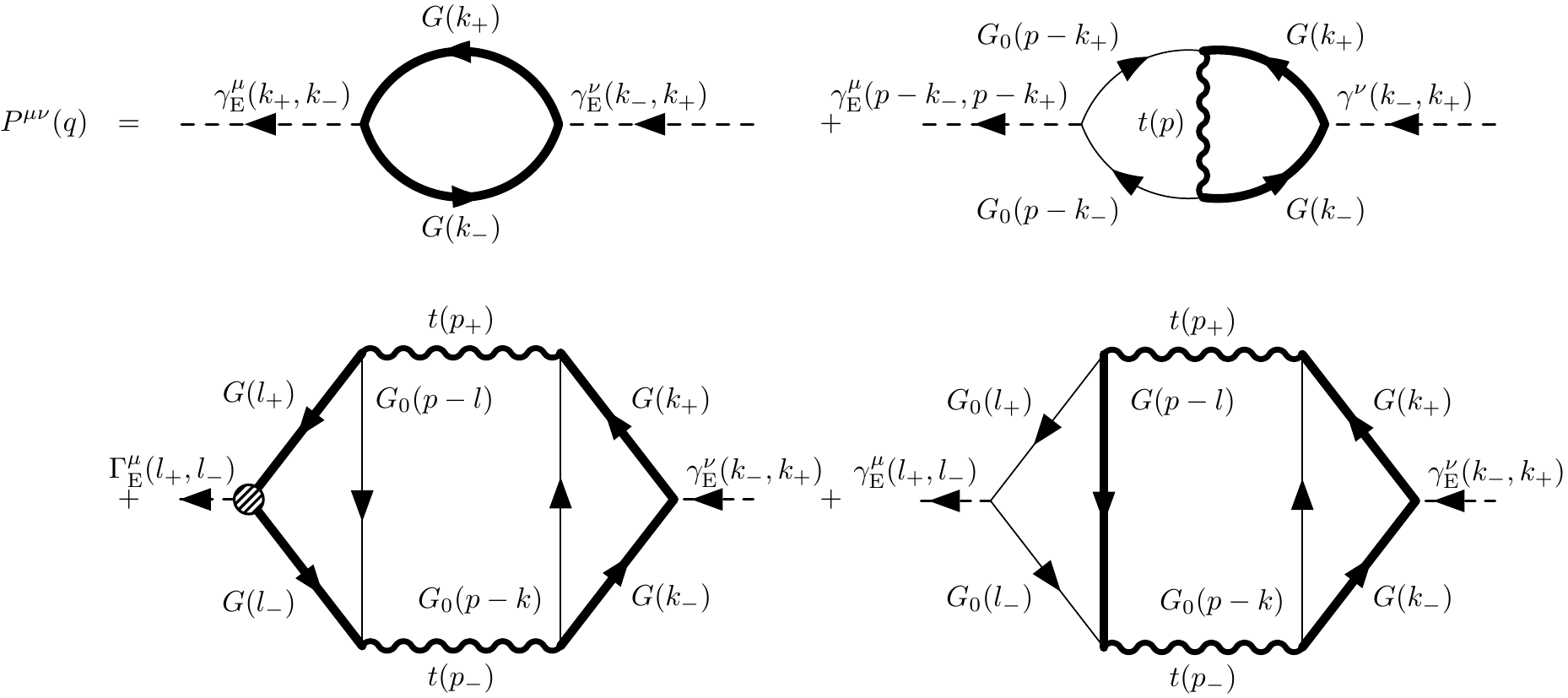}
\caption{Feynman diagrams for the exact EM response functions in the $GG_{0}$ pair-fluctuation theory. In order from left to right, and top to bottom, there is one ``bubble'', one Maki-Thompson, one Aslamazov-Larkin $[\AL_{\E,1}]$, 
and another (non-identical) Aslamazov-Larkin $[\AL_{\E,2}]$ diagram.}
\label{fig:Response_function_diagrams_GG0}
\end{figure*}

\begin{figure*}[t]
\centering\includegraphics[width=14.25cm,height=7cm,clip]{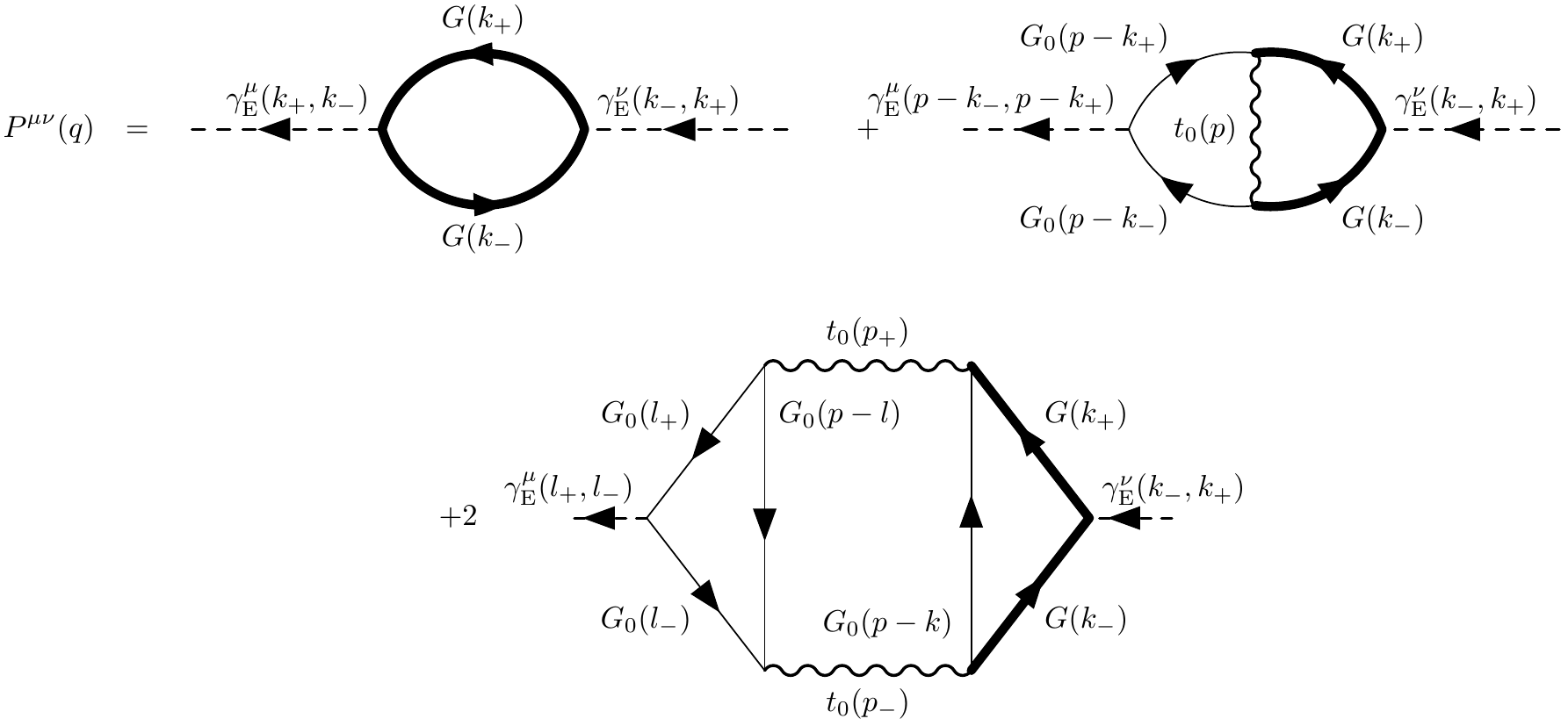}
\caption{Feynman diagrams for the exact EM response functions in the $G_{0}G_{0}$ pair-fluctuation theory. In order from left to right, and top to bottom, there is one ``bubble'', one Maki-Thompson, and two identical Aslamazov-Larkin diagrams.}
\label{fig:Response_function_diagrams_G0G0}
\end{figure*}

Once the full EM vertex has been determined, the exact EM response functions can be computed via
\begin{equation}
P^{\mu\nu}(q)=2\sum_{k}G(k_{+})\Gamma_{\E}^{\mu}(k_{+},k_{-})G(k_{-})\gamma_{\E}^{\nu}(k_{-},k_{+}).
\end{equation}
The Feynman diagrams for the exact EM response functions are given in Fig.~(\ref{fig:Response_function_diagrams_GG0}).
Since the full EM vertex itself appears in the first Aslamazov-Larkin diagram [$\AL_{\E,1}$], 
the explicit closed form of the full EM vertex for the $GG_{0}$ pair-fluctuation theory cannot be obtained. 
The result is a complicated integral equation that is theoretically intractable.

\subsection{\texorpdfstring{$G_{0}G_{0}$}{G0G0} pair-fluctuation theory}

The previous section derived the full EM response for the $GG_{0}$ pair-fluctuation theory. 
A similar derivation can be performed for the $G_{0}G_{0}$ pair-fluctuation theory. 
Since the calculation is almost identical to the one performed in the previous section, only the final results are given. 
The self energy for the $G_{0}G_{0}$ pair-fluctuation theory is $\Sigma(k)=\sum_{p}t(p)G_{0}(p-k)=\sum_{p}t(p+k)G_{0}(p)$,
where the $t$-matrix is defined through the pair-susceptibility by $t^{-1}(p)=g^{-1}+\Pi(p)$, with $\Pi(p)=\sum_{l}G_{0}(p-l)G_{0}(l)$ the pair susceptibility.
The pair susceptibility in this theory has two identical bare Green's functions. 
This means that when all possible vertex insertions in the self energy are performed, two identical Aslamazov-Larkin diagrams will arise from the two bare vertex insertions in the Green's functions in the pair susceptibility. 
Following the same procedure in the previous section, the full EM vertex is given by
\begin{align}
\Gamma_{\E}^{\mu}(k_{+},k_{-})&=\gamma_{\E}^{\mu}(k_{+},k_{-})\nonumber\\
&\quad+\sum_{p}t(p)G_{0}(p-k_{-})\gamma_{\E}^{\mu}(p-k_{-},p-k_{+})G_{0}(p-k_{+})\nonumber\\
&\quad-2\sum_{p}\sum_{l}t(p_{-})t(p_{+})G_{0}(p-k)G_{0}(p-l)G_{0}(l_{+})\gamma_{\E}^{\mu}(l_{+},l_{-})G_{0}(l_{-}).
\end{align}
The Feynman diagrams for the exact EM response functions are given in Fig.~(\ref{fig:Response_function_diagrams_G0G0}).
Notice that, in contrast to the $GG_{0}$ pair-fluctuation theory, the full EM vertex itself does not appear in either the Maki-Thompson or Aslamazov-Larkin diagrams. 
Therefore, provided the $t$-matrix is known exactly, in principle the explicit closed form of the full EM vertex for the $G_{0}G_{0}$ pair-fluctuation theory can be obtained.

Note the distinction between Fig.~(\ref{fig:Response_function_diagrams_GG0}) and Fig.~(\ref{fig:Response_function_diagrams_G0G0}); 
the $GG_{0}$ pair-fluctuation theory contains dressed $t$-matrices, and the Aslamazov-Larkin diagrams $\AL_{\E,1}$ and $\AL_{\E,2}$ have two and one dressed Green's functions, respectively, appearing in the left-most triangle vertex. 
In contrast the $G_{0}G_{0}$ pair-fluctuation theory has $t$-matrices constructed solely of bare Green's functions, and two identical Aslamazov-Larkin diagrams with only bare Green's functions and bare vertices in the left-most triangle vertex.

\begin{figure*}[t]
\centering\includegraphics[width=14.5cm,height=10cm,clip]{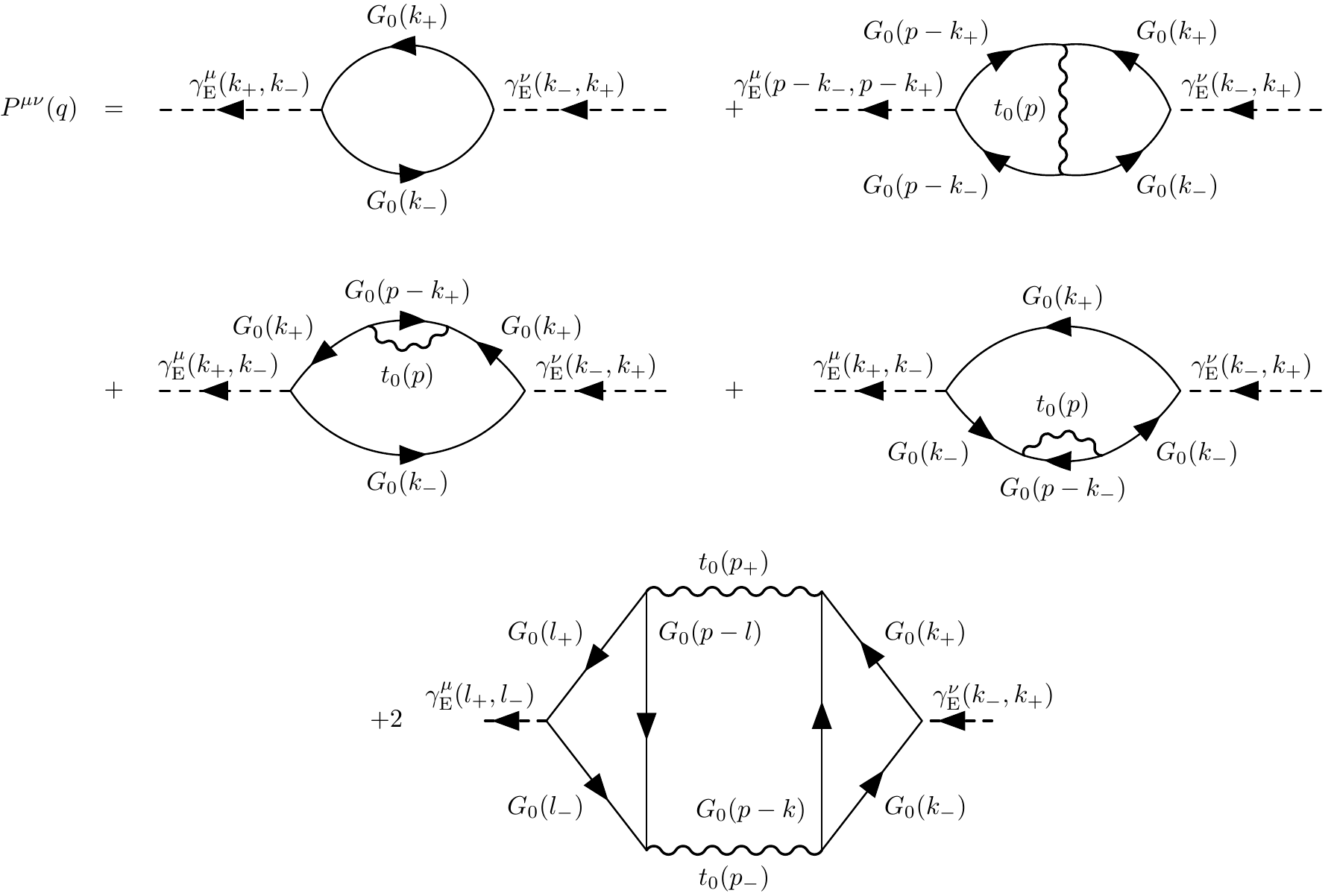}
\caption{Lowest order Feynman diagrams for the EM response functions in the $G_{0}G_{0}$ pair-fluctuation theory. 
In order from left to right, and top to bottom, there is a normal state ``bubble'', one Maki-Thompson, two density of states, and two identical Aslamazov-Larkin diagrams.}
\label{fig:Response_function_diagrams_approximate_G0G0}
\end{figure*}

The standard response functions considered in the weak-fluctuation literature arise by expanding the diagrams in Fig.~(\ref{fig:Response_function_diagrams_G0G0}) to lowest order; 
that is, by expanding the Green's functions according to a truncated Dyson's equation: $G\approx G_{0}+G_{0}\Sigma G_{0}$. 
Performing this expansion on the diagrams in Fig.~(\ref{fig:Response_function_diagrams_G0G0}), the result is the normal state ``bubble'', two density of states, one Maki-Thompson, 
and two identical Aslamazov-Larkin diagrams, all with bare Green's functions and bare vertices. 
This reproduces the Feynman diagrams in Ref.~\onlinecite{Aslamazov_1975}. 
However, it is not a gauge-invariant set of diagrams, and only satisfies the WTI to $\mathcal{O}\left(\Sigma\right)$. 
These diagrams are shown in Fig.~(\ref{fig:Response_function_diagrams_approximate_G0G0}).

\section{Diamagnetic susceptibility in the small \texorpdfstring{$|\mu_{\pair}|$}{mu-pair} limit}
\label{sec:Diamagnetism_Approx_Matsubara}

In Sec.~(\ref{sec:Diamagnetism_Approx}) of the main text, the contribution to diamagnetic susceptibility from the Aslamazov-Larkin diagrams was obtained in the small  $|\mu_{\pair}|$ limit. 
The calculation was simplified by ignoring the Matsubara frequency summation and just treating the integrand with zero bosonic Matsubara frequency. 
This is justified in the small $|\mu_{\pair}|$ limit, and was also performed in the seminal paper~\cite{Aslamazov_1975} of Aslamazov and Larkin. 
In this section the Matsubara frequency summation is performed, and a complete calculation of diamagnetic susceptibility in the small $|\mu_{\pair}|$ limit is presented. 
The starting point is Eq.~(\ref{eq:chi_approx1}) of the main text:
\begin{equation}
\chi_{\dia}=e^2\sum_{p}\left[\frac{\partial\Pi(p)}{\partial p^{x}}\right]^{2}\left\{t(p)\frac{\partial^{2}t(p)}{\partial (p^{y})^2}-\left[\frac{\partial t(p)}{\partial p^{y}}\right]^{2}\right\}.
\end{equation}
The $t$-matrix is related to the pair susceptibility by $t^{-1}(p)=g^{-1}+\Pi(p)$;
thus it follows that $\partial t(p)/\partial p^{y}=-t^{2}(p)\left[\partial t^{-1}(p)/\partial p^{y}\right]=-t^{2}(p)\left[\partial\Pi(p)/\partial p^{y}\right]$. 
Similarly $\partial^{2}t(p)/\partial (p^{y})^{2}=$ $2t^{3}(p)\left[\partial\Pi(p)/\partial p^{y}\right]^{2}-t^{2}(p)\left[\partial^{2}\Pi(p)/\partial (p^{y})^{2}\right]$.
Therefore, using these identities gives
\begin{equation}\label{eq:chi_dia_AL0}
\chi_{\dia}=e^{2}\sum_{p}\left[\frac{\partial\Pi(p)}{\partial p^{x}}\right]^{2}\left\{t^{4}(p)\left[\frac{\partial\Pi(p)}{\partial p^{y}}\right]^{2}-t^{3}(p)\frac{\partial^{2}\Pi(p)}{\partial (p^{y})^{2}}\right\}.
\end{equation}
The term that is quartic in the $t$-matrix can be simplified as follows
\begin{align}
\sum_{p}\left[\frac{\partial\Pi(p)}{\partial p^{x}}\right]^{2}t^{4}(p)\left[\frac{\partial\Pi(p)}{\partial p^{y}}\right]^{2}&=
-\sum_{p}\left[\frac{\partial\Pi(p)}{\partial p^{x}}\right]^{2}\frac{\partial\Pi(p)}{\partial p^{y}}\frac{\partial t(p)}{\partial p^{y}}t^{2}(p),\nonumber\\
&=-\frac{1}{3}\sum_{p}\left[\frac{\partial\Pi(p)}{\partial p^{x}}\right]^{2}\frac{\partial\Pi(p)}{\partial p^{y}}\frac{\partial t^{3}(p)}{\partial p^{y}},\nonumber\\
&=\frac{1}{3}\sum_{p}t^{3}(p)\left\{\left[\frac{\partial\Pi(p)}{\partial p^{x}}\right]^{2}\frac{\partial^{2}\Pi(p)}{\partial (p^{y})^{2}}
+2\frac{\partial\Pi(p)}{\partial p^{y}}\frac{\partial\Pi(p)}{\partial p^{x}}\frac{\partial^{2}\Pi(p)}{\partial p^{y}\partial p^{x}}\right\}.
\end{align}
Inserting this result into Eq.~(\ref{eq:chi_dia_AL0}) and then simplifying gives the diamagnetic susceptibility as
\begin{equation}\label{eq:chi_dia_AL1}
\chi_{\dia}=-\frac{2e^2}{3}\sum_{p}t^{3}(p)\left\{\left[\frac{\partial\Pi(p)}{\partial p^{x}}\right]^{2}
\frac{\partial^{2}\Pi(p)}{\partial (p^{y})^{2}}-\frac{\partial\Pi(p)}{\partial p^{y}}\frac{\partial\Pi(p)}{\partial p^{x}}\frac{\partial^{2}\Pi(p)}
{\partial p^{y}\partial p^{x}}\right\}.
\end{equation}
This is a general expression for the diamagnetic susceptibility due to the dominant contribution in the Aslamazov-Larkin diagrams. 
Note that, for the case of the weak-fluctuation theory, where the pair susceptibility is $\Pi(p)=\sum_{l}G_{0}(p-l)G_{0}(l)$, Eq.~(\ref{eq:chi_dia_AL1}) is equivalent to Eq.~(17) of the Aslamazov-Larkin paper~\cite{Aslamazov_1975}.

To obtain the diamagnetic susceptibility in the small $|\mu_{\pair}|$ limit, we now use the small momentum form of the pair-propagator, as defined in Eq.~(\ref{eq:tmat_propagator}) of the main text:
$t^{-1}_{\R}(\varpi,\mb{p})=Z\left[\kappa\varpi-\mb{p}^2/(2M_{\pair})-|\mu_{\pair}|+i\Gamma\varpi\right]$, and the definition $t^{-1}(p)=g^{-1}+\Pi(p)$. 
Evaluating the spatial derivatives in Eq.~(\ref{eq:chi_dia_AL1}) gives: $\p\Pi(p)/\p p^{x}=\p t^{-1}(p)/\p p^{x}=-Zp^{x}/M_{\pair}$, 
$\p^{2}\Pi(p)/\p (p^{y})^{2}=-Z/M_{\pair}$, and $\p^{2}\Pi(p)/\p p^{x}\p p^{y}=0$. 
Inserting these results into Eq.~(\ref{eq:chi_dia_AL1}) 
then gives the first line of Eq.~(\ref{eq:chi_dia_AL}) of the main text, where the Matsubara frequency summation has been restored:
\begin{equation}
\chi_{\dia}=\frac{2e^2}{3M_{\pair}}\sum_{p}\left(\frac{p^{x}}{M_{\pair}}\right)^{2}\left[Zt(p)\right]^{3}.
\end{equation}
To simplify this expression, note that $\p t^{-1}(p)/\p p^{x}=-Zp^{x}/M_{\pair}$ and $\p t(p)/\p p^{x}=-t^{2}(p)\left[\p t^{-1}(p)/\p p^{x}\right]=Zt^{2}(p)p^{x}/M_{\pair}$. 
Using these identities, along with performing integration by parts, the diamagnetic susceptibility becomes
\begin{align}\label{eq:app_chi_dia}
\chi_{\dia}&=\frac{2e^2}{3M_{\pair}}\sum_{p}\frac{Z^2p^{x}}{M_{\pair}}t(p)\frac{\p t(p)}{\p p^{x}},\nonumber\\
&=\frac{e^2}{3M_{\pair}}\sum_{p}\frac{Z^2p^{x}}{M_{\pair}}\frac{\p t^{2}(p)}{\p p^{x}},\nonumber\\
&=-\frac{e^2}{3M^{2}_{\pair}}\sum_{p}\left[Zt(p)\right]^{2}.
\end{align}

To perform the Matsubara frequency summation, we use the Eliashberg contour~\cite{VarlamovBook} and the identity (valid for bosonic Matsubara frequencies)~\cite{VarlamovBook, Mahan}:
\begin{equation}
T\sum_{i\varpi_{m}}f\left(i\varpi_{m}\right)=\frac{1}{4\pi i}\oint_{\mathcal{C}}dz\ \coth\left(\frac{1}{2}\beta z\right)f(z),
\end{equation}
where $\beta=1/(k_{B}T)$ and we set $k_{B}=1$. Here $\mathcal{C}$ is a closed contour enclosing the poles of $\coth(\beta z/2)$, which occur at the bosonic Matsubara frequencies $z=i\varpi_{m}=2\pi imT$, where $m\in\mathbb{Z}$. 
Since the semi-circle contribution to the integral in Eq.~(\ref{eq:app_chi_dia}) vanishes, the contour integral can be deformed to an integral above and below the real axis. Thus the diamagnetic susceptibility becomes
\begin{align}
\chi_{\dia}&=-\frac{Z^2e^2}{3M^{2}_{\pair}}\sum_{\mb{p}}\frac{1}{4\pi i}\int_{-\infty}^{\infty}dx\ \coth\left(\frac{1}{2}\beta x\right)
\left[t^{2}_{\R}(x,\mb{p})-t^{2}_{\A}(x,\mb{p})\right],\nonumber\\
&=-\frac{Z^2e^2}{3\pi M^{2}_{\pair}}\sum_{\mb{p}}\int_{-\infty}^{\infty}dx\ \coth\left(\frac{1}{2}\beta x\right)
\Re\left[t_{\R}(x,\mb{p})\right]\Im\left[t_{\R}(x,\mb{p})\right].
\end{align}
Here $t_{\R}$ and $t_{\A}$ denote the retarded and advanced propagators, which are related by $t_{\A}(z,\mb{p})=t_{\R}(z,\mb{p})^{*}$.

In the small $|\mu_{\pair}|$ limit the main contribution to the integral occurs when $\beta x \ll 1$, which allows the coth function to be expanded as 
$\coth(\beta x/2)\approx 2T/x$. Inserting the retarded propagator, defined by $t^{-1}_{\R}(x,\mb{p})=Z\left[\kappa x-\mb{p}^2/(2M_{\pair})-|\mu_{\pair}|+i\Gamma x\right]$, into the above expression then gives
\begin{align}
\chi_{\dia}&=\frac{2e^2T\Gamma}{3\pi M^{2}_{\pair}}\sum_{\mb{p}}\int_{-\infty}^{\infty}dx\ \frac{\left.\kappa x-\mb{p}^2/(2M_{\pair})-|\mu_{\pair}|\right.}{\left[\left(\kappa x-\mb{p}^2/(2M_{\pair})-|\mu_{\pair}|\right)^2+\left(\Gamma x\right)^2\right]^2},\nonumber\\
&=-\frac{4e^2T}{3}\sum_{\mb{p}}\frac{1}{\left(\mb{p}^2+2M_{\pair}|\mu_{\pair}|\right)^2},\nonumber\\
&=-\frac{e^2T}{3\pi^2}\int_{-\infty}^{\infty}dp\ \frac{p^2}{\left(p^2+2M_{\pair}|\mu_{\pair}|\right)^2}.\label{eq:app_chi_dia_AL0}
\end{align}
The remaining $p$-integration is easily performed using a closed contour integral in the upper half plane and evaluating the residue at the pole $p=i\left(2M_{\pair}|\mu_{\pair}|\right)^{1/2}$. 
This gives, after restoring the constants $\hbar$ and $c$:
\begin{equation}\label{eq:app_chi_dia_AL}
\chi_{\dia}=-\frac{T\left(2e\right)^2}{24\pi \hbar c^2}\sqrt{\frac{1/(2M_{\pair})}{\left|\mu_{\pair}\right|}}.
\end{equation}
The diamagnetic susceptibility is written in this form to compare with the free boson result, which is derived in the next section. 
The above expression reproduces Eq.~(\ref{eq:chi_dia_AL2}) of the main text, and validates the approximation made there concerning setting the bosonic Matsubara frequency to zero in the integrand.

\section{Free boson diamagnetic susceptibility}
\label{sec:app_Free_Bosons}

This section derives the diamagnetic susceptibility for free bosons and also gives the limiting form when the chemical potential tends to zero. 
While the free boson diamagnetic susceptibility is well known, the aim here is to express it in an identical form to the Aslamazov-Larkin contribution to diamagnetic susceptibility for the $GG_{0}$ pair-fluctuation theory.

The Kubo formula for diamagnetic susceptibility, $\chi_{\dia}$, is given in Eq.~(\ref{eq:chi_Kubo_dia}) of the main text:
\begin{equation}\label{eq:chi_Kubo_app}
\chi_{\dia}=-\underset{\mb{q}\rightarrow0}{\lim}\left[\frac{P^{xx}(i\Omega_{m}=0,\mb{q})+n/m}{\mb{q}^2}\right]_{q^{x}=q^{z}=0}.
\end{equation}
Recall that $P^{xx}(0)=-n/m$ above $T_{c}$. The response function for spin-0 free bosons is given by 
\begin{equation}\label{eq:Response_FreeBoson}
P^{xx}(q)=-\left(e^{*}\right)^2\sum_{p}G_{0}(p_{+})\gamma_{\E}^{x}(p_{+},p_{-})G_{0}(p_{-})\gamma_{\E}^{x}(p_{-},p_{+}),
\end{equation}
where the bare Green's function is $G^{-1}_{0}(p)=i\varpi_{m}-\xi_{\mb{p}}$, with $\xi_{\mb{p}}=\mb{p}^2/(2m_{\b})-\mu_{\b}$ the free-particle dispersion, and the bare vertex is $\gamma_{\E}^{x}(p_{+},p_{-})=p^{x}/m_{\b}$. 
Here $m_{\b}$ and $\mu_{\b}$ are the free boson mass and chemical potential, respectively. The four-vectors $p^{\mu}, q^{\mu}$ are $p^{\mu}=(i\varpi_{m},\mb{p})$, $q^{\mu}=(i\Omega_{m},\mb{q})$, 
where $\varpi_{m}, \Omega_{m}$ are both bosonic Matsubara frequencies, and $p_\pm\equiv p\pm q/2$. The four-vector summation is defined by $\sum_{p}=T\sum_{i\varpi_{m}}\sum_{\mb{p}}$, where $T$ is the temperature. 
Note the relative sign difference compared to the fermionic response function in Eq.~(\ref{eq:Response_Fermion}) of the main text.

Expanding out the Green's functions in Eq.~(\ref{eq:Response_FreeBoson}) to $\mathcal{O}\left(\mb{q}^2\right)$, and then using Eq.~(\ref{eq:chi_Kubo_app}), the free boson diamagnetic susceptibility becomes
\begin{equation}
\chi_{\b}=\frac{\left(e^{*}\right)^2}{4}\sum_{p}\left\{G_{0}(p)\frac{\partial^{2}G_{0}(p)}{\partial(p^{y})^{2}}-
\left[\frac{\partial G_{0}(p)}{\partial p^{y}}\right]^{2}\right\}\left(\frac{p^{x}}{m_{\b}}\right)^{2}.\label{eq:chi_dia_FreeBoson1}
\end{equation}
Comparing this expression with that appearing in Eq.~(\ref{eq:chi_approx1}) of the main text proves the claim at the beginning of this section, 
namely that the Aslamazov-Larkin contribution to diamagnetic susceptibility is of the free boson form but with modified vertices and propagators. 
Now perform integration by parts on the first term to obtain
\begin{equation}
\chi_{\b}=-\frac{\left(e^{*}\right)^2}{2}\sum_{p}\left[\frac{\partial G_{0}(p)}{\partial p^{y}}\right]^{2}\left(\frac{p^{x}}{m_{\b}}\right)^{2}.
\end{equation}
Another expression involving the product of four Green's functions, which has occurred in the fermion literature~\cite{Fukuyama_1971}, can also be obtained
\begin{align}
\chi_{\b}&=-\frac{\left(e^{*}\right)^2}{2}\sum_{p}\left[\frac{\partial G_{0}(p)}{\partial p^{y}}\right]^{2}\left(\frac{p^{x}}{m_{\b}}\right)^{2},\nonumber\\
&=-\frac{\left(e^{*}\right)^2}{2}\sum_{p}\left[G_{0}^{2}(p)\frac{\partial G_{0}^{-1}(p)}{\partial p^{y}}\right]^{2}\left(\frac{p^{x}}{m_{\b}}\right)^{2},\nonumber\\
&=-\frac{\left(e^{*}\right)^2}{2}\sum_{p}G_{0}^{4}(p)\left(\frac{p^{y}}{m_{\b}}\right)^{2}\left(\frac{p^{x}}{m_{\b}}\right)^{2}\label{eq:chi_dia_FreeBoson2}.
\end{align}
The analogous formula for spin-$\tfrac{1}{2}$ fermions agrees with Ref.~\onlinecite{Fukuyama_1971}. Another equivalent expression for the free boson diamagnetic susceptibility can also be obtained, 
which is identical in form to the Aslamazov-Larkin contribution found in Eq.~(\ref{eq:chi_dia_AL}) of the main text.
By performing integration by parts on the second term in Eq.~(\ref{eq:chi_dia_FreeBoson1}), the diamagnetic susceptibility can be expressed as follows
\begin{align}
\chi_{\b}&=\frac{\left(e^{*}\right)^2}{2}\sum_{p}G_{0}(p)\frac{\partial^{2}G_{0}(p)}{\partial(p^{y})^{2}}
\left(\frac{p^{x}}{m_{\b}}\right)^{2},\nonumber\\
&=\frac{\left(e^{*}\right)^2}{2}\sum_{p}G_{0}(p)\frac{\partial}{\partial p^{y}}\left[G_{0}^{2}(p)\frac{p^{y}}{m_{\b}}\right]
\left(\frac{p^{x}}{m_{\b}}\right)^{2},\nonumber\\
&=\frac{\left(e^{*}\right)^2}{2}\sum_{p}G_{0}(p)\left[\frac{1}{m_{\b}}G_{0}^{2}(p)+2G_{0}(p)\frac{\partial G_{0}(p)}{\partial p^{y}}
\frac{p^{y}}{m_{\b}}\right]\left(\frac{p^{x}}{m_{\b}}\right)^{2}\\
&=\frac{\left(e^{*}\right)^2}{2}\sum_{p}G_{0}(p)\left[\frac{1}{m_{\b}}G_{0}^{2}(p)-2G_{0}^{3}(p)\frac{\partial G_{0}^{-1}(p)}{\partial p^{y}}
\frac{p^{y}}{m_{\b}}\right]
\left(\frac{p^{x}}{m_{\b}}\right)^{2}\\
&=\frac{\left(e^{*}\right)^2}{2}\sum_{p}G_{0}^{3}(p)\left[\frac{1}{m_{\b}}+2G_{0}(p)\left(\frac{p^{y}}{m_{\b}}\right)^{2}\right]
\left(\frac{p^{x}}{m_{\b}}\right)^{2}.
\end{align}
Equating this expression with Eq.~(\ref{eq:chi_dia_FreeBoson2}) then gives
\begin{equation}\label{eq:chi_dia_FreeBoson3}
\chi_{\b}=\frac{\left(e^{*}\right)^2}{6m_{\b}}\sum_{p}\left(\frac{p^{x}}{m_{\b}}\right)^{2}G_{0}^{3}(p).
\end{equation}
Comparison of this equation with that appearing in the first line of Eq.~(\ref{eq:chi_dia_AL}) of the main text shows that, 
the Aslamazov-Larkin contribution to diamagnetic susceptibility is equivalent to the free boson diamagnetic susceptibility with free boson charge $e^{*}=2e$, mass $m_{\b}=M_{\pair}$, and chemical potential $\mu_{\b}=\mu_{\pair}$. 
This shows that, in the small $|\mu_{\pair}|$ limit, the underlying effect of the fermionic interactions in the $GG_{0}$ pair-fluctuation theory is to modify the free boson parameters. 
Such an effect is intuitive in the deep BEC regime, where the paired degrees of freedom behave as fluctuating bosons.

The expression in Eq.~(\ref{eq:chi_dia_FreeBoson3}) can be simplified as follows
\begin{align}
\chi_{\b}&=\frac{\left(e^{*}\right)^2}{6m_{\b}}\sum_{p}\left(\frac{p^{x}}{m_{\b}}\right)^{2}G_{0}^{3}(p),\nonumber\\
&=\frac{\left(e^{*}\right)^2}{6m_{\b}}\sum_{p}\left(\frac{p^{x}}{m_{\b}}\right)G_{0}(p)\frac{\partial G_{0}(p)}{\partial p^{x}},\nonumber\\
&=-\frac{\left(e^{*}\right)^2}{12m_{\b}^2}\sum_{p}G_{0}^{2}(p).
\end{align}
After performing the Matsubara frequency summation, and then integrating by parts, the free boson diamagnetic susceptibility becomes
\begin{align}
\chi_{\b}&=\frac{\left(e^{*}\right)^2}{12m_{\b}^2}\sum_{\mb{p}}\frac{\p b(\xi_{\mb{p}})}{\p \xi_{\mb{p}}},\nonumber\\
&=-\frac{\left(e^{*}\right)^2}{24\pi^2m_{\b}}\int_{0}^{\infty}dp\ b(\xi_{\mb{p}}).
\label{eq:chib}
\end{align}
Here $b(x)=\left[\exp(\beta x)-1\right]^{-1}$ is the Bose-Einstein distribution function. 
In the limit that $|\mu_{\b}|/T\ll1$, the above expression reduces to
\begin{equation}\label{eq:chi_dia_FreeBoson5}
\chi_{\b}=-\frac{T\left(e^{*}\right)^2}{24\pi \hbar c^2}\sqrt{\frac{1/(2m_{\b})}{\left|\mu_{\b}\right|}}.
\end{equation}
The constants $\hbar$ and $c$ have been restored to render $\chi_{\b}$ dimensionless. Comparison of Eq.~(\ref{eq:chi_dia_FreeBoson5}) and 
Eq.~(\ref{eq:chi_dia_AL2}) of the main text proves the result stated in the main text, namely that the small $|\mu_{\pair}|$ limit of the Aslamazov-Larkin contribution to diamagnetic susceptibility is equivalent to the free boson diamagnetic susceptibility, 
but with bosonic charge $e^{*}=2e$, mass $m_{\b}=M_{\pair}$, and chemical potential $\mu_{\b}=\mu_{\pair}$.

\section{Thermoelectric response}
\subsection{Ward-Takahashi identity for energy conservation}
\label{sec:WTI_Heat_Vertex}

In Sec.~(\ref{sec:Heat_Diagrammatic}) of the main text the full heat vertex was presented along with a discussion of the WTI for energy conservation. Here we explicitly prove that the full heat vertex satisfies the WTI for energy conservation. 
The full heat vertex, as given in 
Eqs.~(\ref{eq:Full_heat_vertex}-\ref{eq:L_heat_vertex}) of the main text, is 
\begin{equation}\label{eq:app_Full_heat_vertex}
\Gamma^{\mu}_{\H}(k_{+},k_{-})=\gamma^{\mu}_{\H}(k_{+},k_{-})+\MT^{\mu}_{\H}(k_{+},k_{-})+\lambda^{\mu}_{\H}(k_{+},k_{-})
+\AL_{\H,1}^{\mu}(k_{+},k_{-})+\AL_{\H,2}^{\mu}(k_{+},k_{-}),
\end{equation}
where the Maki-Thompson, Aslamazov-Larkin, and $\lambda^{\mu}_{\H}$ heat vertices are 
\begin{align}
\MT^{\mu}_{\H}(k_{+},k_{-})&=\sum_{p}t(p)G_{0}(p-k_{-})\gamma^{\mu}_{\H}(p-k_{-},p-k_{+})G_{0}(p-k_{+}),\label{eq:app_MT_heat_vertex}\\
\AL_{\H,1}^{\mu}(k_{+},k_{-})&=-\sum_{p}\sum_{l}t(p_{-})t(p_{+})G_{0}(p-k)G_{0}(p-l)G(l_{+})\Gamma^{\mu}_{\H}(l_{+},l_{-})G(l_{-}),\label{eq:app_AL1_heat_vertex}\\
\AL_{\H,2}^{\mu}(k_{+},k_{-})&=-\sum_{p}\sum_{l}t(p_{-})t(p_{+})G_{0}(p-k)G(p-l)G_{0}(l_{+})\gamma^{\mu}_{\H}(l_{+},l_{-})G_{0}(l_{-}),\label{eq:app_AL2_heat_vertex}\\
\lambda^{\mu}_{\H}(k_{+},k_{-})&=\sum_{p}g^{-1}\delta^{\mu 0}t(p_{-})t(p_{+})G_{0}(p-k).\label{eq:app_L_heat_vertex}
\end{align}
The WTI for energy conservation is expressed as
\begin{equation}\label{eq:app_energy_WTI}
q_{\mu}\Gamma^{\mu}_{\H}(k_{+},k_{-})=\omega_{-}G^{-1}(k_{+})-\omega_{+}G^{-1}(k_{-}).
\end{equation}
For convenience, this can be equivalently written as $q_{\mu}\Gamma^{\mu}_{\H}(k_{+},k_{-})=\left(k_{-}\right)^{0}G^{-1}(k_{+})
-\left(k_{+}\right)^{0}G^{-1}(k_{-})$, where $k^{0}$ denotes the time component of the four-vector $k^{\mu}=(\omega,\mb{k})$.

The contractions of the vertices in Eqs.~(\ref{eq:app_MT_heat_vertex}-\ref{eq:app_L_heat_vertex}) are computed as follows. Using the bare WTI for energy conservation, the contraction of the MT vertex is
\begin{align}\label{eq:MT_heat_vertex_contraction}
q_{\mu}\MT^{\mu}_{\H}(k_{+},k_{-})&=\sum_{p}t(p)G_{0}(p-k_{-})\left[(p-k_{+})^{0}G^{-1}_{0}(p-k_{-})-(p-k_{-})^{0}G^{-1}_{0}(p-k_{+})\right]G_{0}(p-k_{+}),\nonumber\\
&=\sum_{p}t(p)\left[(p-k_{+})^{0}G_{0}(p-k_{+})-(p-k_{-})^{0}G_{0}(p-k_{-})\right],\nonumber\\
&=\sum_{p}(p-k)^{0}G_{0}(p-k)\left[t(p_{+})-t(p_{-})\right].
\end{align}
The contraction of the AL vertices is
\begin{align}\label{eq:AL_heat_vertex_contraction0}
&q_{\mu}\left[\AL_{\H,1}^{\mu}(k_{+},k_{-})+\AL_{\H,2}^{\mu}(k_{+},k_{-})\right]\nonumber\\
&=-\sum_{p}\sum_{l}t(p_{-})t(p_{+})G_{0}(p-k)G_{0}(p-l)G(l_{+})\left[(l_{-})^{0}G^{-1}(l_{+})-(l_{+})^{0}G^{-1}(l_{-})\right]G(l_{-})\nonumber\\
&\quad-\sum_{p}\sum_{l}t(p_{-})t(p_{+})G_{0}(p-k)G(p-l)G_{0}(l_{+})\left[(l_{-})^{0}G^{-1}_{0}(l_{+})-(l_{+})^{0}G^{-1}_{0}(l_{-})\right]G_{0}(l_{-}),\nonumber\\
&=-\sum_{p}\sum_{l}t(p_{-})t(p_{+})G_{0}(p-k)G_{0}(p-l)\left[(l_{-})^{0}G(l_{-})-(l_{+})^{0}G(l_{+})\right]\nonumber\\
&\quad-\sum_{p}\sum_{l}t(p_{-})t(p_{+})G_{0}(p-k)G(p-l)\left[(l_{-})^{0}G_{0}(l_{-})-(l_{+})^{0}G_{0}(l_{+})\right].
\end{align}
Now consider Eq.~(\ref{eq:AL_heat_vertex_contraction0}). In the first term, on the first line, let $l\rightarrow p_{+}-l$, and in the second term, on the first line, let $l\rightarrow p_{-}-l$, to obtain:
\begin{align}
q_{\mu}\left[\AL_{\H,1}^{\mu}(k_{+},k_{-})+\AL_{\H,2}^{\mu}(k_{+},k_{-})\right]&=-\sum_{p}\sum_{l}t(p_{-})t(p_{+})G_{0}(p-k)G_{0}(l_{-})G(p-l)(p_{-})^{0}\nonumber\\
&\quad+\sum_{p}\sum_{l}t(p_{-})t(p_{+})G_{0}(p-k)G_{0}(l_{+})G(p-l)(p_{+})^{0}.
\end{align}
The sum over $l$ can be computed, using the definition of the pair susceptibility: $\Pi(p_{\pm})=\sum_{l}G_{0}(l_{\pm})G(p-l)$. 
Performing the $l$-summation, then using the definition $t^{-1}(p)=g^{-1}+\Pi(p)$ and simplifying, it follows that 
\begin{align}\label{eq:AL_heat_vertex_contraction1}
&q_{\mu}\left[\AL_{\H,1}^{\mu}(k_{+},k_{-})+\AL_{\H,2}^{\mu}(k_{+},k_{-})\right]\nonumber\\
&=-\sum_{p}t(p_{-})t(p_{+})G_{0}(p-k)\Pi(p_{-})(p_{-})^{0}+\sum_{p}t(p_{-})t(p_{+})G_{0}(p-k)\Pi(p_{+})(p_{+})^{0},\nonumber\\
&=-\sum_{p}t(p_{+})G_{0}(p-k)(p_{-})^{0}+\sum_{p}t(p_{-})G_{0}(p-k)(p_{+})^{0}-q^{0}g^{-1}\sum_{p}t(p_{-})t(p_{+})G_{0}(p-k).
\end{align}

The contraction of the $\lambda^{\mu}_{\H}$ vertex is easily computed to give
\begin{equation}\label{eq:L_heat_vertex_contraction}
q_{\mu}\lambda^{\mu}_{\H}(k_{+},k_{-})=q^{0}g^{-1}\sum_{p}t(p_{-})t(p_{+})G_{0}(p-k).
\end{equation}
Combining Eqs.(\ref{eq:AL_heat_vertex_contraction1}-\ref{eq:L_heat_vertex_contraction}) then produces
\begin{equation}\label{eq:AL_heat_vertex_contraction2}
q_{\mu}\left[\AL_{\H,1}^{\mu}(k_{+},k_{-})+\AL_{\H,2}^{\mu}(k_{+},k_{-})+\lambda^{\mu}_{\H}(k_{+},k_{-})\right]=
\sum_{p}G_{0}(p-k)\left[t(p_{-})(p_{+})^{0}-t(p_{+})(p_{-})^{0}\right].
\end{equation}
Adding Eq.~(\ref{eq:MT_heat_vertex_contraction}) together with Eq.~(\ref{eq:AL_heat_vertex_contraction2}) and simplifying gives
\begin{align}
q_{\mu}\left[\MT^{\mu}_{\H}(k_{+},k_{-})+\AL_{\H,1}^{\mu}(k_{+},k_{-})+\AL_{\H,2}^{\mu}(k_{+},k_{-})+\lambda^{\mu}_{\H}(k_{+},k_{-})\right]&=
\sum_{p}G_{0}(p-k)\left[t(p_{-})(k_{+})^{0}-t(p_{+})(k_{-})^{0}\right],\nonumber\\
&=(k_{+})^{0}\Sigma(k_{-})-(k_{-})^{0}\Sigma(k_{+}).
\end{align}
The contraction of the full heat vertex is thus
\begin{align}
q_{\mu}\Gamma^{\mu}_{\H}(k_{+},k_{-})&=q_{\mu}\left[\gamma^{\mu}_{\H}(k_{+},k_{-})+\MT^{\mu}_{\H}(k_{+},k_{-})+\AL_{\H,1}^{\mu}(k_{+},k_{-})+\AL_{\H,2}^{\mu}(k_{+},k_{-})+\lambda^{\mu}_{\H}(k_{+},k_{-})\right],\nonumber\\
&=(k_{-})^{0}G^{-1}_{0}(k_{+})-(k_{+})^{0}G^{-1}_{0}(k_{-})+(k_{+})^{0}\Sigma(k_{-})-(k_{-})^{0}\Sigma(k_{+}),\nonumber\\
&=(k_{-})^{0}G^{-1}(k_{+})-(k_{+})^{0}G^{-1}(k_{-}).
\end{align}
As claimed in Sec.~(\ref{sec:Heat_Diagrammatic}) of the main text, the full heat vertex in Eq.~(\ref{eq:app_Full_heat_vertex}) satisfies the WTI for energy conservation [Eq.~(\ref{eq:app_energy_WTI})]. 
Note that it is crucial to include the vertex $\lambda^{\mu}_{\H}(k_{+},k_{-})$ to satisfy the WTI.

\subsection{Bosonic electromagnetic vertex}
\label{sec:Boson_Electric_Vertex}

This section presents a derivation of the bosonic electromagnetic vertex appearing in the Aslamazov-Larkin diagrams. 
The derivation is based solely on the form of the ``triangle'' vertices appearing in these diagrams, along with the constraint imposed by the Ward-Takahashi identity. 
Without loss of generality, consider the $\hat{x}$-component. Equivalent results hold for the $\hat{y}$- and $\hat{z}$-components. 
This vertex appears in Eq.~(\ref{eq:chi_approx}) of the main text.

The electromagnetic ``triangle'' vertex appearing in the $\AL^{x}_{\E,1}$ diagram is shown in Fig.~(\ref{fig:AL1_electriccurrent_diagram}). Mathematically this is given by
\begin{equation}\label{eq:AL1_EM_triangle}
\Lambda^{x}_{\E,1}(p,p)=-\sum_{l}G_{0}(p-l)G(l)\Gamma^{x}_{\E}(l,l)G(l).
\end{equation}
The minus sign arises from the fermion loop in the Aslamazov-Larkin triangle vertex.
The WTI for global particle number conservation is $q_{\mu}\Gamma^{\mu}_{\E}(k_{+},k_{-})=G^{-1}(k_{+})-G^{-1}(k_{-})$.
In the $q\rightarrow0$ limit, this produces the Ward identity: $\Gamma^{\mu}_{\E}(k,k)=\partial G^{-1}(k)/\partial k_{\mu}$. 
For the $\hat{x}$-component, it follows that $\Gamma^{x}_{\E}(l,l)=-\partial G^{-1}(l)/\partial l^{x}$. Inserting this into Eq.~(\ref{eq:AL1_EM_triangle}),
and then performing integration by parts, gives
\begin{align}\label{eq:AL1_EM_triangle1}
\Lambda^{x}_{\E,1}(p,p)&=\sum_{l}G_{0}(p-l)G(l)\frac{\p G^{-1}(l)}{\p l^{x}}G(l),\nonumber\\
&=-\sum_{l}G_{0}(p-l)\frac{\p G(l)}{\p l^{x}},\nonumber\\
&=\sum_{l}\frac{\p G_{0}(p-l)}{\p l^{x}}G(l),\nonumber\\
&=-\frac{\p}{\p p^{x}}\sum_{l}G_{0}(p-l)G(l).
\end{align}
By definition, the pair susceptibility is $\Pi(p)=\sum_{l}G_{0}(p-l)G(l)=t^{-1}(p)-g^{-1}$. Therefore, it follows that the bosonic electromagnetic vertex for $\AL^{x}_{\E,1}$ is
\begin{equation}\label{eq:AL1_EM_triangle2}
\Lambda^{x}_{\E,1}(p,p)=-\frac{\p\Pi(p)}{\p p^{x}}=-\frac{\p t^{-1}(p)}{\p p^{x}}.
\end{equation}
For comparison, the fermionic electromagnetic vertex, derived above, is $\Gamma^{x}_{\E}(k,k)=-\p G^{-1}(k)/\p k^{x}.$

\begin{figure*}[t]
\centering\includegraphics[width=6.5cm,height=4cm,clip]{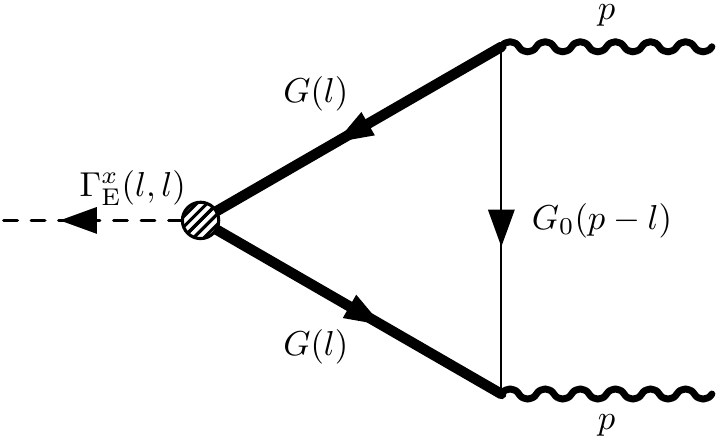}
\caption{The electromagnetic ``triangle'' vertex appearing in the $\AL^{x}_{\E,1}$ Aslamazov-Larkin diagram. 
The external momentum $p$ is the momentum of the pair-propagators. 
The triangle vertex represents a bosonic electromagnetic vertex, $\Lambda^{x}_{\E,1}(p,p)$, which is computed in the text.}
\label{fig:AL1_electriccurrent_diagram}
\end{figure*}

The electromagnetic vertex appearing in the $\AL^{x}_{\E,2}$ diagram is derived in exactly the same manner. This diagram is shown in 
Fig.~(\ref{fig:AL2_electriccurrent_diagram}). Mathematically this is given by
\begin{equation}\label{eq:AL2_EM_triangle}
\Lambda^{x}_{\E,2}(p,p)=-\sum_{l}G(p-l)G_{0}(l)\gamma^{x}_{\E}(l,l)G_{0}(l).
\end{equation}
The minus sign arises from the fermion loop in the Aslamazov-Larkin triangle vertex.
Following the same steps as before, this vertex can be simplified as follows:
\begin{align}\label{eq:AL2_EM_triangle1}
\Lambda^{x}_{\E,2}(p,p)&=\sum_{l}G(p-l)G_{0}(l)\frac{\p G_{0}^{-1}(l)}{\p l^{x}}G_{0}(l),\nonumber\\
&=-\sum_{l}G(p-l)\frac{\p G_{0}(l)}{\p l^{x}},\nonumber\\
&=\sum_{l}\frac{\p G(p-l)}{\p l^{x}}G_{0}(l),\nonumber\\
&=-\frac{\p}{\p p^{x}}\sum_{l}G(p-l)G_{0}(l).
\end{align}
The pair susceptibility can also be written as $\Pi(p)=\sum_{l}G(p-l)G_{0}(l)=t^{-1}(p)-g^{-1}$. Therefore, it follows that the bosonic electromagnetic vertex for $\AL^{x}_{\E,2}$ is
\begin{equation}\label{eq:AL2_EM_triangle2}
\Lambda^{x}_{\E,2}(p,p)=-\frac{\p\Pi(p)}{\p p^{x}}=-\frac{\p t^{-1}(p)}{\p p^{x}}.
\end{equation}
Thus, the bosonic electromagnetic vertex appearing in $\AL^{x}_{\E,1}$ and $\AL^{x}_{\E,2}$ is identical: $\Lambda^{x}_{\E,1}(p,p)=\Lambda^{x}_{\E,2}(p,p)$. It follows that the bosonic EM vertex for 
the combination of $\AL^{x}_{\E,1}+\AL^{x}_{\E,2}$ is
\begin{equation}\label{eq:AL_EM_triangle3}
\Lambda^{x}_{\E,1}(p,p)+\Lambda^{x}_{\E,2}(p,p)\equiv\Lambda^{x}_{\E}(p,p)=-2\frac{\p\Pi(p)}{\p p^{x}}=-2\frac{\p t^{-1}(p)}{\p p^{x}}.
\end{equation}

\begin{figure*}[t]
\centering\includegraphics[width=6.5cm,height=4cm,clip]{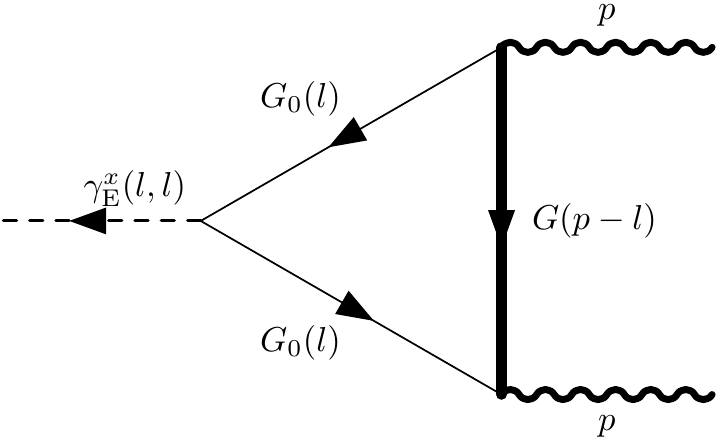}
\caption{The electromagnetic ``triangle'' vertex appearing in the $\AL^{x}_{\E,2}$ Aslamazov-Larkin diagram. The external momentum $p$ is the momentum of the pair-propagators. The triangle vertex represents a bosonic electromagnetic vertex, $\Lambda^{x}_{\E,2}(p,p)$, which is computed in the text.}
\label{fig:AL2_electriccurrent_diagram}
\end{figure*}

\subsection{Bosonic heat vertex}
\label{sec:Boson_Heat_Vertex}

The form of the bosonic heat vertex requires a more lengthy analysis, which is presented in what follows. Without loss of generality, consider the $\hat{x}$-component. Equivalent results hold for the $\hat{y}$- and $\hat{z}$-components. The heat ``triangle'' vertex for the $\AL^{x}_{\H,1}$ Aslamazov-Larkin diagram is shown in Fig.~(\ref{fig:AL1_heatcurrent_diagram}). 
Mathematically this is given by
\begin{equation}\label{eq:AL1_HE_triangle}
\Lambda^{x}_{\H,1}(p,p)=-\sum_{l}G_{0}(p-l)G(l)\Gamma^{x}_{\H}(l,l)G(l).
\end{equation}
The minus sign arises from the fermion loop in the Aslamazov-Larkin triangle vertex.
The WTI for global spacetime translation symmetry (conservation of energy) is $q_{\mu}\Gamma^{\mu}_{\H}(k_{+},k_{-})=\omega_{-}G^{-1}(k_{+})-\omega_{+}G^{-1}(k_{-})$. 
In the $q\rightarrow0$ limit, this gives the heat vertex analogue of the EM Ward identity. 
For the $\hat{x}$-component, it follows that $\Gamma^{x}_{\H}(l,l)=-\epsilon\left[\partial G^{-1}(l)/\partial l^{x}\right]$. 
Inserting this into Eq.~(\ref{eq:AL1_HE_triangle}), and then performing integration by parts, gives
\begin{align}\label{eq:AL1_HE_triangle1}
\Lambda^{x}_{\H,1}(p,p)&=\sum_{l}i\epsilon_{n}G_{0}(p-l)G(l)\frac{\p G^{-1}(l)}{\p l^{x}}G(l),\nonumber\\
&=-\sum_{l}i\epsilon_{n}G_{0}(p-l)\frac{\p G(l)}{\p l^{x}},\nonumber\\
&=\sum_{l}i\epsilon_{n}\frac{\p G_{0}(p-l)}{\p l^{x}}G(l),\nonumber\\
&=-\frac{\p}{\p p^{x}}\sum_{l}i\epsilon_{n}G_{0}(p-l)G(l).
\end{align}
Now write this in a symmetric form, by letting $l\rightarrow l+p/2$, and also $l\rightarrow -l+p/2$, and summing one half of each of the resulting expressions; this produces
\begin{align}\label{eq:AL1_HE_triangle2}
\Lambda^{x}_{\H,1}(p,p)&=-\frac{1}{2}\frac{\p}{\p p^{x}}\sum_{l}\left(i\epsilon_{n}+i\varpi_{m}/2\right)G_{0}(p/2-l)G(l+p/2)\nonumber\\
&\quad-\frac{1}{2}\frac{\p}{\p p^{x}}\sum_{l}\left(-i\epsilon_{n}+i\varpi_{m}/2\right)G_{0}(p/2+l)G(-l+p/2).
\end{align}

The heat ``triangle'' vertex for the $\AL^{x}_{\H,2}$ Aslamazov-Larkin diagram is shown in Fig.~(\ref{fig:AL2_heatcurrent_diagram}). Mathematically this is given by
\begin{equation}\label{eq:AL2_HE_triangle}
\Lambda^{x}_{\H,2}(p,p)=-\sum_{l}G(p-l)G_{0}(l)\gamma^{x}_{\H}(l,l)G_{0}(l).
\end{equation}
The minus sign arises from the fermion loop in the Aslamazov-Larkin triangle vertex.
Following the same steps as before, this vertex can be simplified as follows:
\begin{align}\label{eq:AL2_HE_triangle1}
\Lambda^{x}_{\H,2}(p,p)&=\sum_{l}i\epsilon_{n}G(p-l)G_{0}(l)\frac{\p G^{-1}_{0}(l)}{\p l^{x}}G_{0}(l),\nonumber\\
&=-\sum_{l}i\epsilon_{n}G(p-l)\frac{\p G_{0}(l)}{\p l^{x}},\nonumber\\
&=\sum_{l}i\epsilon_{n}\frac{\p G(p-l)}{\p l^{x}}G_{0}(l),\nonumber\\
&=-\frac{\p}{\p p^{x}}\sum_{l}i\epsilon_{n}G(p-l)G_{0}(l).
\end{align}
Now write this in a symmetric form, by letting $l\rightarrow l+p/2$, and also $l\rightarrow -l+p/2$, and summing one half of each of the resulting expressions; this produces
\begin{align}\label{eq:AL2_HE_triangle2}
\Lambda^{x}_{\H,2}(p,p)&=-\frac{1}{2}\frac{\p}{\p p^{x}}\sum_{l}\left(i\epsilon_{n}+i\varpi_{m}/2\right)G(p/2-l)G_{0}(l+p/2)\nonumber\\
&\quad-\frac{1}{2}\frac{\p}{\p p^{x}}\sum_{l}\left(-i\epsilon_{n}+i\varpi_{m}/2\right)G(p/2+l)G_{0}(-l+p/2).
\end{align}

\begin{figure*}[t]
\centering\includegraphics[width=6.5cm,height=4cm,clip]{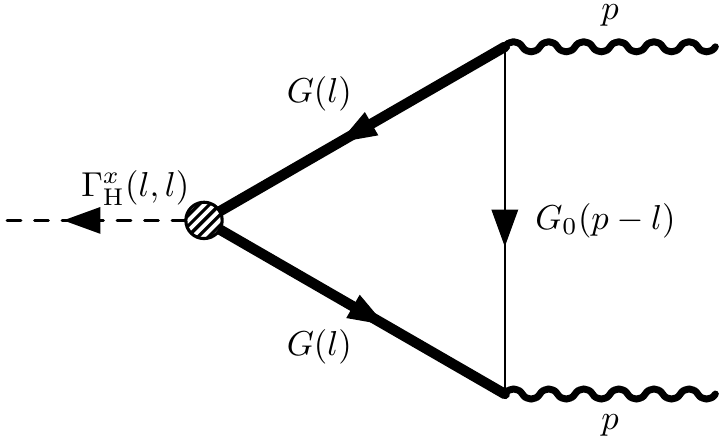}
\caption{The heat ``triangle'' vertex appearing in the $\AL^{x}_{\H,1}$ Aslamazov-Larkin diagram. The external momentum $p$ is the momentum of the pair-propagators. The triangle vertex represents a bosonic heat vertex $\Lambda^{x}_{\H,1}(p,p)$, which is computed in the text.}
\label{fig:AL1_heatcurrent_diagram}
\end{figure*}

Adding the results in Eq.~(\ref{eq:AL1_HE_triangle2}) and Eq.~(\ref{eq:AL2_HE_triangle2}) together and simplifying then produces
\begin{align}
\Lambda^{x}_{\H,1}(p,p)+\Lambda^{x}_{\H,2}(p,p)\equiv\Lambda^{x}_{\H}(p,p)
&=-\frac{i\varpi_{m}}{2}\frac{\p}{\p p^{x}}\sum_{l}G_{0}(p/2-l)G(l+p/2)\nonumber\\
&\quad-\frac{i\varpi_{m}}{2}\frac{\p}{\p p^{x}}\sum_{l}G_{0}(p/2+l)G(-l+p/2),\nonumber\\
&=-i\varpi_{m}\frac{\p}{\p p^{x}}\sum_{l}G_{0}(p-l)G(l).
\end{align}
Using the definition of the pair susceptibility, $\Pi(p)=\sum_{l}G_{0}(p-l)G(l)=t^{-1}(p)-g^{-1}$, it follows that the bosonic heat vertex for 
the combination of $\AL^{x}_{\H,1}+\AL^{x}_{\H,2}$ is
\begin{equation}\label{eq:AL_HE_triangle3}
\Lambda^{x}_{\H,1}(p,p)+\Lambda^{x}_{\H,2}(p,p)\equiv\Lambda^{x}_{\H}(p,p)=-\varpi\frac{\p\Pi(p)}{\p p^{x}}=-\varpi\frac{\p t^{-1}(p)}{\p p^{x}}.
\end{equation}
For comparison, the fermionic full heat vertex, derived above, is $\Gamma^{x}_{\H}(k,k)=-\omega\left[\p G^{-1}(k)/\p k^{x}\right].$ 
Notice that it is the sum of two heat triangle vertices which produces a bosonic heat vertex in a form similar to its fermionic counterpart. This factor of two has caused a lot of controversy in the literature~\cite{Sergeev_1994,Sergeev_2008,Ussishkin_2003,Serbyn_2009,Levchenko_2011}. The factor of two difference between the fermionic and bosonic result is due to charge, as will be explained in detail in the next paragraph.

If we restore the electric charge $e$ appearing in the EM vertex, then the relation between the fermionic heat and EM vertices is
\begin{equation}
\Gamma^{x}_{\H}(k,k)=\frac{\omega}{e}\Gamma^{x}_{\E}(k,k).
\end{equation}
The bosonic heat vertex $\Lambda^{x}_{\H}$ is defined by $\Lambda^{x}_{\H,1}(p,p)+\Lambda^{x}_{\H,2}(p,p)\equiv\Lambda^{x}_{\H}(p,p)=\varpi\left[\p\Pi(p)/\p p^{x}\right]=\varpi\left[\p t^{-1}(p)/\p p^{x}\right]$. Similarly the bosonic EM vertex $\Lambda^{x}_{\E}$ obeys $\Lambda^{x}_{\E,1}(p,p)+\Lambda^{x}_{\E,2}(p,p)\equiv\Lambda^{x}_{\E}(p,p)=2[\p\Pi(p)/\p p^{x}]=2[\p t^{-1}(p)/\p p^{x}]$. Thus, the relation between the bosonic heat and EM vertices is
\begin{equation}
\Lambda^{x}_{\H}(p,p)=\frac{\varpi}{2e}\Lambda^{x}_{\E}(p,p)=\frac{\varpi}{e^{*}}\Lambda^{x}_{\E}(p,p).
\end{equation}
Here $e^{*}=2e$, and the factor of two appears due to the composite bosons comprising of paired fermions. Thus, the fundamental relation between the heat and EM vertices is that the heat vertex equals the matter current multiplied by energy (frequency) whereas the EM vertex equals the matter current multiplied by charge; thus the heat vertex equals the EM vertex multiplied by the ratio of frequency to charge. The normalization by the corresponding charge ($e$ for fermions and $e^{*}$ for bosons) causes there to be a factor of two difference between microscopic fermions and composite bosons, because the composite bosons are formed from the pairing of two fermions and thus have charge 
$e^{*}=2e$~\cite{Comment3}. This result has also been derived independently in Refs.~\onlinecite{Narikiyo1_2011,Narikiyo2_2011}.

\begin{figure*}[t]
\centering\includegraphics[width=6.5cm,height=4cm,clip]{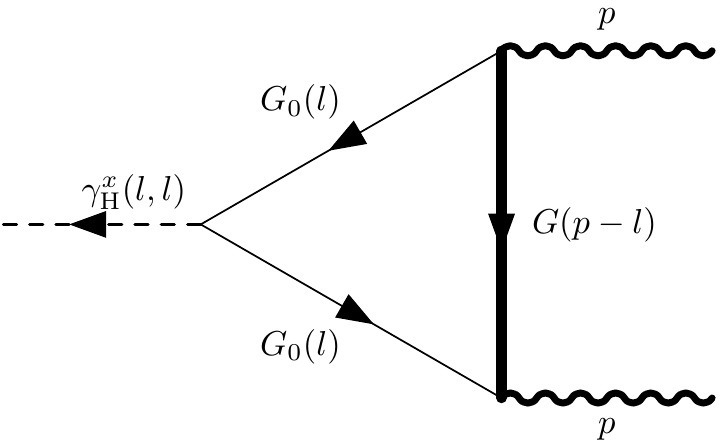}
\caption{The heat ``triangle'' vertex for the $\AL^{x}_{\H,2}$ Aslamazov-Larkin diagram. The external momentum $p$ is the momentum of the pair-propagators. The triangle vertex represents a bosonic heat vertex $\Lambda^{x}_{\H,2}(p,p)$, which is computed in the text.}
\label{fig:AL2_heatcurrent_diagram}
\end{figure*}

\subsection{Transverse thermoelectric coefficient }
\label{sec:Thermoelectric_Coefficient}

The result for the transverse thermoelectric coefficient  is given in Eq.~(\ref{eq:thermoelectric_coeff3}) of the main text. Here further details of the calculation are presented. As stated in Sec.~(\ref{sec:Heat_Diagrammatic}) of the main text, the transverse thermoelectric coefficient  is computed by performing all possible electromagnetic vertex insertions in the heat-current-electric-current correlation function. In the small $|\mu_{\pair}|$ limit, only the electromagnetic vertex insertions in the Aslamazov-Larkin diagram contribution to this correlation function are of interest~\cite{Ussishkin_2003}. The reason for this is similar to what occurs in the case of the diamagnetic susceptibility, which in the small $|\mu_{\pair}|$ limit has a singular contribution arising from only the Aslamazov-Larkin diagrams. Finally, only the electromagnetic vertex insertions in the pair-propagator need to be considered. Again, the resulting correlation function gives the most singular contribution because it contains three pair propagators~\cite{Ussishkin_2003}. For a given pair propagator, there are two electromagnetic triangle vertices that can be inserted into the propagator. Since the $\AL^{x}_{\E,1}$ and $\AL^{x}_{\E,2}$ electromagnetic vertices are equivalent to the bosonic electromagnetic vertex $\Lambda^{x}_{\E,1}$ [see Eq.~(\ref{eq:AL1_EM_triangle2}) and Eq.~(\ref{eq:AL2_EM_triangle2})], these two insertions give a symmetry factor of two. There is also the symmetry factor of two arising from spin-degeneracy for a spin-$\tfrac{1}{2}$ system of fermions. Finally, the heat vertex for the combination of $\AL^{y}_{\H,1}$ and $\AL^{y}_{\H,2}$ reduces to the bosonic heat vertex $\Lambda^{y}_{\H}$. Thus, the bosonic three-point correlation function has a total symmetry factor of four. This symmetry factor  of four can be absorbed into the EM vertices using $\Lambda_{\E}=2\Lambda_{\E,1}$. The correlation function now becomes
\begin{align}\label{eq:boson_response}
\Lambda^{yyx}(i\Omega_{m},Q)&=
-e^2\sum_{p}\biggl[\Lambda^{y}_{\H}(i\varpi_{m}+i\Omega_{m},\mb{p}_{+};i\varpi_{m},\mb{p}_{-})\Lambda^{y}_{\E}(i\varpi_{m},\mb{p}_{-};i\varpi_{m},\mb{p}_{+})\Lambda^{x}_{\E}(i\varpi_{m},\mb{p}_{+};i\varpi_{m}+i\Omega_{m},\mb{p}_{+})\nonumber\\
&\quad\quad\quad\quad\times t(i\varpi_{m}+i\Omega_{m},\mb{p}_{+})t(i\varpi_{m},\mb{p}_{-})t(i\varpi_{m},\mb{p}_{+})\nonumber\\
&\hspace{1.35cm}+\Lambda^{y}_{\H}(i\varpi_{m}-i\Omega_{m},\mb{p}_{-};i\varpi_{m},\mb{p}_{+})\Lambda^{y}_{\E}(i\varpi_{m},\mb{p}_{+};i\varpi_{m},\mb{p}_{-})
\Lambda^{x}_{\E}(i\varpi_{m},\mb{p}_{-};i\varpi_{m}-i\Omega_{m},\mb{p}_{-})\nonumber\\
&\quad\quad\quad\quad\times t(i\varpi_{m}-i\Omega_{m},\mb{p}_{-})t(i\varpi_{m},\mb{p}_{+})t(i\varpi_{m},\mb{p}_{-})\biggr].
\end{align}
Here the vertex notation is defined by $\Lambda^{x}(p+Q+i\Omega_{m},p)\equiv\Lambda^{x}(i\varpi_{m}+i\Omega_{m},\mb{p}+\mb{Q};i\varpi_{m},\mb{p})$, and $\mb{p}_{\pm}\equiv
\mb{p}\pm\mb{Q}/2$. The vector $\mb{Q}$ is along the $\hat{x}$-direction: $\mb{Q}=Q\hat{\mb{x}}$. The diagrams for this bosonic three-point correlation function are shown in Fig.~(\ref{fig:Thermoelectric_diagram}).

\begin{figure*}[t]
\centering\includegraphics[width=16cm,height=5cm,clip]{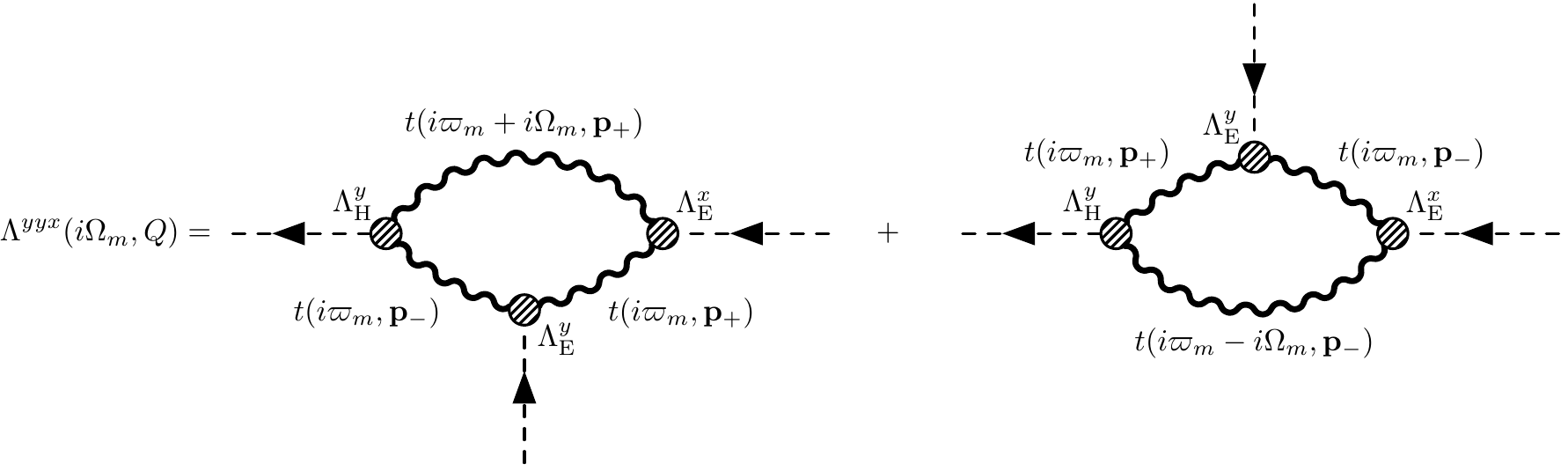}
\caption{The Aslamazov-Larkin diagrams that give the singular contribution to the transverse thermoelectric coefficient . 
The vertices $\Lambda_{\E}$ and $\Lambda_{\H}$ represent bosonic electromagnetic and heat vertices, respectively. 
The bosonic vertices have been computed in appendix~(\ref{sec:Boson_Electric_Vertex}) and appendix~(\ref{sec:Boson_Heat_Vertex}) from the triangle vertices in the Aslamazov-Larkin diagrams.}
\label{fig:Thermoelectric_diagram}
\end{figure*}

The bosonic electromagnetic and heat vertices are given in Eq.~(\ref{eq:AL_EM_triangle3}) and Eq.~(\ref{eq:AL_HE_triangle3}), respectively. Using these results to express the bosonic vertices in Eq.~(\ref{eq:boson_response}) then gives
\begin{align}
\Lambda^{yyx}(i\Omega_{m},Q)&=-4e^2\sum_{p}\biggl[\frac{Zp^{x}_{+}}{M_{\pair}}\left(\frac{Zp^{y}}{M_{\pair}}\right)^{2}
\left(i\varpi_{m}+i\Omega_{m}/2\right)t(i\varpi_{m}+i\Omega_{m},\mb{p}_{+})t(i\varpi_{m},\mb{p}_{-})t(i\varpi_{m},\mb{p}_{+})\nonumber\\
&\hspace{1.55cm}+\frac{Zp^{x}_{-}}{M_{\pair}}\left(\frac{Zp^{y}}{M_{\pair}}\right)^{2}\left(i\varpi_{m}-i\Omega_{m}/2\right)
t(i\varpi_{m}-i\Omega_{m},\mb{p}_{-})t(i\varpi_{m},\mb{p}_{+})t(i\varpi_{m},\mb{p}_{-})\biggr].
\end{align}
The Matsubara frequency summation is performed by using the Eliashberg contour~\cite{VarlamovBook, Mahan, Ussishkin_2003} [see appendix~(\ref{sec:Diamagnetism_Approx_Matsubara}) for details]. After performing the Matsubara frequency summation, and analytically continuing to real frequencies: $i\Omega_{m}\rightarrow\Omega+i0^{+}$, the result is 
\begin{align}\label{eq:Lambda_function}
\Lambda^{yyx}(\Omega,Q)&=-4e^2\sum_{\mb{p}}\int_{-\infty}^{\infty}\frac{dx}{2\pi}\ \coth\left(\frac{1}{2}\beta x\right)
\biggl\{\frac{Zp^{x}_{+}}{M_{\pair}}\left(\frac{Zp^{y}}{M_{\pair}}\right)^{2}\left(x+\Omega/2\right)
t_{\R}(x+\Omega,\mb{p}_{+})\Im\left[t_{\R}(x,\mb{p}_{-})t_{\R}(x,\mb{p}_{+})\right]\nonumber\\
&\quad+\frac{Zp^{x}_{+}}{M_{\pair}}\left(\frac{Zp^{y}}{M_{\pair}}\right)^{2}\left(x-\Omega/2\right)
t_{\A}(x-\Omega,\mb{p}_{+})t_{\A}(x-\Omega,\mb{p}_{-})\Im\left[t_{\R}(x,\mb{p}_{+})\right]\nonumber\\
&\quad+\frac{Zp^{x}_{-}}{M_{\pair}}\left(\frac{Zp^{y}}{M_{\pair}}\right)^{2}\left(x-\Omega/2\right)
t_{\A}(x-\Omega,\mb{p}_{-})\Im\left[t_{\R}(x,\mb{p}_{+})t_{\R}(x,\mb{p}_{-})\right]\nonumber\\
&\quad+\frac{Zp^{x}_{-}}{M_{\pair}}\left(\frac{Zp^{y}}{M_{\pair}}\right)^{2}\left(x+\Omega/2\right)
t_{\R}(x+\Omega,\mb{p}_{-})t_{\R}(x+\Omega,\mb{p}_{+})\Im\left[t_{\R}(x,\mb{p}_{-})\right]\biggr\}.
\end{align}
The transverse thermoelectric coefficient  can now be computed using the Kubo formula:
\begin{equation}\label{eq:app_thermoelectric_coeff1}
\frac{j_{y}}{EB}=-\underset{\Omega,Q\rightarrow0}{\mathrm{lim}}\frac{1}{\Omega Qc}\mathrm{Re}
\left[\left.\Lambda^{yyx}(\Omega,Q)\right|_{i\Omega_{m}\rightarrow\Omega+i0^{+}}\right].
\end{equation}
Inserting the retarded pair-propagator, defined by $t^{-1}_{\R}(x,\mb{p})=Z\left[\kappa x-\mb{p}^{2}/(2M_{\pair})-|\mu_{\pair}|+i\Gamma x\right]$, into Eq.~(\ref{eq:Lambda_function}), and then taking the limits $Q\rightarrow0$ followed by $\Omega\rightarrow0$ in Eq.~(\ref{eq:app_thermoelectric_coeff1}) gives $j_{y}/EB$. In the small $|\mu_{\pair}|$ limit, the main contribution to the integral occurs 
when $\beta x\ll 1$, which allows the coth function to be expanded as $\coth(\beta x/2)\approx2T/x$. In this limit, the current becomes
\begin{align}
\frac{j_{y}}{EB}&=\frac{4Te^2}{c}\sum_{\mb{p}}\left(\frac{Zp^{x}}{M_{\pair}}\right)^{2}\left(\frac{Zp^{y}}{M_{\pair}}\right)^{2}\int_{-\infty}^{\infty}\frac{dx}{\pi}\frac{1}{x}\left[\Re\left(t^{3}_{\R}(x,\mb{p})\right)\Im\left(t_{\R}(x,\mb{p})\right)-\Im\left(t^{3}_{\R}(x,\mb{p})\right)\Re\left(t_{\R}(x,\mb{p})\right)\right],\nonumber\\
&=\frac{4Te^2}{c}\sum_{\mb{p}}\left(\frac{p^{x}}{M_{\pair}}\right)^{2}\left(\frac{p^{y}}{M_{\pair}}\right)^{2}\int_{-\infty}^{\infty}\frac{dx}{\pi}\frac{1}{x}\frac{2\Gamma x\left(\kappa x-\mb{p}^2/2M_{\pair}-|\mu_{\pair}|\right)}{\left[\left(\kappa x-\mb{p}^2/(2M_{\pair})-|\mu_{\pair}|\right)^{2}+\left(\Gamma x\right)^{2}\right]^{3}},\nonumber\\
&=\frac{-3Te^2}{c}\left(\frac{\kappa^2+\Gamma^2}{\Gamma^2}\right)\sum_{\mb{p}}\left(\frac{p^{x}}{M_{\pair}}\right)^{2}\left(\frac{p^{y}}{M_{\pair}}\right)^{2}\frac{1}{\left(\mb{p}^2/(2M_{\pair})+|\mu_{\pair}|\right)^4},\nonumber\\
&=\frac{-2Te^2}{c}\left(\frac{\kappa^2+\Gamma^2}{\Gamma^2}\right)\sum_{\mb{p}}\frac{1}{\left(\mb{p}^2+2M_{\pair}|\mu_{\pair}|\right)^2}.
\end{align}
The momentum integral is the same as that performed in Eq.~(\ref{eq:app_chi_dia_AL0}). Using that result for the momentum integration 
then gives the result in Eq.~(\ref{eq:thermoelectric_coeff2}) of the main text:
\begin{equation}\label{eq:app_thermoelectric_coeff2}
\frac{j_{y}}{EB}=-\frac{Te^2}{4\pi\hbar^2 c}\sqrt{\frac{1/(2M_{\pair})}{\left|\mu_{\pair}\right|}}
\,\left(\frac{\kappa^2+\Gamma^2}{\Gamma^2}\right).
\end{equation}
The constants $\hbar$ and $c$ have been restored in this expression. The transverse thermoelectric coefficient  is determined from 
$\widetilde{\alpha}_{xy}=B\left[c\chi_{\dia}/\hbar-j^{y}/(EB)\right]$; using Eq.~(\ref{eq:app_chi_dia_AL}) and  Eq.~(\ref{eq:app_thermoelectric_coeff2}) then gives the result stated in Eq.~(\ref{eq:thermoelectric_coeff3}) of the main text:
\begin{equation}
\widetilde{\alpha}_{xy}=\frac{BTe^2}{12\pi\hbar^2 c}\sqrt{\frac{1/(2M_{\pair})}{\left|\mu_{\pair}\right|}}\,\left(\frac{3\kappa^2+\Gamma^2}{\Gamma^2}\right).
\end{equation}
\end{widetext}

\bibliography{Review}

\begin{thebibliography}{57}
\bibitem [{Com({\natexlab{a}})}]{Comment1}%
  \BibitemOpen
  \href@noop {} {}\bibinfo {note} {There has been some
  misunderstanding that applying the BCS-BEC crossover picture to the cuprates
  implies a negative fermionic chemical potential. We emphasize that this is
  not true. In fact, the BEC regime cannot be reached in $d$-wave
  superconductors except at unphysical small electron densities, as discussed
  in Ref.~\protect\onlinecite{Chen_99}.}\BibitemShut {Stop}%
\bibitem [{Com({\natexlab{c}})}]{Comment3}%
  \BibitemOpen
  \href@noop {} {}\bibinfo {note} {In the expression given
  for the response function, the electron charge $e$ has been explicitly
  incorporated [see Eq.~(\ref{eq:Response_Fermion})]. Thus, the full vertex
  $\Gamma^{\mu}_{\E}(k_{+},k_{-})$ is in fact the matter-current vertex.
  Technically the EM vertex is given by $e\Gamma^{\mu}_{\E}(k_{+},k_{-})$. It
  is convenient to explicitly put the charge in the response function, however,
  one must then be aware that when we refer to the ``EM" vertex
  $\Gamma^{\mu}_{\E}$ it is in fact the matter-current vertex. When the heat
  vertex $\Gamma^{\mu}_{\H}$ is also considered, the relation
  $\Gamma^{y}_{\H}(k,k)=\omega\Gamma^{y}_{\E}(k,k)$ states that the
  heat-current vertex is equal to energy multiplied by the matter-current
  vertex. Combining these two relations, it follows that the relation between
  the heat-current vertex and the actual EM vertex is thus
  $\Gamma^{y}_{\H}(k,k)=(\omega/e)\Gamma^{y}_{\E}(k,k)$. Incorporating the
  charge here is important, due to factors of two that have caused confusion in
  the literature. For fermions it is found that
  $\Gamma^{y}_{\H}(k,k)=(\omega/e)\Gamma^{y}_{\E}(k,k)$, whereas for the
  composite bosons considered in this work
  $\Lambda^{y}_{\H}(p,p)=(\varpi/e^{*})\Lambda^{y}_{\E}(p,p)$, where
  $e^{*}=2e$. The relation between the heat-current vertex and the
  electromagnetic-current vertex is thus a universal relation, however, the
  explicit charge appearing in this relation is dependent on the particle
  constituents, i.e., whether they are fundamental particles or composite
  particles.}\BibitemShut {Stop}%
\bibitem [{not()}]{noteonPauli}%
  \BibitemOpen
  \href@noop {} {}\bibinfo {note} {In the calculation of $\chi_{\Pauli}$, the
  free electron mass $m$ is replaced by the effective band mass of the
  cuprates. This effect leads to an extra factor of roughly 0.3 in
  $\chi_{\Pauli}$.}\BibitemShut {Stop}%
\bibitem [{Com({\natexlab{e}})}]{Comment4}%
  \BibitemOpen
  \href@noop {} {}\bibinfo {note} {At this critical doping
  the pair mass becomes infinitely heavy, leading to a quartic ($q^4$) pair
  dispersion, which suppresses both $T_c$ and $T_\chi$ down to
  zero.}\BibitemShut {Stop}%
\end{thebibliography}
\end{document}